\documentclass{aa}                             
\usepackage[varg]{txfonts}
\usepackage{graphicx}

\usepackage{natbib}

\newcommand{\kms} {km\,s$^{-1}$}

\newcommand{\logq} {log\,{\em $Q$}}
\newcommand{\grav} {log\,{\em $g$}}

\newcommand{\Mdot}{\ensuremath{\dot{M}}}
\newcommand{\Rstar}{\ensuremath{R_\ast}}
\newcommand{\vinf}{\ensuremath{v_{\infty}}}
\newcommand{\vmic}{\ensuremath{v_{\rm mic}}}
\newcommand{\vmac}{\ensuremath{V_{\rm mac}}}
\newcommand{\vesc}{\ensuremath{v_{\rm esc}}}
\newcommand{\Msun}{\ensuremath{M_\odot}}
\newcommand{\Rsun}{\ensuremath{R_\odot}}
\newcommand{\Teff}{\ensuremath{T_{\rm eff}}}
\newcommand{\vsini}{\ensuremath{V\sin i}}
\newcommand{\fcl}{\ensuremath{f_{\rm cl}}}
\newcommand{\fclmax}{\ensuremath{f_{\rm cl}^{\rm max}}}
\newcommand{\yhe}{\ensuremath{Y_{\rm He}}}
\begin{document}

%
\title{To clump or not to clump}

\subtitle{The impact of wind inhomogeneities on the optical and NIR spectroscopic
analysis of massive OB stars
} 

\author{K.~R\"ubke\inst{1,2}, A.~Herrero\inst{3,4}, J.~Puls\inst{5}}
             \institute{
             Departamento de F\'isica, Ingeniería de Sistemas y Teoría de la Señal, Universidad de Alicante, Carretera de San Vicente s/n, E03690, San Vicente del Raspeig
             \and
             Departamento de F\'isica Aplicada, Facultad de Ciencias, Universidad de Alicante, Carretera de San Vicente s/n, E03690, San Vicente del Raspeig, Spain
             \and
             Instituto de Astrof\'isica de Canarias, 38206 La Laguna, Tenerife, Spain.
             \and
             Departamento de Astrof\'isica, Universidad de La Laguna, E-38205 La Laguna, Tenerife, Spain.
             \and
             LMU M\"unchen, Universit\"ats-Sternwarte, Scheinerstr. 1, 81679, M\"unchen, Germany           
             }
                   
\offprints{ahd@iac.es}

\date{Date}

\titlerunning{To clump or not to clump}
\authorrunning{R\"ubke et al.}


\abstract
{Winds of massive stars have density inhomogeneities (clumping) that
may affect the formation of spectral lines in
different ways, depending on their formation region. Most of
previous and current spectroscopic analyses have been
performed in the optical or ultraviolet domain. However, massive
stars are often hidden behind dense clouds rendering near-infrared 
observations necessary. It is 
thus inevitable to compare the results
of such analyses and the effects of clumping in the
optical and the near-infrared, where lines share most of the line
formation region.} 
{ Our objective is to investigate whether a spectroscopic
analysis using either optical or infrared observations results in
the same stellar parameters with comparable accuracy, and whether
clumping affects them in different ways.}
{We analyzed optical and near-infrared observations of a set of massive
O stars with spectral types O4-O9.5 and all luminosity classes. We used
{\sc Fastwind} model atmospheres with and without
optically thin clumping. We first studied the
differences in the stellar parameters derived from
the optical and the infrared using unclumped models.
Based on a coarse model grid,
different clumping stratifications were tested.
A subset of four linear clumping laws was selected to study the
differences in the stellar parameters derived from
clumped and unclumped models, and from the optical and the infrared
wavelength regions.} 
{We obtain similar stellar parameters in the optical and the infrared,
although with larger uncertainties in the near-infrared, both with and
without clumping, albeit with some individual deviating cases. We
find that the inclusion of clumping improves the fit to H$_\alpha$ or
\ion{He}{ii} 4686 in the optical for supergiants, as well as that of
Br$_\gamma$ in the near-infrared, but it sometimes worsens the fit to
\ion{He}{ii} 2.18$\mu$m. Globally, there are no
significant differences when using the clumping laws tested in this
work. We also find that the high-lying Br lines in the infrared should
be studied in more detail in the future.}
{The infrared can be used for spectroscopic analyses, giving similar
parameters as from the optical, though with
larger uncertainties. The best fits to different lines are obtained
with different (linear) clumping laws, indicating that the wind
structure may be more complex than adopted in the
present work. No clumping law results in a better
global fit, or improves the consistency between optical and infrared
stellar parameters. Our work shows that the optical and infrared lines
are not sufficient to break the dichotomy between the mass-loss rate and
clumping factor.}
%
\keywords{Stars: early-type -- Stars: mass-loss}

%
\maketitle
%
%


\section{Introduction}

The evolution of massive stars is an intricate subject. These relatively
scarce objects evolve through various and sometimes extreme stages
such as blue super- and hypergiants, luminous blue variables,
Wolf-Rayet stars, and red supergiants, reaching (in most cases) their maximum
luminosity when dying as supernovae before becoming compact objects
such as neutron stars and black holes, or just a diffuse remnant
evidencing the explosion \citep{Langer12}. Moreover, they are usually born in double or multiple systems \citep{Sana12} whose components may interact along their evolution, adding new possibilities to the evolutionary  zoo: stars that have
been spun up, stars stripped from their outer layers, stars that have
been violently ejected from their system and travel through space as
walk- or runaways, high-mass X-ray and $\gamma$-ray binaries, or
combinations of neutron stars and black holes in binary systems that
may emit gravitational waves \citep[e.g.,][]{deMink13, Gotberg18,
Renzo19, Langer20, Bodensteiner20, Sander19, Abbott22}

Being powerful sources of energy and matter, these stars have a strong
impact on their surroundings and even on their host galaxy, whose
chemical and mechanical evolution is affected. Moreover, our
interpretation of the spectra or the population diagrams of the host
galaxy depends on our correct understanding of its present
and past massive star population \citep{Wang20, Menon21}.

Advances in our modeling of the different evolutionary stages require
that the physical parameters of the stars are accurately known, which
means correctly modeling the main relevant processes that dominate the
evolution is necessary. It has long been realized that the process of mass loss has
a strong impact on the evolution of these stars from the early phases
onward \citep{Chiosi86}. Thus accurate knowledge of their mass-loss
rates is crucial. For hot stars, the dominant mechanism
producing the stellar wind is the scattering and absorption of energetic
photons via spectral line transitions, and the corresponding momentum
transfer onto the stellar plasma. The  line-radiation-driven wind
theory \citep{CAK75, Pauldrach1986} has been quite successful in
explaining how mass is driven away from the stellar surface by the
radiation field. The actual size of the mass-loss rate, however, is
still debated to date, and there might be uncertainties within a
factor of about three, with significant discrepancies regarding the
derived values when using different diagnostic tools 
\citep[e.g.,][]{Fullerton06}

The main reason for  these uncertainties (at least in the earlier
phases of massive stellar evolution) is the wind structure. Because of
the intrinsic instability of the line-driving process -- the so-called
line deshadowing instability (LDI: e.g., \citealt{ORI, Feldmeier95,
Sundqvist2013}, and already \citealt{LS1970})--, the stellar wind is
predicted to deviate from homogeneity. Most likely, it
is strongly structured, forming clumps of high density separated by an
interclump medium which is rarefied or even almost void. The effect of
this structure on the line profiles used as diagnostic tools is
different for resonance lines (usually observed in the ultraviolet,
and with an opacity that depends linearly on density) and for
recombination lines (usually observed in the optical or near-infrared,
with an opacity that depends on density quadratically). In addition,
and due to the Doppler effect, the spatial distribution of the
velocity plays also a role in allowing photons to escape
\citep[``vorosity'' effect,][]{owocki08}. 

This density structure, or clumping, is currently modeled within two
flavors of approximation. In the first one, known as micro- or
optically thin clumping, and firstly implemented (in
its current description) into a non-local thermodynamic equilibrium  (NLTE) 
atmosphere code by
\citet{Schmutz1995}, the light interacts with the wind-plasma only
within the overdense clumps, which are adopted to be optically thin for {all} considered processes. This assumption is usually
justified for recombination line processes such as H$_\alpha$ in not
too dense winds. In the alternative approximation, known as macro- or
optically thick clumping (see \citealt{OGS04, Oskinova07, owocki08,
Surlan13, SPI, SPII, SPIII}), the actual optical depth of the clumps
for the considered process is (or needs to be) accounted for; for example,
even if a clump may be optically thin in H$_\alpha$, it is most likely
optically thick for a (UV) resonance line. 

In the optically thick case, the light is affected by porosity
effects (both in physical space for continua and in velocity space
for lines), which usually allow for increased photon escape
through the interclump medium\footnote{a very instructive 
illustration can be found in \citet{Brands22}}. 
Compared to the average
opacity resulting from the assumption of optically thin clumping, the
effective opacity in optically thick clumps decreases\footnote{though
on an absolute scale, the effective opacity also increases with
increasing absorber density, until a certain saturation threshold is
reached \citep{OGS04, SPI, SPII}}, leading to potential de-saturation
effects, particularly in UV resonance lines \citep{Oskinova07}. 
Moreover, in such a situation, a non-void interclump medium also plays
a decisive role, not only for opening porosity channels, but also for
providing additional opacity to allow for saturated UV resonance lines
which would otherwise become (in contrast to observations)
desaturated \citep{SPI}.

Clumping has a severe effect on the derived mass-loss
rates. When recombination lines are used as diagnostics, their
emission (and absorption) increases in the clumps with the square of
the density.\ In addition, since the average of the square is larger than the
square of the average, the actual mass-loss rate is { lower} than
the one obtained when adopting a homogeneous medium. When resonance
lines are used, the effect of over- and underdensities (almost)
cancels out the microclumping approach, and the derived mass-loss rate
remains unaffected. When, for resonance lines, optically thick
clumping is accounted for, the actual mass-loss rate may be {larger} than the one obtained from both a micro-clumped and a
homogeneous medium.

The distribution of clumping as a function of distance from the star
or velocity (which is usually adopted to increase monotonically, but
see \citealt{Sander23}) has been studied by several authors using
different diagnostics that probe different wind regions, broadly
moving to longer wavelengths to probe outer regions
\citep[e.g.,][]{Puls06, Najarro2011, Bouret12, Rubio22}. They agree
that clumping starts close to the photosphere and increases up to a
maximum, remaining constant or decreasing in the intermediate and
outermost regions. The degree of clumping, that is the maximum contrast
between the density in the clumps and the density in an homogeneous
medium with the same mean density, has also been studied by these and
other authors \citep[e.g.,][]{Hawcroft21, Brands22} with values that
range from three to 20 for Galactic stars, or at least for stars with high
mass-loss rates (when analyzing lower-metallicity stars, \citealt{Brands22}).

Massive stars are often hidden behind dense clouds of gas and dust,
either local to them and their star-forming regions, or as a result of
the accumulated matter in their direction. Therefore, it is often
necessary to observe them in the near-infrared (NIR), where extinction
is less severe than in the optical. This is particularly true for our
Galaxy, where the high extinction in the Galactic Plane hides a
significant number of massive stars, rendering NIR observations a key
tool for their study.

In this paper we aim to study the effect of clumping onto the
stellar parameter determination when using optical  or NIR diagnostic
lines, as well as the consistency of the parameters obtained from the
two wavelength domains. Our study has been done in the approximation of
micro-clumping, as recent studies have shown that macro-clumping has
no significant effect on the recombination lines \citep{Sundqvist2018,
Hawcroft21, Brands22}. Moreover, we used different clumping
distributions that have been proposed in the literature. To this end,
we analyzed Galactic O-type stars with spectral types O4-O9.5 and
luminosity classes from I to V. The stars have been observed in the
optical and infrared with a high resolving power and high signal-to-noise ratio (S/N).

We present the data used for our study in Sect.~\ref{Data}, and our
methodology in Sect.~\ref{method}. In Sect.~\ref{nocl}, we explain how we derived the
stellar parameters when adopting a homogeneous wind, both in the
optical and the NIR. In Sect.~\ref{sec_clump}, we explore the
effects of clumping on the stellar parameters, using different clumping
distributions on a test model grid. In Sect.~\ref{FWgrid_clump}, we
analyze the observed stars again, now with clumping.
Sect.~\ref{discussion} discusses the impact of clumping on the
analysis results. Conclusions are presented in
Sect.~\ref{conclusions}.


\section{The data}
\label{Data}

\begin{figure*}[h!]
\centering 
\includegraphics[width=16cm,angle=0]{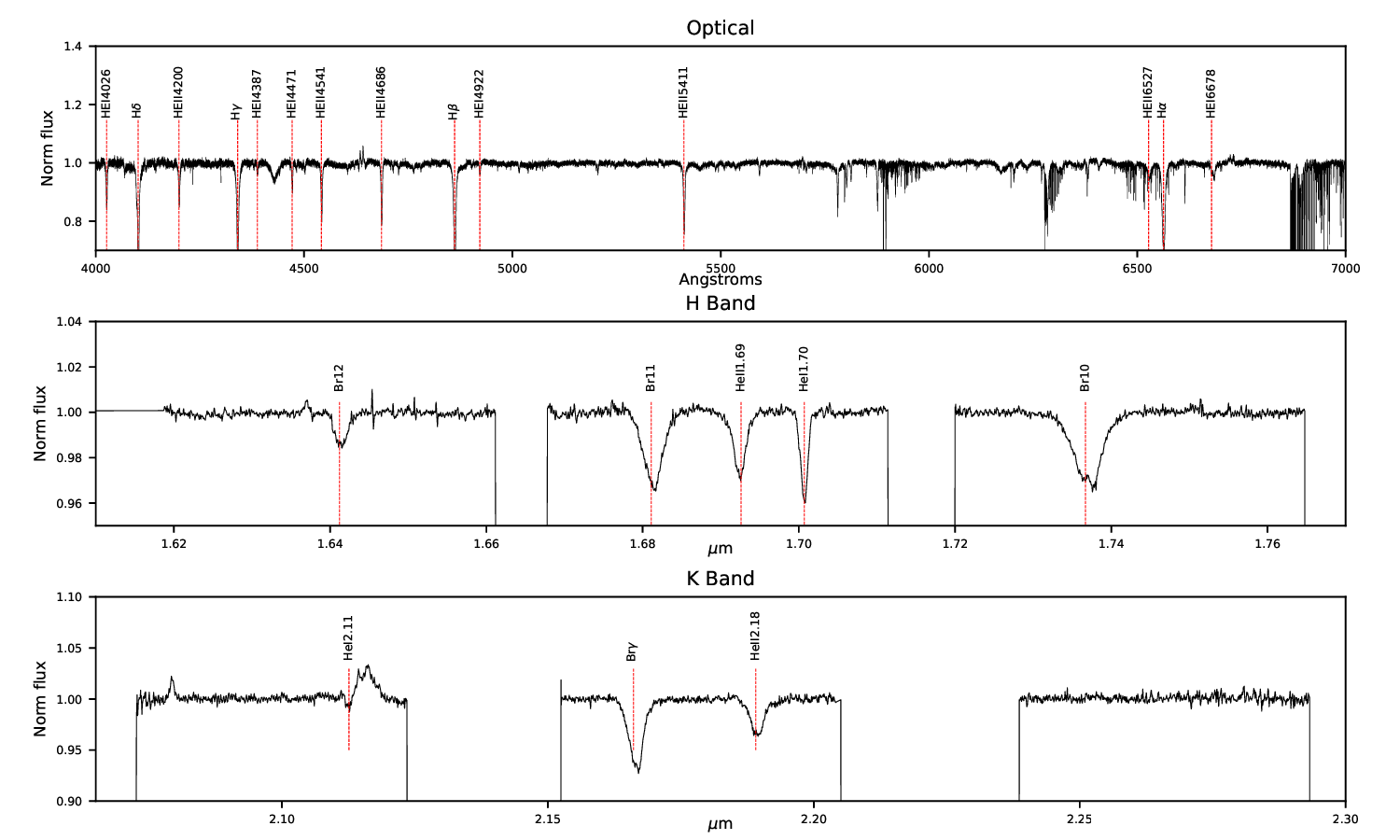}
\caption{Example for the spectra available in our sample
(HD\,46150). Upper panel: optical spectrum from the IACOB database;
lower panels: NIR spectra from the  \cite{Hanson2005} catalog.}
\label{fig:exampleHD46150}
\end{figure*}

For our work, we selected those O stars from the NIR catalog by
\cite{Hanson2005} that were also present in the IACOB Spectroscopic
Database \citep{Ssimon2011} at the beginning of our project (see Table
\ref{tbl:stars_sample}). The Hanson et al. spectra were obtained with
the Infra-Red Camera and Spectrograph (IRCS) mounted at the Cassegrain
focus of the 8.2 m Subaru Telescope at Mauna Kea, Hawaii, in the H and
K bands with a resolving power R$\sim$12\,000 and signal-to-noise
S/N$\sim$200-300. They cover specific regions of the H and K bands: 
1.618--1.661, 1.667--1.711, 1.720--1.765, 2.072--2.123, 2.152--2.205, 
2.238--2.293 and 2.331--2.388 $\mu$m. 
Although the wavelength coverage is not complete,
the main H and He lines in the NIR are present. IACOB spectra were
obtained with the Fibre-fed Echelle Spectrograph (FIES) attached to
the Nordic Optical Telescope (NOT) with S/N $\geq$ 150 and $R
\sim46\,000$, covering the full range from 3710 to 7270 \AA. Details
on the observations and data reduction can be found in the references
above. The sample covers the range of O spectral types from O4 to O9.5
and all luminosity classes. According to \cite{Gholgado2022}, most
stars show line profile variations, but only one is classified as SB1
(HD 30614, $\alpha$ Cam). Thus we assume that the spectra are 
not significantly contaminated by companions. Although some spectral types are
under-represented (like mid-type supergiants or cool late-type dwarfs),
the sample as a whole provides a good testbed for the global behavior
of O-type stars (see Fig.~\ref{fig:exampleHD46150} for an example of
the available data).

\begin{table}
\begin{center}
\caption{O stars selected for the analysis. Spectral types are from
the Galactic O Star Catalog (GOSC, \cite{JMaiz2013}, accessible at
https://gosc.cab.inta-csic.es/gosc.php).
The last column displays the variability classification
by \cite{Gholgado2022}: Line Profile Variations (LPV), Single
Spectroscopic Binary (SB1) or no evidence of radial velocity
variations (--)} 
\label{tbl:stars_sample}
\begin{tabular}{ rlll } \hline \hline
$\#$ & STAR ID   &Spectral Type & Variability \\ \hline 
1 & HD46223   & O4 V((f))    & LPV  \\   
2 & HD15629   & O4.5 V((fc)) &   LPV  \\   
3 & HD46150   & O5 V((f))z   & LPV \\   
4 & HD217086  & O7 Vnn((f))z  & -- \\   
5 & HD149757  & O9.2 IVnn    &  LPV  \\  
6 & HD190864  & O6.5 III(f)  & --  \\   
7 & HD203064  & O7.5 IIIn((f)) &  LPV \\   
8 & HD15570   & O4 If       & -- \\   
9 & HD14947   & O4.5 If     & LPV  \\   
10 & HD30614   & O9 Ia      & SB1  \\   
11 & HD210809  & O9 Iab    & LPV  \\   
12 & HD209975  & O9.5 Ib   & LPV  \\    
 \hline \hline
\end{tabular} 
\end{center}
\end{table}

\section{Methodology}
\label{method}

To determine optical and NIR parameters, we use two main tools:
a full grid of synthetic optical and near-infrared spectra, and an
automatic tool that allows us to determine the parameters for a large
sample of stars. We generate the first one using the code {\sc Fastwind} \citep[version 10.1]{Puls2005}, covering the range of
massive OB star parameters, with a grid of $\sim$100\,000 models detailed
below. To create this grid of models, we used the distributed computation system 
HTCondor\footnote{http://research.cs.wisc.edu/htcondor/. The
supercomputer facility HTCondor@Instituto de Astrofisica de Canarias
consists of a cluster of 914 cores, each capable of running in
parallel, enabling us to create a full grid of models within
roughly one week.}.  The second ingredient is \texttt{iacob\_gbat}
(\citealt{Ssimon2011, Sabin2014}, and \citealt{Holgado2018}, Appendix
A), an automatic tool that allows us to fit the observed spectrum,
returning the stellar/wind parameters corresponding to the
best-fitting model (as defined by the methodology described in
Sect.~\ref{sec:gbat}). Since our version of this algorithm has been
designed for the optical range, we needed to expand it to the NIR.

\subsection{A model grid for optical/NIR FASTWIND analyses}
\label{sect_model_grid}

The NLTE, line-blanketed and unified model atmosphere code {\sc Fastwind} 
requires as input the atmospheric parameters. I.e., for the
description of the photosphere, we have to provide effective
temperature, \Teff, gravity, $\log g$, radius, \Rstar, microturbulent
velocity, \vmic, and surface abundances. Wind parameters are mass-loss
rate, \Mdot, terminal velocity, \vinf, and the exponent $\beta$ of the
canonical $\beta$-velocity law, as well as a description of the
inhomogeneous wind structure (``clumps''). Since for the considered
parameter space, all investigated features remain optically thin in the
clumps \citep{Sundqvist2018},
we need to provide ``only'' the spatial stratification of the clumping
factor\footnote{under the simplifying assumption of a void interclump
medium, the inverse of the volume-filling factor}, $f_{\rm cl}$, that
describes the overdensities of the clumps with respect to the average
wind density. Setting $f_{\rm cl}$ to unity everywhere results in a
smooth wind model.

It is obvious that the combination of all these parameters would
result in a huge amount of models. To reduce that number, we constrain
the stellar radius and the terminal velocity (from $\vesc$, see
\citealt{Kudritzki2000}) using prototypical values
\citep[see][]{Holgado2018}, and calculate the mass-loss rate from the
condition that the wind strength parameter (or optical depth
invariant), $Q = \dot M/(\Rstar \vinf)^{3/2}$, results in one of the
grid-values as denoted in Table \ref{tbl:range_grid} 
for which the units are \Msun\,${\rm a}^{-1}$ for $\Mdot$,  \kms for \vinf, and \Rsun~for $\Rstar$.
The quantity $Q$ combines mass-loss rate, stellar radius, and wind terminal velocity in
such a way that the emission in H$_\alpha$ (and other wind diagnostics
lines, as long as recombination-dominated) can be shown to vary
(almost) as a function of $Q$ alone (see \citealt{Puls1996},
\citealt{Repolust2005}, Fig. 12, and \citealt{Holgado2018}, Appendix
B). 

\begin{figure}
\centering 
\includegraphics[height=8cm,angle=90]{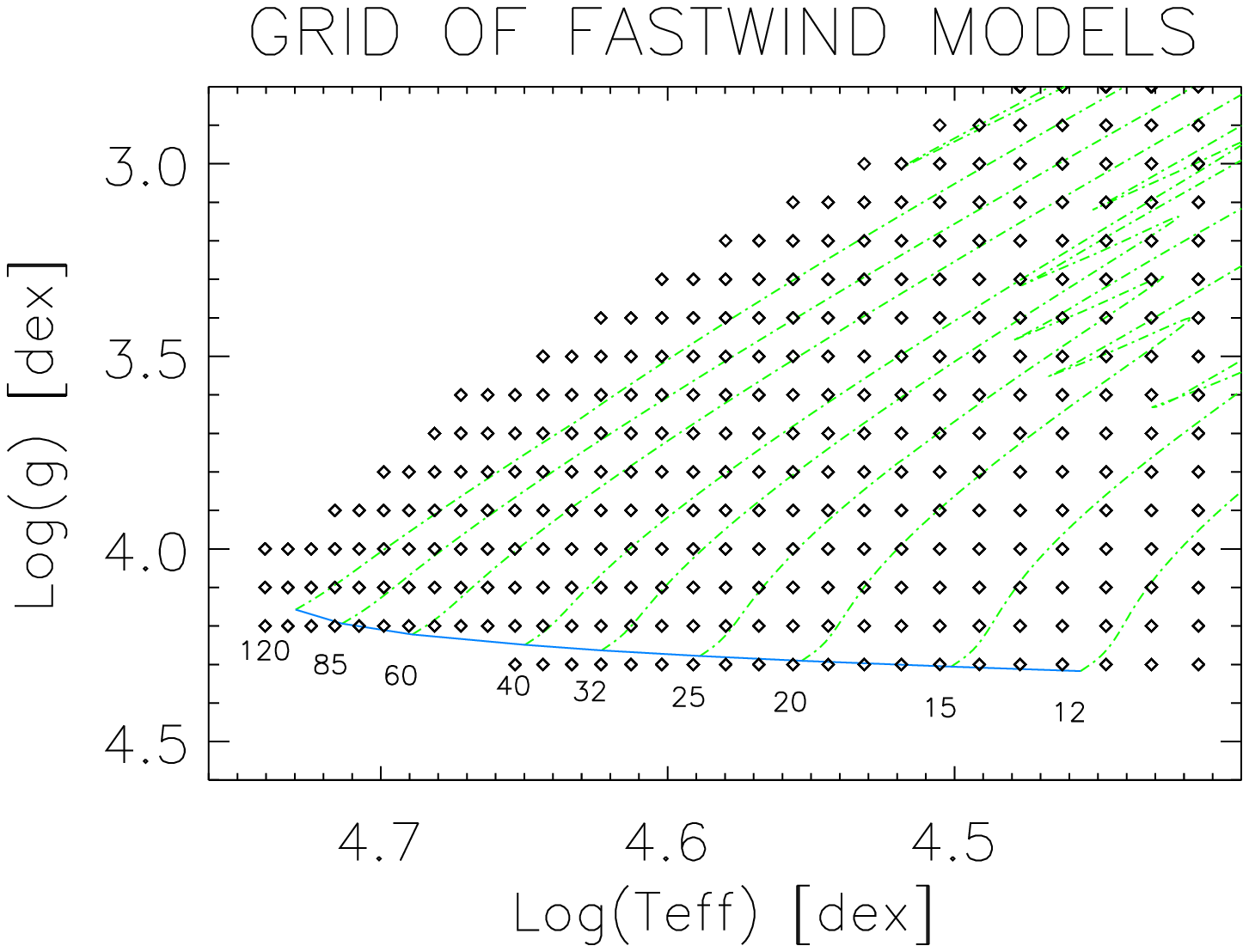}
\caption{Location of models from the {\sc Fastwind} grid in the $\log
g$ vs. $\log$ \Teff\ plane. Nonrotating Geneva evolutionary tracks
\citep{Ekstrom2012} are plotted in green, and the blue line defines the corresponding Zero-Age Main Sequence (ZAMS). The numbers indicate the tracks' initial
stellar masses in units of \Msun.}
\label{fig:grid}
\end{figure}

Table \ref{tbl:range_grid} displays more information about our model
grid (here for the case of unclumped models), where \yhe~
denotes the He-abundance as $N_{\rm He}/N_{\rm H}$, with $N$ the
corresponding particle density.
Figure \ref{fig:grid} illustrates the distribution of grid models in
the $\log g$ vs. $\log \Teff$ (Kiel) diagram, together with the
Geneva evolutionary tracks for 5, 10, 15, 20, 25, 40, 50, 60, 85 and 120
M$_{\odot}$, and for ``solar'' conditions ($Z=0.014$), as
published by \citet{Ekstrom2012}. The final grid contains a total of
107\,547 models. In Sect.~\ref{sec_clump} we will calculate
additional grids, with various clumping laws as described there.

\begin{table}
\begin{center}
\caption{Parameter ranges for the grid models. 
The metallicity composition follows the solar values
provided by \citet{Asplund2009}, and $Q$ is calculated in units of
\Msun\,${\rm a}^{-1}$ for $\Mdot$,  \kms for \vinf, and \Rsun~for $\Rstar$.
}
\label{tbl:range_grid}
\begin{tabular}{ l l} 
\hline  \hline
Parameter      & Range of values \\ \hline
\Teff [K]  &[22000--55000], stepsize 1000 K\\
\grav [$g$ in cgs] &[2.6--4.3], stepsize 0.1 dex\\
$\vmic$ [km s$^{-1}$]  &5,10,15,20 \\
\yhe   & 0.06, 0.10, 0.15, 0.20, 0.25, 0.30\\
\logq  &-15.0, -14.0, -13.0, -12.7, -12.5, -12.3,\\                  
               & -12.1,-11.9, -11.7\\
$\beta$        & 0.8, 1.0, 1.2, 1.5\\ \hline \hline
\end{tabular} 
\end{center}
\end{table}

Previous model-grid spectra used by our working group have been
calculated for the optical range  \citep[e.g.,][]{Sabin2017, Holgado2018}. 
For our current study, we needed to
extend them to the near infrared. 
Table~\ref{tbl:lines} lists the H and He lines included in our
synthetic spectra. This list refers only to the diagnostic lines
covered in the formal solution; for the solution of the rate equation
system, all decisive lines are considered. The table also
indicates additional blends of the major component. For example, the
total Br$_\gamma$ complex comprises four different transitions. Blends
from additional elements, such as nitrogen, have been neglected. As well, 
Br$_{12}$ was finally not included among the diagnostics (see comments
in Sect.~\ref{nir_extension}).


\begin{table*}
\caption{Diagnostic H/He optical and NIR lines used in the current
work (regarding Br$_{12}$ see Sect.~\ref{nir_extension}). The
\ion{He}{i} line at 2.11 $\mu$m is severely contaminated by \ion{N}{iii} 
2.1155 $\mu$m, and \ion{He}{i} 2.05 $\mu$m is not present
in the \cite{Hanson2005} spectra. These two lines are not included in
our analysis. Wavelengths are given in air.}
\label{tbl:lines}
\begin{center}
\begin{tabular}{|l|l|l|} \hline
Line & Wavelength [\AA] & number of H/He components \& identification\\ \hline
  \multicolumn{3}{|c|}{Optical} \\
H$_\alpha$        & 6562 &2 -  \ion{H}{i} (2-3) \& \ion{He}{ii} (4-6)      \\
H$_\beta$         & 4861 &2 -  \ion{H}{i} (2-4) \& \ion{He}{ii} (4-8)      \\
H$_\gamma$        & 4340 &2 -  \ion{H}{i} (2-5) \& \ion{He}{ii} (4-10)     \\
H$_\delta$        & 4101 &2 -  \ion{H}{i} (2-6) \& \ion{He}{ii} (4-12)     \\
\ion{He}{i} 4387         & 4387 &1 -  \ion{He}{i} (2p1-5d1)              \\
\ion{He}{i} 4922         & 4922 &1 -  \ion{He}{i} (2p1-4d1)              \\
\ion{He}{i} 4026         & 4026 &2 -  \ion{He}{i} (2p3-5d3) \& \ion{He}{ii} (4-13)\\
\ion{He}{i} 4471         & 4471 &1 -  \ion{He}{i} (2p3-4d3)              \\
\ion{He}{i} 6678         & 6678 &2 -  \ion{He}{i} (2p1-3d1) \& \ion{He}{ii} (5-13)\\
\ion{He}{ii} 4200        & 4200 &1 -  \ion{He}{ii} (4-11)                \\
\ion{He}{ii} 4541        & 4541 &1 -  \ion{He}{ii} (4-9)                 \\
\ion{He}{ii} 4686        & 4686 &1 -  \ion{He}{ii} (3-4)                 \\
  \multicolumn{3}{|c|}{H-band}                    \\
Br$_{10}$        & 17362 &1 -  \ion{H}{i} (4-10)                  \\
Br$_{11}$        & 16807 &1 -  \ion{H}{i} (4-11)                  \\
Br$_{12}$        & 16407 &1 -  \ion{H}{i} (4-12)                  \\
\ion{He}{i} 1.70         & 17000 &1 -  \ion{He}{i} (3p3-4d3)              \\
\ion{He}{ii} 1.69        & 16900 &1 -  \ion{He}{ii} (7-12)                \\
  \multicolumn{3}{|c|}{K-band}                   \\
Br$_\gamma$       & 21660 &4 -  \ion{H}{i} (4-7), \ion{He}{i} (4d1-7f1), \ion{He}{i} (4d3-7f3) \& \ion{He}{ii} (8-14) \\ 
\ion{He}{ii} 2.18        & 21880 &1 -  \ion{He}{ii} (7-10)                  \\ \hline 
\end{tabular}
\end{center}
\end{table*}

\subsection{Automatic fitting and extension to the NIR 
\label{iacob_gbat}}

\subsubsection{\texttt{iacob\_gbat}}
\label{sec:gbat}
\texttt{iacob\_gbat} is a grid-based automatic tool (\citealt{Ssimon2011, Sabin2014}, 
and \citealt{Holgado2018}, Appendix A)
developed to compare a large amount of synthetic spectra with the
observed ones. It calculates the fitness of the individual synthetic
spectra, and provides us with the best fit (following specific
criteria, see below), and the corresponding stellar parameters
 including appropriate error bars as described below. Before running the tool, one needs
to determine the rotational and macroturbulent velocities (\vsini, and
\vmac, respectively). A wrong determination of these velocities can
 result in an erroneous value for all stellar parameters \citep[][Fig.
2.13]{CarolinaT}. Rotational and macroturbulent
velocities are obtained with the \texttt{iacob\_broad} tool, developed
by \citet{Ssimon2014}. Details can be found in Sections
\ref{optical_nocl} and \ref{sec:vrot_ir}.

In the next steps, we define interactively the wavelength range of the
considered lines, correct for radial velocity, in case renormalize the
continuum, and/or clip nebular lines. Finally, we run
\texttt{iacob\_gbat} to determine the six stellar and wind parameters
(see Sect.~\ref{sect_model_grid}). The basic strategy of
\texttt{iacob\_gbat} is to find the minimum $\chi^2$ from the sum of
the corresponding individual $\chi^2_i$ for each considered line $i$,
i.e., the optimal solution.

The weight given to each line, $\frac{1}{\sigma_i}$, is iteratively
determined, either from the photon noise in the neighboring continuum
of the line, or, if larger, from the minimum average deviation between
the synthetic and the observed line $i$, for the overall best-fitting
model\footnote{ Since the best-fitting model is not known in advance,
an iterative procedure needs to be invoked.}.

This strategy ensures that systematic errors are accounted for (in
case where the synthetic profiles are outside the noise-level compared
to the observed ones), and that such lines obtain a low weight in the
overall $\chi^2$. A detailed
description of the total procedure can be found in \citet[Appendix
A]{Holgado2018}\footnote{In this appendix, Holgado et al. provide
relations based on a reduced $\chi^2$, though all previous and current
versions of \texttt{iacob\_gbat} apply the standard, non-reduced
quantity.}.

\begin{figure}
\centering
\includegraphics[width=6cm,angle=90]{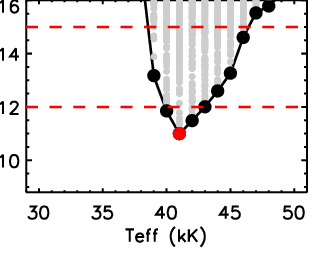}
\caption{An example of the distribution of $\chi^2$ ($y$-axis) versus
effective temperature (HD\,15\,629). The minimum of $\chi^2$ is
indicated by a red dot, and $1-$ and $2-\sigma$ ranges are found from
the intersection between the dashed lines and the distribution.
\label{sigma1}}
\end{figure}

As a result, a distribution of $\chi^2$ values is obtained that can be
used to identify the best-fitting model and the corresponding
values/uncertainties for each of the stellar and wind parameters. In
Fig.~\ref{sigma1}, we plot the distribution of $\chi^2$ versus \Teff\
for HD\,15\,629. The minimum $\chi^2$ value (resulting from an
interpolation of the lower envelope) provides us with the
appropriate value for \Teff, and the 1-$\sigma$ uncertainty is
estimated from the range where $\chi^2 = \chi^2_{\rm Min} +1$.

Sometimes, the distributions present specific difficulties: cases in
which we cannot determine a given parameter with sufficient accuracy,
or values that are at the border of the grid parameter range.
Thus and always, the final output has to be examined individually,
to identify these cases and at least to minimize corresponding
problems. A more detailed discussion of the different problems can be 
found in \cite{CarolinaT}.


%


\subsubsection{Extension to the near infrared}
\label{nir_extension}

To extend the \texttt{iacob\_gbat} tool toward the NIR, we added
several modules to the code. In addition to including all the NIR
lines from Table~\ref{tbl:lines} for the determination of the best fit
model, we 
performed several tests to check the extended version. 

\begin{figure}
\begin{center}
\includegraphics[width=8.5cm]{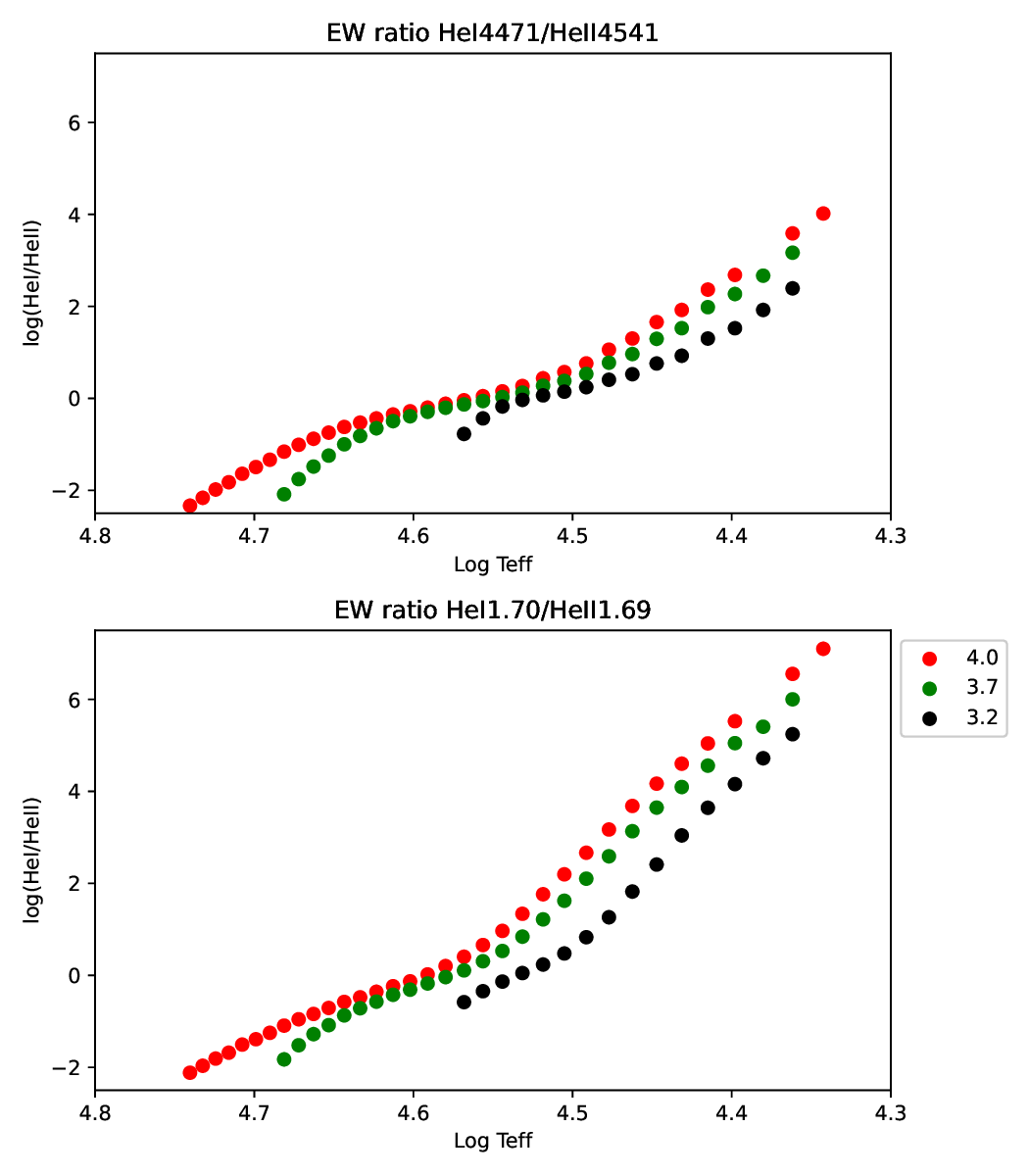}
\caption{Equivalent width (EW) ratios for selected optical and NIR
\ion{He}{i}/\ion{He}{ii} lines, as a function of $\log\Teff$ and $\log
g$ (see legend), resulting from our model-grid calculations.}
\label{fig:ratio_temp}
\end{center}
\end{figure}

The ratio between the strengths of
\ion{He}{i} 4471 and \ion{He}{ii} 4541 is a good temperature diagnostics
in the optical range.  
As shown in Fig.~\ref{fig:ratio_temp}, 
the ratio between \ion{He}{i} 1.70 and
\ion{He}{ii} 1.69 in the NIR  yields a similar diagnostic. Here we show their equivalent width ratio
for a series of models ranging from 25\,000 to 55\,000 K, and for
three values of $\log g$. Obviously, these H-band lines can be as
sensitive to the temperature as the optical ones, and with a very
analogous behavior.

Similar to the Balmer lines in the optical, the shape and wings of the
Brackett lines in the NIR are sensitive to gravity\footnote{A
discussion of specific dependencies which are different from the
behavior of the optical lines can be found in \citet{Repolust2005}}.
However, during our test calculations, we realized a peculiar behavior
of the different Brackett lines, making it difficult or even
impossible to simultaneously fit the observed spectra. Indeed,
particularly the higher members of the Brackett series (starting
around Br$_{12}$) are only poorly represented by our synthetic
profiles. We carried out a series of tests, grouping the lines in
pairs (Br$_{10}$ \& Br$_{11}$, Br$_{11}$ \& Br$_{12}$, Br$_{10}$ \&
Br$_{12}$), i.e., skipping always one of the lines in our parameter
determination. This way, we checked which pair was more consistent with
the rest of the NIR lines. Our tests indicated that the highest
member considered here, Br$_{12}$, gave the poorest agreement.

Currently, the origin of this disagreement remains unclear, but might
be related to insufficient accuracy of line-broadening data,
collision strengths for hydrogen transitions with higher upper levels, difficulties in the reduction process, or a combination of all of them all 
(see also \citealt{Repolust2005} and Sect.~\ref{discussion}). 
Forthcoming work needs to identify the region in stellar parameter
space where the problem appears most strongly, its physical origin,
and potential solutions. Meanwhile, and since this problem becomes
particularly worrisome only from Br$_{12}$ on, we decided to skip this
line from our line list when applying the \texttt{iacob\_gbat} tool
for our IR analysis.

\section{First results: Parameter determinations adopting smooth winds}
\label{nocl}

We divide our stellar sample in three groups according to the 
luminosity class of the stars (i.e., [I-II], [III] and [ IV-V]). Each of the
three groups presents a particular behavior w.r.t. the fits obtained.
Dwarf stars show the best fits to the observed spectrum, whereas fit
difficulties increase for giants and are usually largest for the
luminosity class I stars, those with the strongest winds.

\begin{figure*}
\begin{center}
\includegraphics[width=13cm,angle=90]{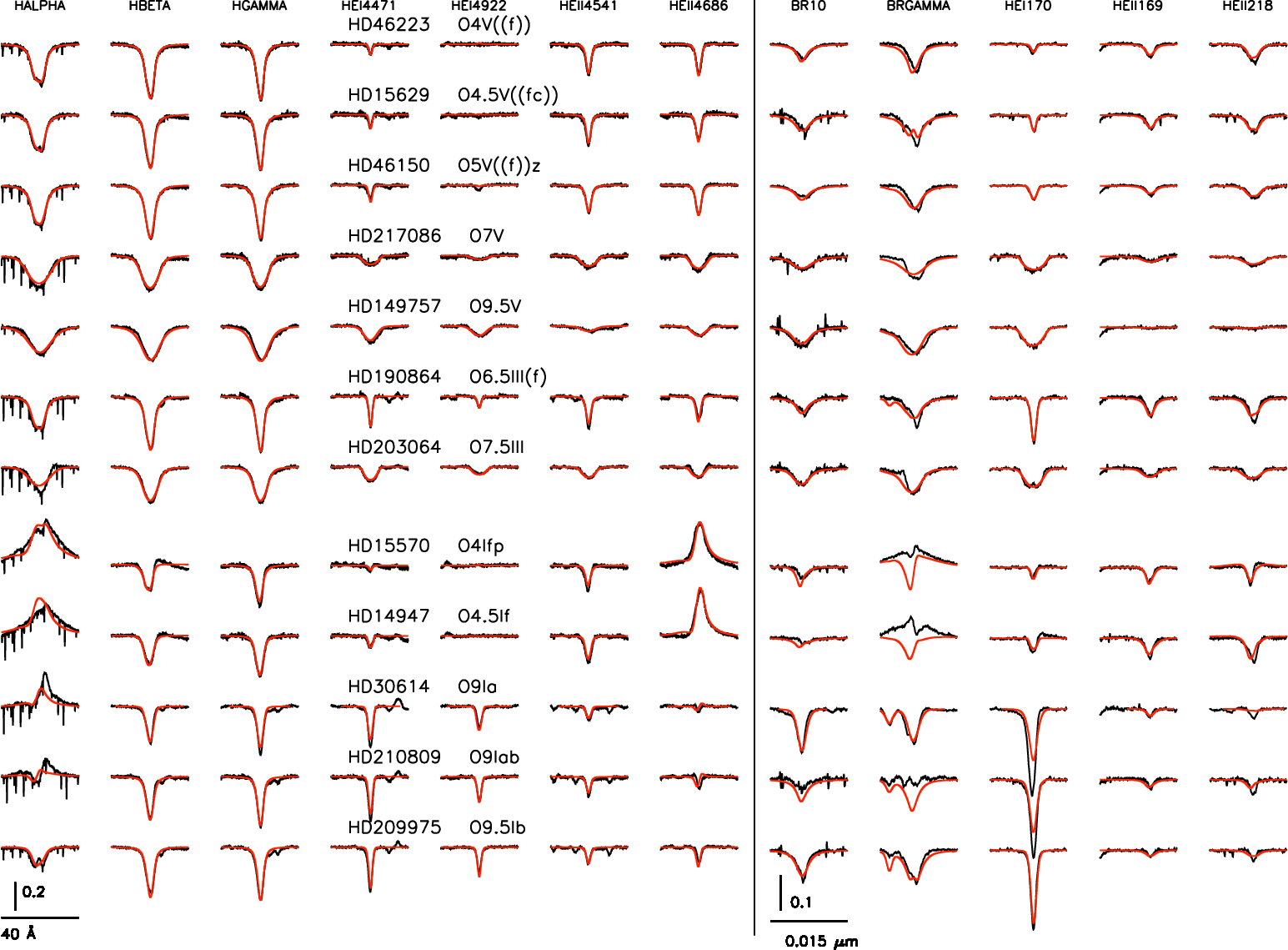}
\caption{Spectral fits for selected optical (left) and NIR (right) lines using unclumped models.
Observations are shown in black, and best fit model profiles in red.
We stress that the individual model parameters for the best fitting
optical and NIR profiles differ (to various extents) since the analyses have been
performed separately for both ranges (cf. Table~\ref{tbl:iacob_OP} vs.
Table~\ref{tbl:iacob_IR_bli}). The horizontal bar gives the wavelength scale
for each range, and the scale of the ordinate axis is given by the
vertical bar (at the bottom of the H$_\alpha$
column for the optical range, and at the bottom of
the Br$_{10}$ column for the NIR.)}
\label{fig:OP_profiles}
\end{center}
\end{figure*}

\begin{figure*}
\begin{center}
\includegraphics[width=15cm,angle=0]{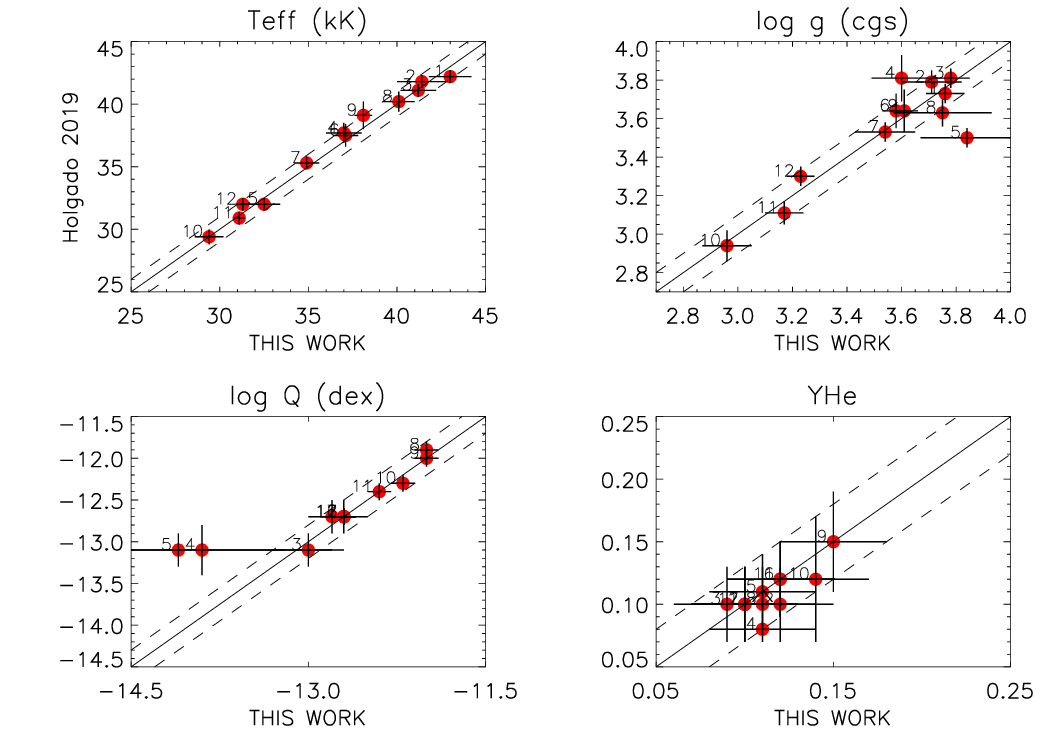}     
\caption{Comparison between the stellar parameters obtained by
\cite{GonzaloT} (see also \citealt{Holgado2018}) and our work. 
Upper left panel: effective
temperature. The dashed lines represent $\pm$ 1000~K; upper right
panel: logarithmic gravity  ($\pm$0.1 dex);
lower left panel: $\log Q$  ($\pm$0.2 dex);
lower right panel: helium abundance \yhe ($\pm$0.03). 
Numbers indicate the stars as listed in
Tab.~\ref{tbl:stars_sample}.} 
\label{fig:op_gholgado}
\end{center}
\end{figure*}

\subsection{Stellar parameters from the optical spectrum}
\label{optical_nocl}

We first determine the stellar parameters using only the optical
spectra secured in the IACOB database. We determine the \vsini\ and
\vmac\ values using the \texttt{iacob-broad} package
\citep{Ssimon2014}. Our values for the optical, presented in
Table~\ref{tbl:stars_vsini} together with their NIR
counterparts\footnote{we only discuss here the results for the optical. 
For a further discussion, see Sect.~\ref{sec:vrot_ir}}, 
agree with (and have errors similar to) those
from \cite{Ssimon2014} within 20~\kms\ or $\pm$20$\%$, except for the
\vmac\ of the fast rotators. 
\begin{table}
\begin{center}
\caption{Comparison between \vsini\ and \vmac\ values obtained from 
optical (``OP'') metal lines and from the NIR \ion{He}{i} $\lambda1.70\mu$m
line. Typical uncertainties are $\pm$10$\%$ in the optical and $\pm$15$\%$ 
in the infrared. All velocities are given in \kms. 
\label{tbl:stars_vsini}}
\small
\begin{tabular}{rllllll}
\hline 
$\#$ & Star ID &Type & \vsini & \vmac &\vsini & \vmac \\ \hline
        &             &         & OP   &    OP   &   NIR &  NIR\\    \hline 
1 & HD46223   &O4 V((f))   &52    &97      & 70  & 100  \\   
2 & HD15629   &O4.5 V((fc))&70    &69      & 68  & 96   \\   
3 & HD46150   &O5 V((f))z  &69    &107     & 107 & 114  \\   
4 & HD217086  &O7 Vnn((f))z&382   &104     & 372 & 18 \\   
5 & HD149757  &O9.2 IVnn   &290   &290     & 366 & 165 \\  
6 & HD190864  &O6.5 III(f) &65    &90      & 73  & 113  \\   
7 & HD203064  &O7.5 IIIn((f))    &315   &98      & 344 & 103 \\  
8 & HD15570   &O4 If      &38    &120     & 74  & 92  \\   
9 & HD14947   &O4.5 If     &117   &49      & 132 & 25  \\   
10 & HD30614   &O9 Ia       &115   &72      & 78  & 213 \\   
11 & HD210809  &O9 Iab      &76    &79      & 72  & 167 \\  
12 & HD209975  &O9.5 Ib     &52    &95      & 73  & 113 \\   \hline 
\end{tabular} 
\end{center}
\end{table}

However, because of the high rotational velocities, this has no impact
on the final results (within the uncertainties). Updated values have
been recently presented by \cite{Gholgado2022}. For most stars, the
differences are well within the adopted uncertainties. Only HD
149\,757 and HD 15\,570 show a larger difference. For the first
object, \cite{Gholgado2022} estimate 385 and 94~\kms\ for \vsini\ and
\vmac, respectively, compared to a value of 290~\kms\ for both
quantities as derived here. This is a consequence of the degeneracy between
rotational and macroturbulent velocities when both reach high values.
For the second star, we find 38 and
120~\kms, whereas \cite{Gholgado2022} estimate 81 and 115~\kms. We attribute
this large difference to the use of different spectra and different lines. Holgado et al. have
used the \ion{N}{v} 4605 line, which is in a region of complicate normalization due to
the nearby strong \ion{N}{iii} emission, whereas we have used the \ion{O}{iii} 5592 line. 
To ensure that these differences will not affect our results,
we have repeated our optical and infrared analyses described
below with the values from \cite{Gholgado2022}, without any
significant differences. This finding results from the combined rotational and 
macroturbulence broadening, producing similar profiles in these cases.

Table~\ref{tbl:iacob_OP} summarizes the parameters obtained from our
optical analysis after running the \texttt{iacob\_gbat} tool. Here and in the following similar tables, upper and lower limits refer
to the corresponding parameter ranges of our model grid(s) only. As an
example, $\beta > 1.$0 would mean that $\beta$ ranges, within its
1-$\sigma$ uncertainties, from 1.0 to 1.5, when consulting
Table~\ref{tbl:range_grid}. In Table~\ref{tbl:iacob_OP}, such limits
frequently occur for the parameters $\beta$ and $v_{\rm mic}$.  
For the strong H$_\alpha$ and/or \ion{He}{ii} 4686 wind emission from
our supergiants (which actually should allow for quite a precise
determination of $\beta$), this simply means that the contribution of
these lines to the global $\chi^2$ is low when counted with equal
weights as done here. The additional information contained in
the other optical \ion{H}{} and \ion{He}{} lines is usually not
sufficient to constrain these parameters more accurately. The
inclusion of information from UV P Cygni lines would be very helpful
in these cases. On the other hand, more precise values for the
micro-turbulent velocity can be only obtained from the analyses of
metal lines from species with more than one ionization stage visible (e.g.,
\citealt{Markova2008}); however, in addition, such values might depend on the
chosen atom.

In Table~\ref{tbl:iacob_OP}, gravities are not corrected for the
effects of centrifugal acceleration, as we here are only
interested in the formal fits and do not compare with evolutionary
models. Errors were obtained from \texttt{iacob-gbat} as described
above, but following the arguments from \cite{Sabin2017} we set a
lower limit of 0.1 in $\log Q$ and 0.03 in \yhe\ for these
uncertainties when the automatically derived formal errors turned out
to be lower\footnote{sometimes, the \texttt{iacob-gbat} tool may
deliver unrealistically low errors, as it does not take into account
uncertainties like the continuum normalization}.
Fig.~\ref{fig:OP_profiles}, left side, displays a comparison
between selected observed optical profiles and the synthetic
lines from the best fit model for each star.

\begin{table*}
\begin{center}
\caption{Stellar parameters obtained from the optical analysis using
unclumped models. Gravities do not include a centrifugal correction. 
Upper and lower limits refer to the corresponding parameter ranges of our
model grid only (see Table~\ref{tbl:range_grid}).}
\label{tbl:iacob_OP}    
\small
\begin{tabular}{llllllll} \hline
      Star  &spectral type&   \Teff(kK)     &    $\log g$ (dex)      &
      $ \log Q$      & \yhe         &     \vmic\ (\kms)   &    $\beta$ \\ \hline 
   HD46223 & O4 V((f))      & 43.0 $\pm$ 1.2 & 3.76 $\pm$ 0.07   & -12.8 $\pm$ 0.2  & 0.10 $\pm$ 0.03   &        $>$ 9.1 &  1.0 $\pm$ 0.2  \\
   HD15629 & O4.5 V((fc))   & 41.4 $\pm$ 1.4 & 3.71 $\pm$ 0.11   & -12.7 $\pm$ 0.2  & 0.12 $\pm$ 0.03   &       $<$ 19.9 &  1.0 $\pm$ 0.2  \\
   HD46150 & O5 V((f))z     & 41.2 $\pm$ 1.0 & 3.78 $\pm$ 0.07   & -13.0 $\pm$ 0.3  & 0.09 $\pm$ 0.03   &        $>$ 5.0 &        $>$ 0.8    \\
  HD217086 & O7 Vnn((f))z   & 37.0 $\pm$ 1.0 & 3.60 $\pm$ 0.11   & -13.9 $\pm$ 1.1  & 0.11 $\pm$ 0.03   & 12.4 $\pm$ 7.4 &        $<$ 1.2  \\
  HD149757 & O9.2 IVnn      & 32.5 $\pm$ 0.9 & 3.84 $\pm$ 0.17   & -14.1 $\pm$ 0.9  & 0.11 $\pm$ 0.03   & 12.2 $\pm$ 7.2 &        $<$ 1.2  \\
  HD190864 & O6.5 III(f)    & 37.1 $\pm$ 0.7 & 3.58 $\pm$ 0.05   & -12.7 $\pm$ 0.1  & 0.12 $\pm$ 0.03   & 15.1 $\pm$ 3.4 &  0.9 $\pm$ 0.1\\
  HD203064 & O7.5 IIIn((f)) & 34.9 $\pm$ 0.7 & 3.54 $\pm$ 0.11   & -12.7 $\pm$ 0.1  & 0.10 $\pm$ 0.03   &       $>$ 15.2 &  0.9 $\pm$ 0.1  \\
   HD15570 & O4 If          & 40.1 $\pm$ 0.9 & 3.75 $\pm$ 0.18   & -12.0 $\pm$ 0.1  & 0.11 $\pm$ 0.03   &        $>$ 5.0 &        $>$ 1.0    \\
   HD14947 & O4.5 If        & 38.1 $\pm$ 0.5 & 3.61 $\pm$ 0.05   & -12.0 $\pm$ 0.1  & 0.15 $\pm$ 0.03   &        $>$ 9.5 &        $>$ 1.2    \\
   HD30614 & O9 Ia          & 29.4 $\pm$ 0.8 & 2.96 $\pm$ 0.09   & -12.2 $\pm$ 0.1  & 0.14 $\pm$ 0.03   &       $>$ 15.9 &        $>$ 0.8    \\
  HD210809 & O9 Iab         & 31.1 $\pm$ 0.3 & 3.17 $\pm$ 0.07   & -12.4 $\pm$ 0.1  & 0.12 $\pm$ 0.03   &       $>$ 16.2 &        $>$ 1.0    \\
  HD209975 & O9.5 Ib        & 31.3 $\pm$ 0.4 & 3.23 $\pm$ 0.05   & -12.7 $\pm$ 0.1  & 0.10 $\pm$ 0.03   &       $>$ 12.2 &        $>$ 1.1    \\ \hline
\end{tabular} 
\end{center}
\end{table*}                                                                                                                 

From the fits shown in Fig.~\ref{fig:OP_profiles}  (left side) we draw the
following conclusions:
 
\begin{itemize}

\item Except for one object (see below), all dwarfs show excellent fits. Even 
the fast rotators do not show any significant problems, despite of
potential effects not considered here, like gravity darkening or geometrical
deformation; the fit for HD\,149\,757 is poorer, as the model yields too broad 
wings in some of the Balmer lines.

\item The two giants within our sample are mid-types. HD\,190\,864 shows small
differences in the cores of the \ion{He}{i} lines, with slightly too
shallow theoretical profiles for \ion{He}{ii}\,4200 and 4541
complemented by a slightly too deep profile for \ion{He}{ii}\,4686. 
HD\,203\,064, a fast rotator, displays a poor fit to H$_\alpha$ and,
to a lesser extent, to \ion{He}{ii}\,4686. 

\item The supergiants display the largest fitting problems, particularly in H$_\alpha$,
sometimes together with problems in H$_\beta$ and \ion{He}{ii}\,4686 (much
less though), which points to some wind influence. This agrees with
the findings by \cite{Holgado2018}. The largest
difficulties are found for the H$_\alpha$ P-Cygni like profile of the late
types, HD\,30\,614 (of Ia luminosity class) and HD\,210\,809. In both
stars the \ion{He}{ii}\,4686 core shows a shift to the red. The best fit in
this group is obtained for the less luminous supergiant, HD\,209\,975
(Ib). Early-type supergiants have an intermediate behavior in
H$_\alpha$ (despite of showing emission), although they present some
difficulties for the red wing of H$_\beta$ that are not seen in
late-type supergiants.
\end{itemize}

We compare our parameters with those recently quoted by
\cite{GonzaloT} (most of the values used here have already been published
in \citealt{Holgado2018}), see Fig.~\ref{fig:op_gholgado}.  All
temperatures agree well within the errors given here and by
\cite{GonzaloT}. For the (uncorrected) gravity, we find significant
differences for the rapidly rotating dwarfs, particularly HD 149\,757,
for which we obtain $\log g= 3.84 \pm 0.17$, whilst \cite{GonzaloT}
inferred 3.50$\pm$0.05. 
%
Although marginally within the uncertainties, HD 217\,086 also shows
differences in $\log g$ (3.60$\pm$0.11 versus 3.81$\pm$0.12). We
attribute these differences to the difficulties with the normalization
and radial velocity correction in fast rotators. As the line wings are
very extended and reach the continuum rather smoothly, a small
difference in the data treatment may result in a relatively large
difference in gravity. In addition, in the case of HD\,149\,757,
variability also plays a role\footnote{We have analyzed a different spectrum than \cite{GonzaloT},
and the Balmer lines are slightly broader in our case.}.

For $\log Q$, the agreement is excellent\footnote{stars 1, 2, 6, 7 and
12 cluster around the same locus in the figure}, except again for the
fast rotating dwarfs. This is basically due to the lack of sensitivity
of the diagnostics (mainly, the H$_\alpha$ line) at these low values
of $Q$, combined with high rotational velocities. The helium abundances
agree also well\footnote{here, stars 1, 2, 3, 7, 8, and 12 overlap in
the figure, as do 6 and 11}.

\subsection{Analysis in the near infrared}

In this subsection, we derive the stellar parameters solely from the
near infrared, following a similar methodology as we did in the
previous subsection. This will tell us how far results obtained for
stars in heavily obscured clusters can be compared to those provided
in the extensive literature of optical analyses. While this exercise
has already been carried out by other authors (e.g., \citealt{Repolust2005},
or more recently within investigations when fitting simultaneously
optical and infrared spectra, e.g., \citealt{Najarro2011} or
\citealt{Bestenlehner2014}), we have to check whether our automatic
procedure extended to the near infrared results in reliable stellar
parameters. 

\subsubsection{Determination of \vsini\ and \vmac\ in the NIR}
\label{sec:vrot_ir} 
We start again by deriving \vsini\ and \vmac\ using
\texttt{iacob-broad}. In the optical, these values were derived using
metal lines, whose broadening is dominated by processes determining
these quantities. However, the metal NIR lines are too weak in our
spectra and are not available for all stars. For this reason, we are
forced to use \ion{He}{i} lines, which are affected by the Stark
effect, limiting our ability to measure the rotational velocity for
slow rotators (or the macroturbulent velocity when this is low).
\ion{H}{} and \ion{He}{ii} lines are even less well suited, since they
are dominated by the strong linear Stark effect. Thus we decided to
use the \ion{He}{i} $\lambda1.70\mu$m line, which is strong enough for
all the stars. \cite{ramirezagudelo2013} have shown that it is
possible to derive accurate rotational velocities from the
(quadratically) Stark broadened optical \ion{He}{i} lines. However,
the Stark broadening increases toward the infrared, and thus it could
place a lower limit (see below) on the derived \vsini\ values.

\begin{figure}
\begin{center}
\includegraphics[height=9cm,angle=90]{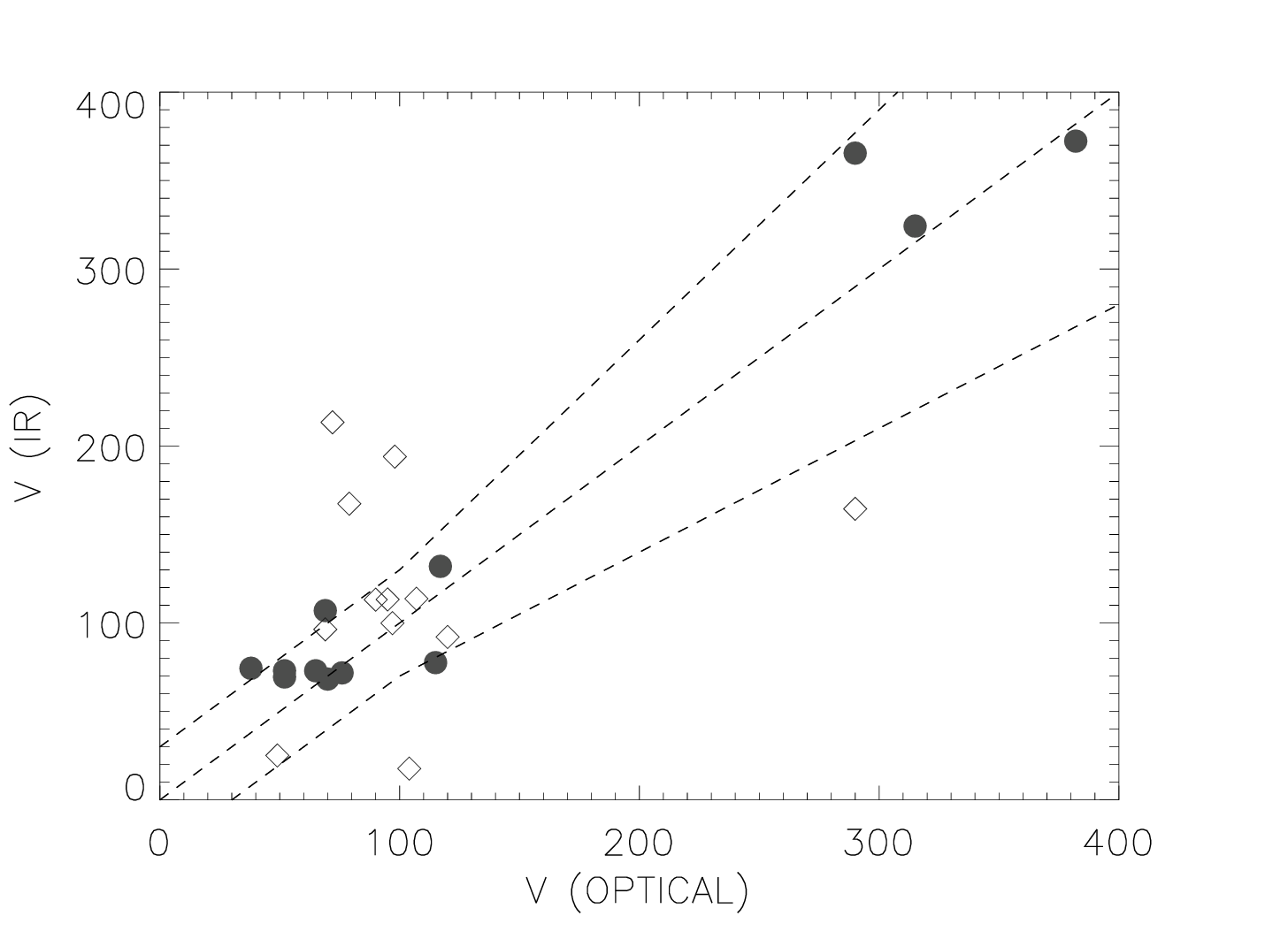}
\caption{\vsini\ (filled circles) and \vmac\ (open diamonds) values
obtained from the optical metal lines and from \ion{He}{i}
$\lambda1.70\mu$m in the NIR. The dashed lines give the band
$\pm$ 30 \kms (for low \vsini) or 30\% of \vsini~ (optical), whatever is
larger.}
\label{fig:vsini_op_ir}
\end{center}
\end{figure}

Figure \ref{fig:vsini_op_ir} compares the projected rotational
velocities obtained from both wavelengths ranges (filled circles),
whereas Table~\ref{tbl:stars_vsini} gives the numerical values. In the
figure, dashed lines indicate the region that departs by $\pm$30 \kms\
or $\pm$30$\%$ (whatever is larger) from the 1:1 relationship. This
band marks the region where stellar parameters are not affected beyond
errors by changes in the adopted rotational velocity \citep{CarolinaT}.
It does not indicate the uncertainties in the determinations,
which sometimes are larger than the difference between the values obtained 
from the optical and the NIR spectra, as discussed below. 
We see that the \vsini\ pairs are always located within these bands,
and that most values agree reasonably well. Therefore, we do not
expect a significant impact on our results due to these differences.

We also see that there might be a limit to the lowest rotational
velocities determined with \ion{He}{i} 1.70$\mu$m (around 80~\kms),
although this would require more slowly rotating stars to be confirmed
(the points cluster close to the 1:1 relation). The only really
departing point, at \vsini\ (opt) = 115 and \vsini\ (NIR) = 78 \kms,
corresponds to HD\,30\,614, with a strong \ion{He}{i}
$\lambda1.70\mu$m line in absorption. This discrepancy is related to
the large value found for \vmac\ ( see Tab.~\ref{tbl:stars_vsini} and 
open diamonds in Fig.~\ref{fig:vsini_op_ir}). As expected, \vmac\
departs strongly from the 1:1 relation for some objects, especially
for fast rotators. However, the tests we performed for HD\,149\,757
and HD\,15\,570 indicate that no significant changes in the
stellar/wind parameters are expected for these stars.

\begin{table*}
\begin{center}
\caption{Stellar parameters obtained from the NIR analysis using
unclumped models. Gravities do not include a centrifugal correction.
For upper an lower limits see caption of Table~\ref{tbl:iacob_OP}.
\label{tbl:iacob_IR_bli}}
\begin{tabular}{lllllll}\hline
Star       & \Teff(kK)     &    $\log g$ (dex)      & $\log Q$       &
\yhe      &     \vmic\ (\kms)   &  $\beta$ \\ \hline 
   HD46223 & 41.2 $\pm$ 1.4  & 3.79 $\pm$ 0.10  &  -12.7 $\pm$ 0.2  &        $ <$ 0.10    &          $>$ 5.0   &     $>$ 0.9  \\
   HD15629 & 39.5 $\pm$ 1.7  & 3.66 $\pm$ 0.17  &  -13.2 $\pm$ 0.7  &        $< $ 0.09    &  12.4 $\pm$ 7.4  &      $>$ 0.8  \\
   HD46150 & 39.6 $\pm$ 1.0  & 3.85 $\pm$ 0.12  &  -12.9 $\pm$ 0.3  &         $ <$ 0.08   &          $<$ 18.5  &     $>$ 0.8 \\
  HD217086 & 36.9 $\pm$ 1.1  & 3.86 $\pm$ 0.15  &  -13.8 $\pm$ 1.2  &  0.15 $\pm$ 0.07  &           $>$ 5.0  &      $>$ 0.8 \\
  HD149757 & 32.3 $\pm$ 1.7  & 3.58 $\pm$ 0.31  &  -13.7 $\pm$ 1.3  &  0.19 $\pm$ 0.10  &     $>$ 5.0 &        $>$ 0.8 \\
  HD190864 & 36.8 $\pm$ 1.0  & 3.64 $\pm$ 0.14  &  -12.7 $\pm$ 0.3  &   0.21$\pm$0.09   &   12.4 $\pm$ 7.4 &       $>$ 0.9 \\
  HD203064 & 34.3 $\pm$ 1.5  & 3.70 $\pm$ 0.32  &  -12.5 $\pm$ 0.2  &  0.20 $\pm$ 0.10  &           $>$ 5.0  &        $>$ 1.0 \\
   HD15570 & 38.8 $\pm$ 1.8  & 3.55 $\pm$ 0.15  &  -11.9 $\pm$ 0.1  &  0.10 $\pm$ 0.03   &          $<$ 19.9  &       $<$ 1.0 \\
   HD14947 & 43.6 $\pm$ 2.8  & 4.03 $\pm$ 0.36  &  -12.5 $\pm$ 0.5  &         $>$ 0.17   &        12.4$\pm$7.4     $>$ 0.9 \\
   HD30614 & 27.6 $\pm$ 0.8  & 2.78 $\pm$ 0.08  &  -12.0 $\pm$ 0.1  &         $<$ 0.17   &           $>$ 5.0  &   $<$1.2  \\
  HD210809 & 35.4 $\pm$ 1.2  &        $>$ 3.80  &  -12.8 $\pm$ 0.3  &         $>$ 0.23  &         $ >$ 10.4  &    $>$ 0.8 \\
  HD209975 & 32.1 $\pm$ 1.3  & 3.33 $\pm$ 0.18  &         $<$ -13.4   &        $>$ 0.12    &         15.9$\pm$3.9  &   $<$ 1.2 \\ \hline
\end{tabular} 
\end{center}
\end{table*}

We conclude that it is possible to derive the rotational and
macroturbulent velocities from the NIR spectrum alone, although with
larger uncertainties than from the optical spectra, 
 and a presumably lower limit for the derived \vsini.

\subsubsection{Stellar parameters from the NIR spectrum}
\label{ir_nocl}

%
%

We now derive the stellar parameters for the same stars as in
Sect.~\ref{optical_nocl}, using the NIR spectra secured and reduced by
\cite{Hanson2005}. 
The results of the NIR analysis are presented in Table
\ref{tbl:iacob_IR_bli}. Again, $\beta$ and microturbulent
velocities could not be well constrained, indicating that for these
stars the near infrared is not better suited than the optical for this
task. This suggests that the difference in line formation depth
between the optical and the H- and K-band spectra is not sufficient to
provide new information, at least at the resolution and S/N of
the spectra analyzed here. The comparison of the observed profiles
with those from the best fit model is presented in
Fig.~\ref{fig:OP_profiles}, right side. Inspection of these profile fits leads to
the following summary:

\begin{figure}[h!]
\centering
\includegraphics[width=5cm,angle=90]{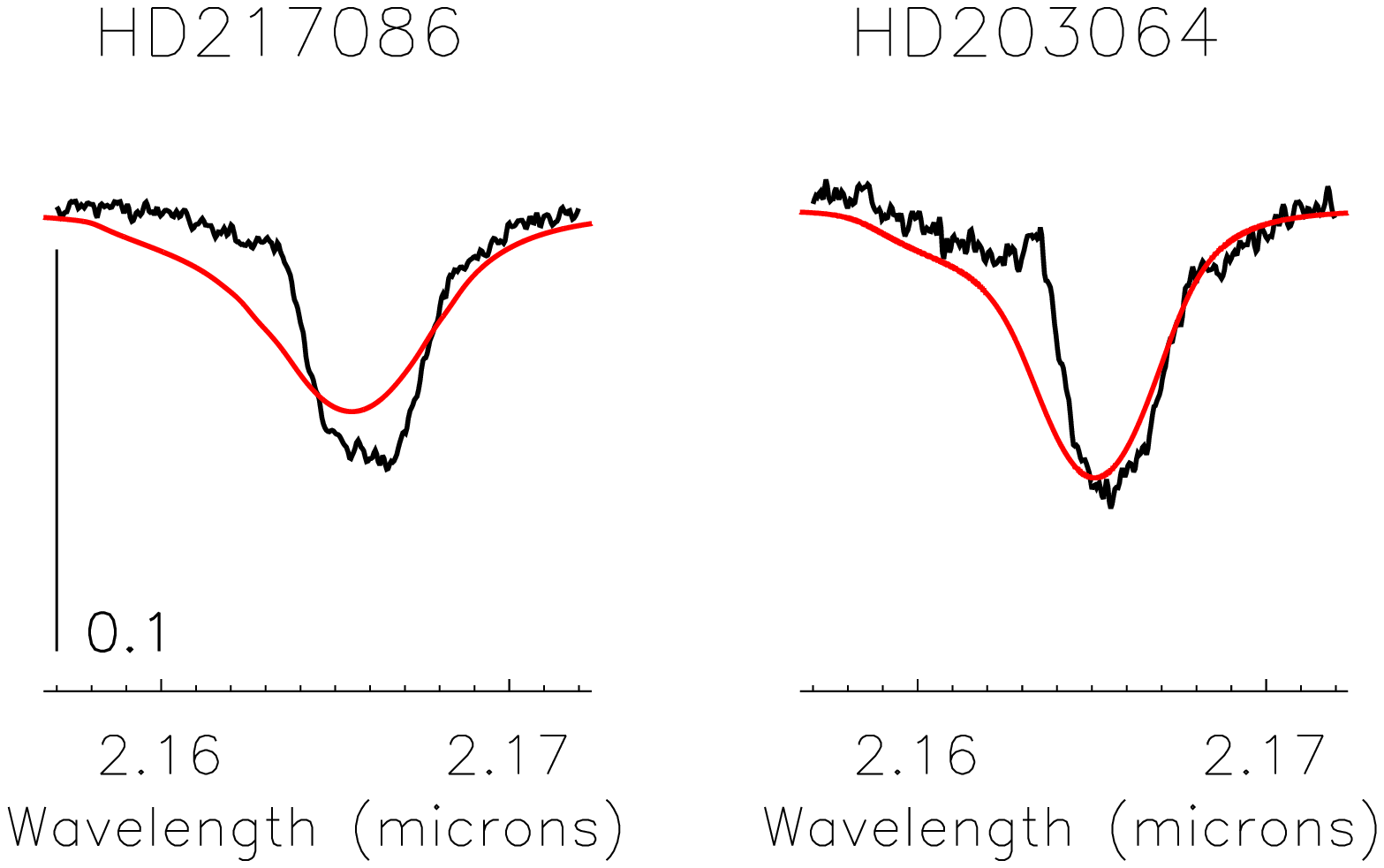}
\caption{Br$_\gamma$ for the rapidly rotating dwarf and giant stars
HD\,217\,086 and HD\,203\,064. Both line profiles show a small blue
emission peak close to the core of the line, resulting in a distorted
blue wing. Red profiles are from best fitting models.
\label{fastrotatorbrg}}
\end{figure}

\begin{figure*}[!h]
\begin{center}
\includegraphics[width= 15.0cm,angle=0]{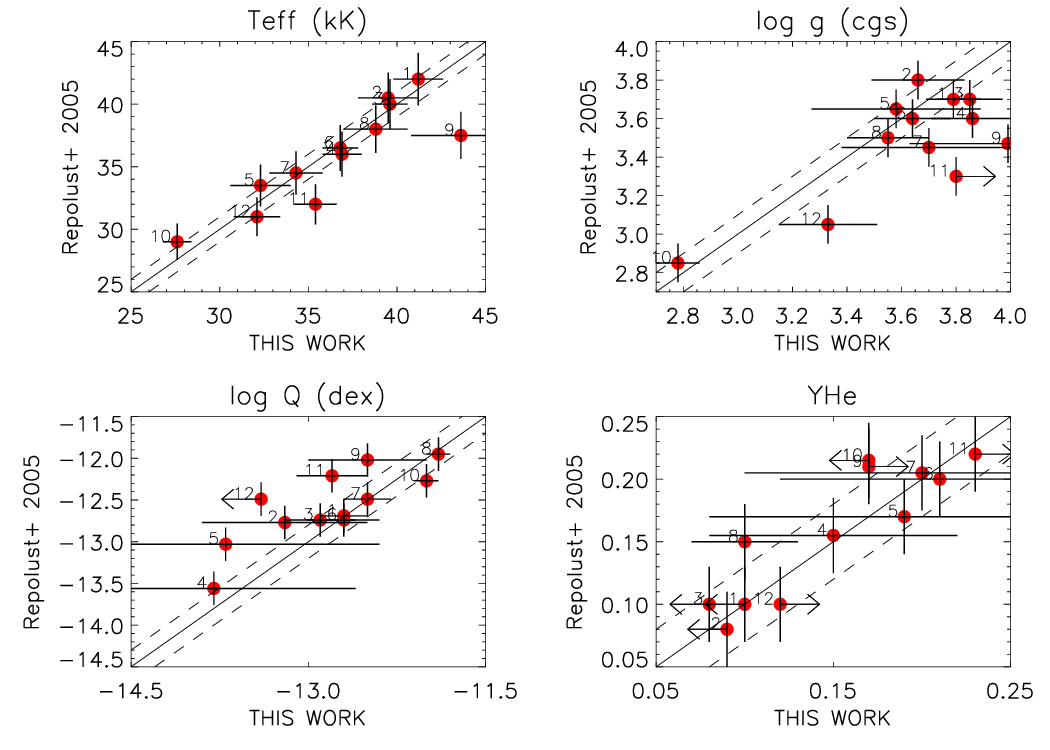}       
\caption
{Comparison between the stellar parameters obtained by
\cite{Repolust2005} and our work, both from the NIR alone. Upper left
panel: effective temperature. The dashed lines represent $\pm$ 1000K;
upper right panel: logarithmic gravity (uncorrected, $\pm$ 0.1 dex).
Star number 9 has been slightly shifted in both axes for clarity;
lower left panel: log Q ($\pm$0.2 dex); lower right panel: \yhe
 ($\pm$0.03), and stars \#7, 9, 10, and 11 have been slightly shifted
upward from its value in \cite{Repolust2005} (\yhe = 0.20) to avoid
overlap, as well as star \#4 (\yhe =0.15). Numbers indicate the stars
as listed in Tab.~\ref{tbl:stars_sample}.}
\label{fig:ir_repol} 
\end{center}
\end{figure*}

\begin{itemize}

\item The best fit quality is again obtained for the dwarfs, 
but now not without significant problems. The best fitted lines are
 the \ion{He} ones, especially \ion{He}{i} $\lambda1.70 \mu$m. 
Br$_{10}$ and Br$_{11}$ also fit reasonably,
but Br$_\gamma$ is not well fitted.
For this line, the fast rotator HD\,217\,086 shows a profile
different from the other dwarfs, with a strong and relatively narrow
absorption in the blue half-line (presumably because of a narrow
emission component, see Fig. \ref{fastrotatorbrg}), and a broad
absorption redward from line center.
 
\item Giants -- The O7.5 III fast rotator HD 203\,064 displays a 
similar Br$_\gamma$ profile as the O7 V fast rotator HD 217\,086, and
a similarly poor fit (see also Fig. \ref{fastrotatorbrg}), pointing to
some process(es) not considered in our models, presumably related to
differential rotation (see \citealt{PetrenzPuls1996} for a discussion
of similar line-shapes of H$_\alpha$). The fit to Br$_{10}$ and
Br$_{11}$, however, is much better. The slower rotating giant,
HD\,190\,864, shows also a good fit for Br$_{10}$ and Br$_{11}$, and a
poor fit for Br$_\gamma$, although without the characteristic shape of
the fast rotators. For both giants, the fit to the He lines is of
varying quality. Globally, the fits are again acceptable,
except for Br$_\gamma$.


\item For almost all supergiants, the Brackett lines, particularly Br$_\gamma$,
show a poor fit quality, except for, surprisingly, HD\,30\,614 (that
had the largest problems in the optical) and, to a lesser extent, the
low luminosity object, HD\,209\,975. 
The early-type supergiants show the poorest fits to the Brackett
spectrum, with the models predicting an absorption profile for
Br$_\gamma$ while the observations show emission instead. The only
exception with a reasonable fit is Br$_{11}$ from HD\,15\,570.
Regarding the He lines, these are also poorly fitted in the late-type
supergiants. Within a given spectral subtype, \ion{He}{i}
$\lambda1.70\mu$m departs more and more from a good fit with
increasing luminosity. Still, for the cooler supergiants,
\ion{He}{ii}$\lambda 2.18\mu$m is always stronger than predicted, and
\ion{He}{ii}$\lambda1.69\mu$m (not shown) only modestly reproduced. The situation
is different for the early-type supergiants, where the fits to the
\ion{He}{ii} lines are acceptable, though far from being perfect.

\end{itemize}
  
We compare again with previous results in the literature, namely those
from \cite{Repolust2005} (Figure~\ref{fig:ir_repol}) 
Globally, there is a fair
agreement\footnote{gravities given in \cite{Repolust2005} are
corrected for centrifugal acceleration. Using their data, we have
uncorrected them and have also calculated the corresponding \logq\ for
the comparison here.} for all stars, except for stars $\#9$ and $\#11$
(HD\,14\,947 and HD\,210\,809).  Here, we obtain a higher \Teff~ and
\grav, which relates to the fact that in both stars the shallow
Br$_{10}$ and Br$_{11}$ lines are well fitted in our approach, whilst
in the fits by Repolust et al. they appear as too strong. Details
about the consequences of such shallow Br$_{10/11}$ lines are
discussed in Sect.~\ref{discussion}.
In the case of the first star, the high temperature
forces an increase in the \ion{He}{} abundance to fit the \ion{He}{i}
line at 1.70$\mu$m (our best model has \yhe = 0.30).

Moreover, the \logq\ of star $\#$12  (HD\,209\,975) shows a large discrepancy, with a
much lower value in our work, due to the reaction of the \ion{He}{ii}
lines to mass-loss. While the \ion{He}{i} line and the Brackett lines
have only a small response to an increased mass-loss, the \ion{He}{ii}
lines (already too shallow in our fit) would become even shallower.
 Indeed, grid models calculated with a \logq\ similar to that of \cite{Repolust2005}lie just
beyond our 1-$\sigma$ uncertainty  from the best-fit model. Finally, the helium abundances
agree well, although a lot of upper or lower values are present.

Part of the larger dispersion (compared to  the optical analysis, see 
Fig.~\ref{fig:op_gholgado})
is attributed not to the effect of the improvements in {\sc Fastwind}
since \cite{Repolust2005} analyses were carried out (indeed, test
calculations by J.P. have shown that the impact of such improvements
on the IR signatures is marginal), but to the differences in the
by-eye (as used by Repolust et al.) and automatic techniques. When the
line fits are poorer, the subjective weight given to a particular fit
increases, pushing the result into a given direction, whereas the
automatic procedure still forces a compromise for all considered
profiles.

An extreme example is given by star number $\#$9 (HD 14\,947). By means
of our automatic fitting procedure, we find acceptable models (those
that contribute to the final parameters values) that extend up to
effective temperatures of 47\,000~K, because of the  uncertainties
 by a very weak \ion{He}{i} line, biasing the final parameters
toward hotter temperatures. As pointed out, the corresponding values
by Repolust et al. are much lower, mainly because they neglected the
deviations between synthetic and observed Br$_{10}$ and Br$_{11}$
lines.
 
\begin{figure*}[!h]
\begin{center}
\includegraphics[width=15.0cm]{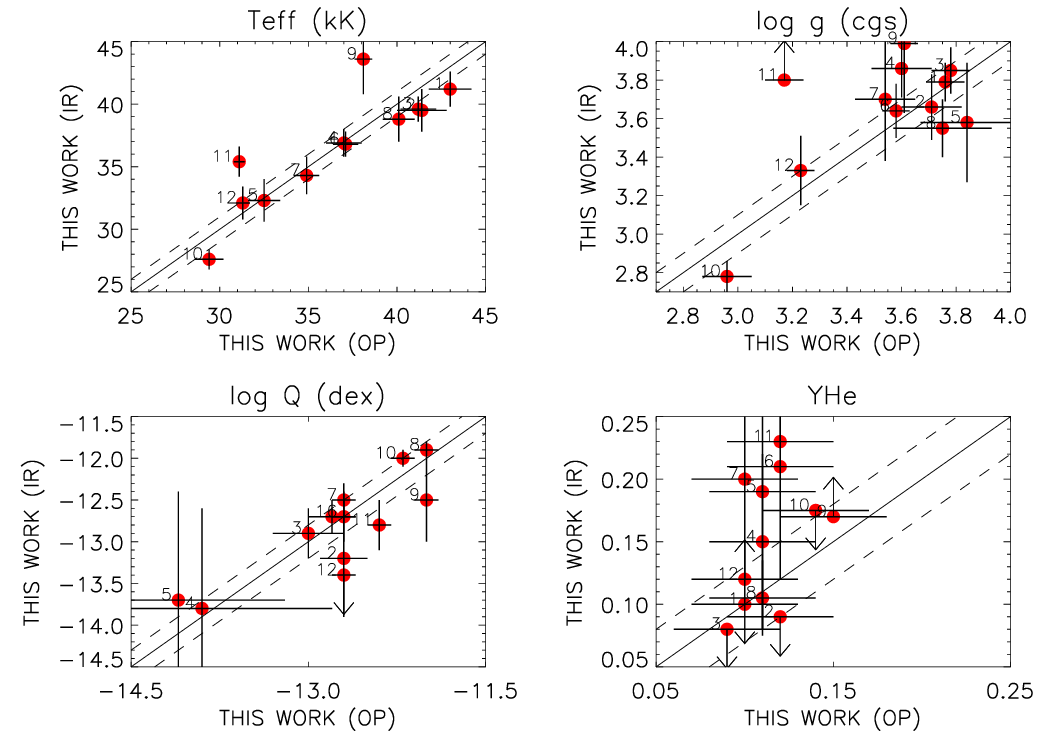}    
\caption{Comparison between stellar parameters obtained in the optical
and the infrared. Upper left panel: effective temperature. The dashed
lines represent $\pm$1000~K; upper right panel: logarithmic gravity
 ($\pm 0.1$~dex). Star $\#$9 has been slightly shifted
from its value in Tab.~\ref{tbl:iacob_IR_bli}; lower left panel: $\log
Q$ ($\pm$0.2~dex); lower right panel: \yhe ($\pm$0.03), and stars \#10
and 8 have been slightly displaced from their values in
Tab.~\ref{tbl:iacob_IR_bli}. Numbers indicate the stars as listed in
Tab.~\ref{tbl:stars_sample}. \label{fig:op_repol}}
\label{fig:comp_opir}
\end{center}
\end{figure*}

The final comparison is that of the parameters derived from the 
optical versus the NIR (Fig.~\ref{fig:comp_opir}), as this will indicate their reliability when 
derived from the infrared alone. Globally,
there is a fair global agreement within the errors, as shown by the
mean values of the differences $<\Delta \Teff>$ = $<\Teff (\rm{Opt}) -
\Teff (\rm{NIR})>$ = $-83\pm 697$~K, $<\Delta \log g>$ =
$-0.08\pm0.07$~dex, and $<\Delta \log Q>$ = $+0.08\pm0.10$~dex. Again,
stars $\#$9 (HD 14\,947) and  $\#$11 (HD 210\,809) show large
differences, produced by the higher gravities and their impact on
nearly all other parameters, and star$\#$12 (HD
209\,975) shows a too low \logq\ value in the infrared.
%
With more \ion{He}{i} lines, this effect does not
appear in the optical.  Finally, for the helium abundances, the average
agreement is poorer than for the other parameters ($<\Delta \yhe>$ =
$-0.04\pm0.01$), but in this case, the statistics is not as good due to
the large number of upper or lower limits present in the results.
Nevertheless, a certain trend to derive higher abundances in the near
infrared might become visible.

Globally, the infrared fits are worse than the optical ones, which
reflects in larger systematic uncertainties (partly related to few dominating
objects).
Moreover, inspection of the $\chi^2$
distributions from \texttt{iacob\_gbat} and the fits from
Fig.~\ref{fig:OP_profiles} indicate that this is not due to the
differences in resolution and S/N between the optical and
infrared spectra, but a consequence of a less accurate reproduction of
the infrared lines given the model-inherent assumptions
(e.g., a smooth wind until now). This finding is different from the
results quoted by \cite{Repolust2005} who found comparable errors in
both wavelength ranges, and reflects the different approach of error
determination and also the different fitting procedure itself.
Finally,  there is a relatively large number of objects for
which only upper or lower limits for the helium abundance could be
derived, suggesting a lack of sensitivity of the infrared spectrum to
that parameter (or a degeneracy because of the larger uncertainties
involved). In our case, the  problem lies partly in the lack
of a sufficient number of suitable \ion{He}{} lines, particularly from
\ion{He}{i}. 

Overall, however, we may conclude that we can use the infrared spectra
to determine stellar parameters in a similar way as we are used to do
with the optical ones, but we observe specific trends and larger
uncertainties that have to be taken into account.

\section{Clumping}
\label{sec_clump}

The line-driven winds from massive stars are prone to instabilities,
in particular the line-deshadowing instability (LDI, e.g.,
\citealt{ORI}), which result in an inhomogeneous outflow (e.g.,
\citealt{OCR88, Owocki91, Feldmeier95}). These density inhomogeneities
(clumps) modify the shape and strength of spectral lines formed in
the wind, and need to be accounted for in corresponding wind
diagnostics (e.g., \citealt{Hillier91, Schmutz1995, Hillier1998, crowther2002, 
Hillier2003, Bouret2003, Puls2006, Puls2008}). 
Particularly affected by
these inhomogeneities is the emission in lines formed through
recombination processes such as H$_\alpha$ or the NIR lines used as
wind diagnostics. 

In these processes, the emission is proportional to $\rho^2$, and it is
the difference between the averaged quantity $\langle\rho^2\rangle$
(integrated over the optical path length) and the corresponding smooth
wind quantity $\langle\rho\rangle^2$ that leads to more emission in an
inhomogeneous structure for the same mean density
$\langle\rho\rangle$, since $\langle\rho^2\rangle \ge
\langle\rho\rangle^2$ always.

Alternatively, for an observed emission, one derives a lower mass-loss
rate when adopting a clumped wind.  Moreover, as clumping may be
radially dependent, it may affect lines formed in different layers in
the atmosphere in a different way, which may help (at least partially)
to explain the discrepancies found in the previous sections when
fitting either optical or NIR lines. 

In the conventional approach considering optically thin clumps
(which is appropriate for the diagnostics investigated in the current
work, e.g., \citealt{Sundqvist2018}), the wind
structure is characterized by the so-called clumping factor, defined
as
\begin{equation}
\fcl = \frac{\langle\rho^2\rangle}{\langle\rho\rangle^2} \geq 1.
\end{equation}
Under the simplifying assumption that the interclump matter is
void, this clumping factor describes the clump overdensity $\rho_{\rm cl} =
\fcl \langle\rho\rangle$.

As long as \fcl\ is spatially constant, the wind emission in lines like
H$_\alpha$ will be the same when adopting either a smooth wind with
$\dot{M}_{\rm unclumped}$ or an inhomogeneous one with $\dot{M}_{\rm
clumped}$, if both mass-loss rates are related via
\begin{equation}
\label{Mdotclvsuncl}
\dot{M}_{\rm clumped} = \frac{\dot{M}_{\rm unclumped}}{\sqrt{\fcl}}.
\end{equation}
Thus, neglecting wind-clumping might lead to overestimated mass-loss
rates, at least if, as adopted, the clumps remain optically thin 
at all considered wavelengths. 

Optically thick clumping (also called ``macro-clumping''
or ``porosity'' -- including porosity in velocity space --, e.g.,
\citealt{OGS04, Oskinova07, owocki08, Surlan13, SPI, SPII, SPIII}) can
lead to additional changes, even when the clumps remain optically thin
for the majority of diagnostics/wavelengths. This is because important
transitions such as the Lyman ionization and/or the Lyman lines become
much easier optically thick than other processes (whenever neutral
hydrogen is non-negligible), and are then desaturated because of
porosity effects (for an instructive visualization of such effects,
see \citealt{Brands22}). Consequently, the hydrogen ionization and
excitation may change, leading to a change in the global radiation
field and (wind) plasma conditions\footnote{As long as clumps are
optically thick only for specific transitions from trace ions or less
abundant atomic species, porosity will affect the corresponding
diagnostics (e.g., the UV PV-diagnostics, see \citealt{Oskinova07,
Surlan13, SPI, SPII}), but not the global atmospheric model and
radiation field.}. Potentially affected are, in particular, the winds
from massive late-type B and A-stars, where this effect might explain
certain shortcomings in the current modeling of important
wind-diagnostics such as H$_\alpha$ from such objects. Test
calculations for O-type stars \citep{Sundqvist2018}, on the other
hand, indicate that in their parameter domain this should pose no
problem, since hydrogen remains highly ionized. Thus, in the
following, we will consider exclusively optically thin clumping.

\begin{figure}
\begin{center}
\includegraphics[height=8cm,angle=90]{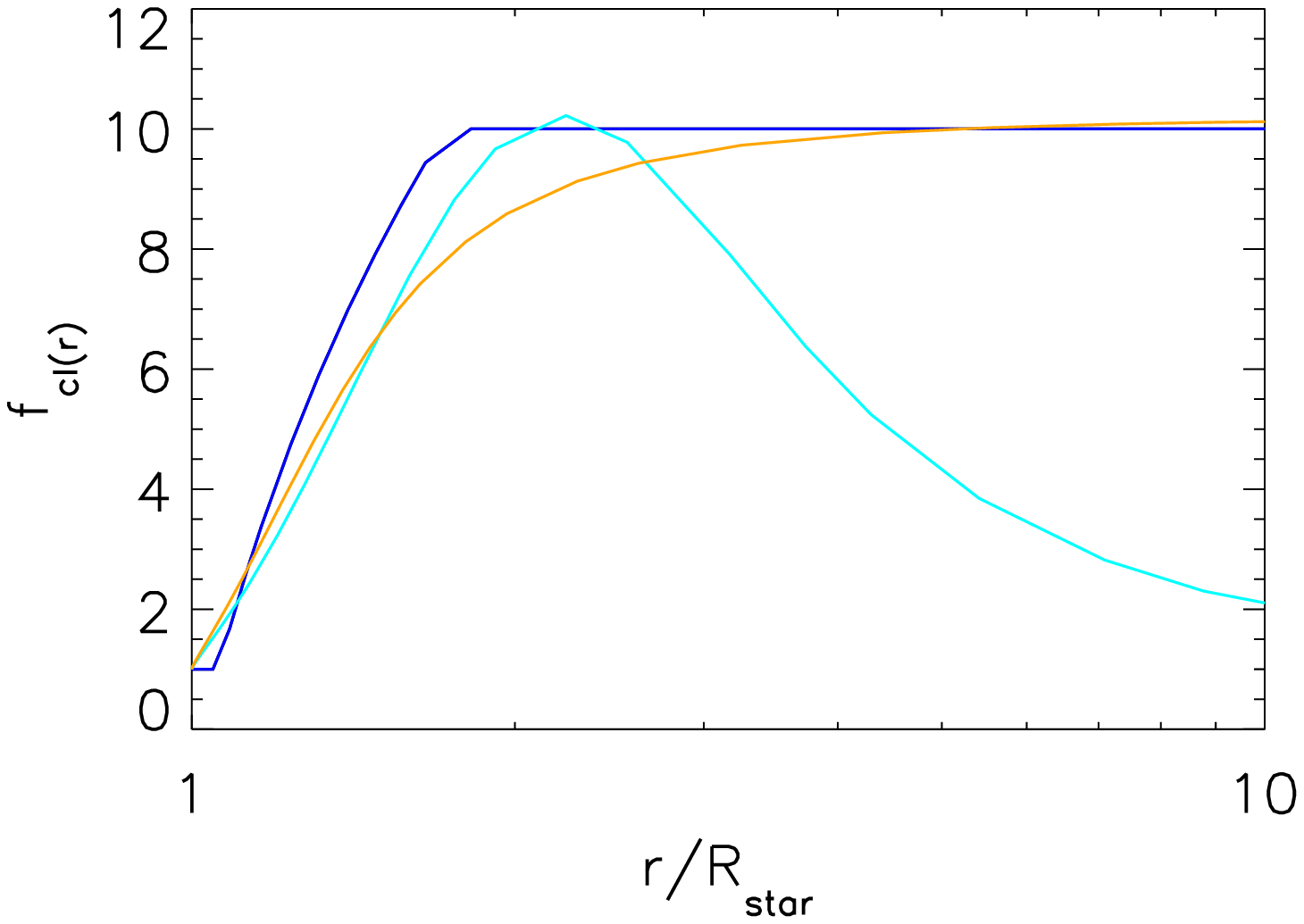}
\caption{Comparison between three different clumping laws investigated
in the current work (see Tab.~\ref{tbl:clf_mod}).  
Blue: Linear$_{10-050}$; orange:
Hillier$_{200}$; cyan: Najarro$_{200}$. The example refers to a
velocity law with \vinf= 1200 \kms and $\beta = 0.8$.}
\label{fig:cl3law}
\end{center}
\end{figure}

To this end, we compare three different clumping laws, $\fcl(r)$.
First, we consider a linear increase of the clumping factor from unity
(smooth density in the photosphere/lowermost wind) to a maximum value
between two points in the wind. After reaching this maximum, the
clumping factor is adopted to remain constant. We call this the
"linear law". The second law is the one suggested by
\cite{Hillier2003} (Hillier law), where the clumping
factor\footnote{in fact, Hillier and coworkers adopt the volume
filling factor as the basic quantity} follows an exponential increase
(as a function of velocity) until it reaches a maximum (and then stays
constant). Finally, our third law bases on \cite{Najarro2011}
(Najarro law) and is similar to the Hillier law in the lower wind,
but includes an exponentially decreasing $\fcl(r)$ beyond its maximum.
Fig~\ref{fig:cl3law} illustrates the different laws. The
"Najarrro law'' is motivated by results from a combined analysis of
UV, optical, NIR and L-band (including Br$_\alpha$) spectra for a
small O-star sample, as well as an NIR analysis of massive
stars in the Quintuplet Cluster \citep{Najarro2009},
and turns out to be quite similar to theoretical predictions from
radiation-hydrodynamic simulations including the LDI (e.g.,
\citealt{RO02, RO05}).

The considered clumping laws are, among others, implemented in
{\sc Fastwind}, and require specific input parameters, as detailed in the
following:

\begin{itemize}
\item[$\bullet$] the linear law is characterized by three parameters, $\fcl^{\rm max}$, 
$v_1$, and  $v_2$,
\begin{eqnarray}
   \fcl (v) & = & 1 \quad {\rm for} \quad v(r) < v_1 \nonumber \\
   \fcl (v) & = & 1 + (\fcl^{\rm max}-1) \times
        \left(\frac{v(r)-v_1}{v_2-v_1}\right)  \quad {\rm for} \quad
        v_1 \le v(r) \le v_2 \nonumber \\
   \fcl (v) & = & \fcl^{\rm max} \quad {\rm for} \quad v_2 < v(r)
\end{eqnarray}
where $\fcl^{\rm max}$ is the maximum value for the clumping factor,
$v_1$ is the wind velocity at clumping onset (restricted to be
larger/equal to the speed of sound), and $v_2$ is the velocity where 
maximum clumping shall be reached.

\item[$\bullet$] The Hillier law requires two input parameters and is 
expressed as
\begin{equation}
\label{hillierlaw}
f_{\rm V}(v) = f_{\rm V}^{\infty}+(1-f_{\rm V}^{\infty}) \cdot
   \exp{\left(-\frac{v(r)}{v_{\rm cl1}}\right)},
\end{equation}
where $f_{\rm V} \le 1$ is the volume filling factor (equal to the inverse
of $\fcl$ when the interclump medium is assumed to be void, as
frequently done). The two parameters defining this relation are
$f_{\rm V}^\infty$, the filling factor when the wind velocity reaches
the terminal velocity (corresponding to $1/\fcl^{\rm max}$ in our
tests), and $v_{\rm cl1}$ which marks the point where clumping begins
to become important and controls how fast the function reaches its
minimum. In this law, clumping begins to increase directly from the
bottom of the photosphere on, but becomes significant only for $v \ga
v_{\rm cl1}$.

\item[$\bullet$] the Najarro law is formulated as               
\begin{eqnarray}
\label{najarrolaw}
f_{\rm V}(v)& = &f_{\rm V}^{\infty}+(1-f_{\rm V}^{\infty}) \cdot
   \exp{\left(-\frac{v(r)}{v_{\rm cl1}}\right)}+ \nonumber \\  
   &+& (1-f_{\rm V}^{\infty}) \cdot 
   \exp{\left(- \frac{v_{\infty}-v(r)}{v_{\rm cl2}}\right)}
\end{eqnarray}
where $f_\infty$ and $v_{\rm cl1}$ are the same quantities as in
Hillier's law, whereas $v_{\rm cl2}$ prescribes how fast the filling
factor increases again after reaching its minimum (i.e., how fast
\fcl\ decreases after reaching its maximum). The
above clumping law has been modified compared to the original
formulation by \citet{Najarro2011}, enforcing an unclumped outermost
wind region with $f_{\rm V}(v \approx \vinf) \rightarrow 1$.
\end{itemize}

\begin{table*}
\begin{center}
\caption{The clumping laws used in our analyses. See text.}
\label{tbl:clf_mod}.
\begin{tabular}{|l|c|c|c|c|} \hline \hline 
Clumping law / label & $\fcl^{\rm max}$ & $v_1/\vinf$ & $v_2/\vinf$ &
discussed/used \\ \hline \hline
Linear$_{10-025}  $  & 10 & 0.1   &  0.25  & Sects. 5-7   \\ \hline 
Linear$_{10-050}  $  & 10 & 0.1   &  0.50  & Sects. 5-7   \\ \hline 
Linear$_{20-040}  $  & 20 & 0.1   &  0.40  & Sect.~5      \\ \hline 
Linear$_{20-094}  $  & 20 & 0.1   &  0.94  & Sect.~5      \\ \hline
Linear$_{20-025}  $  & 20 & 0.1   &  0.25  & Sects.~6-7   \\ \hline 
Linear$_{20-050}  $  & 20 & 0.1   &  0.50  & Sects.~6-7   \\ \hline \hline
Clumping law       & $\fcl^{\rm max}$  & $f_{\rm V}^{\infty}$  
& $v_{\rm cl1}$ [\kms] & $v_{\rm cl2}$ [\kms] \\ \hline \hline
Hillier$_{100}    $  & 10.5 & 0.095 &  100.  & --   \\ \hline 
Hillier$_{200}    $  & 10.5 & 0.095 &  200.  & --   \\ \hline 
Najarro$_{200}    $  & 10.3 & 0.095 &  200.  & 100. \\ \hline \hline 
\end{tabular}
\end{center}
\end{table*}

Table~\ref{tbl:clf_mod} displays the various parameters adopted for
our forthcoming tests. Overall, in the current section, we consider
four different linear laws\footnote{Table~\ref{tbl:clf_mod}
contains also two additional linear laws that will be considered in
Sects.~\ref{FWgrid_clump} and \ref{discussion}.}, two variants of the
Hillier law, and one of the Najarro law. 
For $v_1$ and $v_2$ (linear law) we adopt a compromise based on the
range of values provided by \cite{Najarro2011}, and fix these
quantities in terms of a specific fraction of the terminal velocity.
In this way, our $v_1$ and $v_2$ values (absolute velocities) are
consistent with the ranges obtained by Najarro et al. In summary, all
$v_1$ values have been fixed to 10\% of \vinf (see below),
whilst $v_2$ varies in between 25 to 94\% of \vinf.

When inspecting the current literature, the $\fcl^{\rm max}$
parameter covers a large range, from close to unity (unclumped) to
values as high as 100. Here, we will test the values $\fcl^{\rm max}$
= [10, 20], following Table 2 in \cite{Najarro2011}. Obviously, such
an approach has only an exploratory character, since it is highly
unlikely that all or most stars follow such a restricted combination
of the various parameters. Once we understand better how the profile
fits and the derived stellar parameters react to clumping, we will be
in a good position to consider at least $\fcl^{\rm max}$ as a free
parameter in our fitting approaches covering the IR band. 
Such studies have already started in analyses of the combined optical
and UV regime, cf. \citet{Hawcroft21}, \citet{Brands22}.

\begin{figure}[!htb]
\begin{center}
\includegraphics[width=6cm,angle=90]{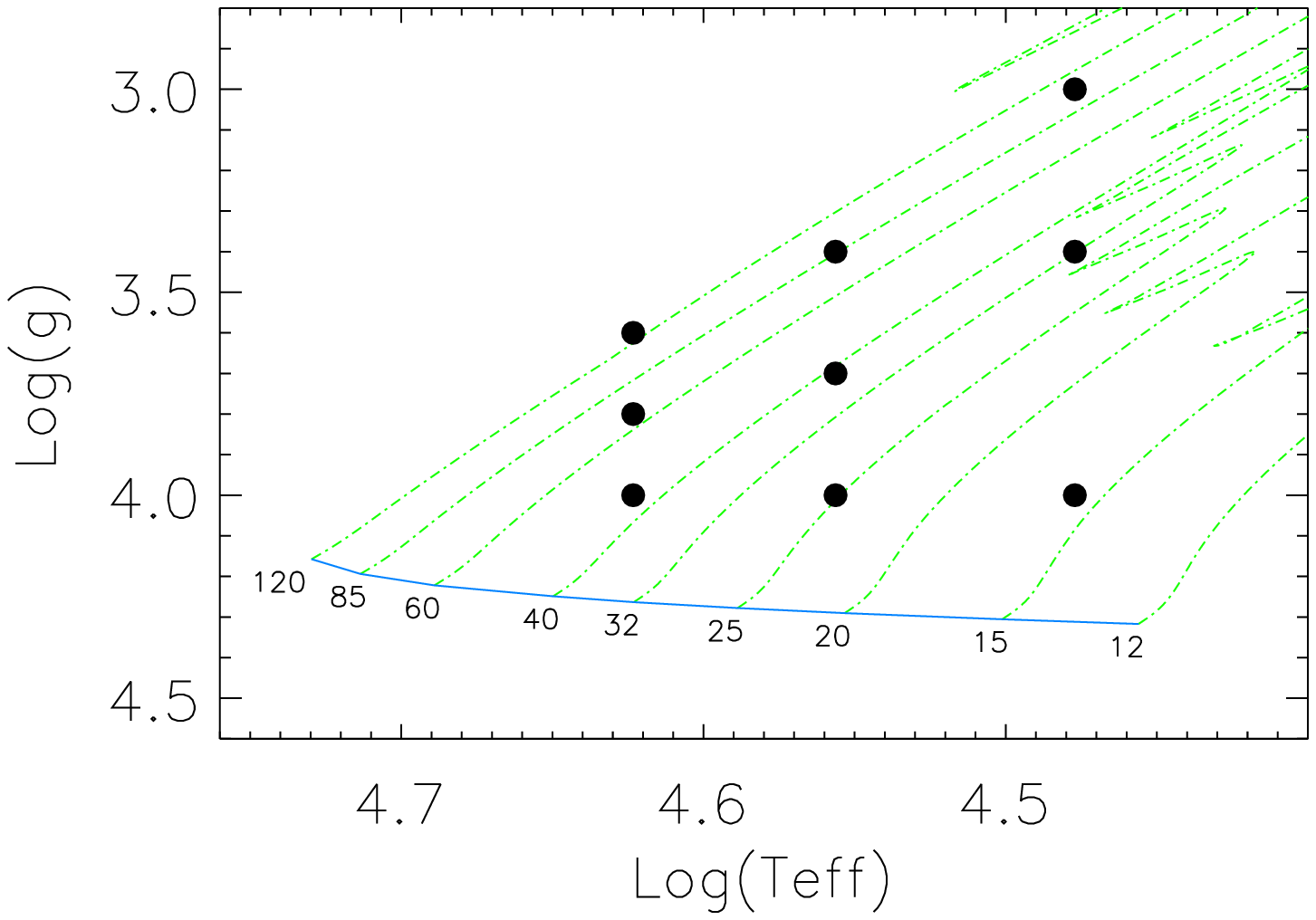}
\caption{Coarse-grid models in the $\log \Teff - \log g$ diagram,
chosen to be representative for hot dwarfs to ``cool'' supergiants in
the O-star regime. Overplotted in green are evolutionary tracks for
Galactic nonrotating stars from \cite{Ekstrom2012}, and the blue line
defines the corresponding ZAMS.  The numbers give the initial stellar
masses in units of \Msun.}
\label{fig:coarse_grid}
\end{center}
\end{figure}

\begin{figure}[!htb]
\includegraphics[width=5cm,angle=90]{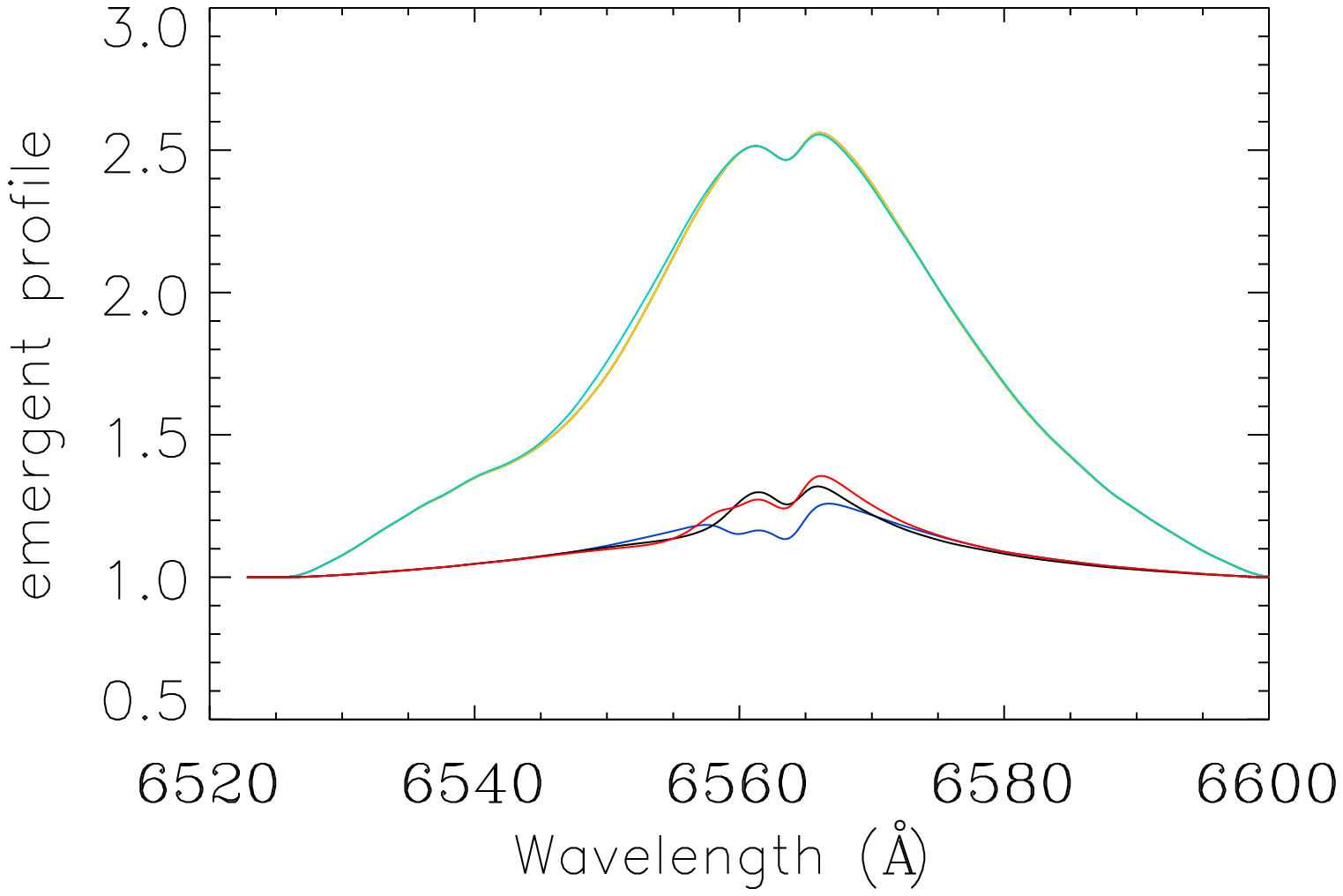}

\caption{
Comparison of synthetic H$_\alpha$ profiles, for models with \Teff\ =
36\,000\,K, $\log g = $3.40, and differing wind-clumping properties.
Black: unclumped wind with $\log Q = $-11.90. Orange: clumped model
with the same mass-loss rate/wind strength parameter, using the
Linear$_{10-025}$ law. Red: clumped model with the
same clumping properties/clumping law, but a mass-loss rate reduced by
$\sqrt{\fcl^{\rm max}}$. Green and blue: same as the
orange and red models, respectively, but using the
Linear$_{10-050}$ law (the orange and green profiles are nearly
coincident). All profiles have been broadened by \vsini\ = \vmac\ =
50\,\kms, adopting a resolving power of 12\,000.}
%
\label{fig:mcompensate}
\end{figure}

In our specific models based on the Hillier and Najarro laws, we
adopt values that result in a similar maximum as the linear law with
$\fcl^{\rm max} = 10$, and a similar increase toward this maximum
(see Fig.~\ref{fig:cl3law}). 
We check two Hillier laws, with the
Hillier$_{100}$ increasing faster toward maximum clumping than
Hillier$_{200}$ (similar to Linear$_{10-025}$ vs. Linear$_{10-050}$).
We finally note that the quantitative behavior of the Hillier and
Najarro laws, when expressed in terms of a radial coordinate, strongly
depends on the adopted velocity law (\vinf\ and $\beta$). 

\begin{table*}
\begin{center}
\caption{ Stellar and wind parameters for the coarse grid models.
For each luminosity class, we display \Teff (in kK), $\log g$, and the 
unclumped $\log Q$ values. For all models, a helium abundance of \yhe
= 0.1 and a velocity exponent $\beta = 0.8$ has been adopted.}
\label{tbl:small_clf_grid}
%

\begin{tabular}{|l|c|c|c||c|c|c||c|c|c|} \hline \hline 
& \multicolumn{3}{c||}{V} & \multicolumn{3}{c||}{III} & \multicolumn{3}{c}{I} \\ 
& \Teff & $\log g$ & $\log Q$ & \Teff & $\log g$ & $\log Q$ & \Teff & $\log g$ & $\log Q$ \\ \hline          
HOT   &  42 & 4.0 & -14. & 42 & 3.8 & -13. & 42 & 3.6 & -11.9  \\  
MID   &  36 & 4.0 & -14. & 36 & 3.7 & -13. & 36 & 3.4 & -11.9  \\
COOL  &  30 & 4.0 & -14. & 30 & 3.4 & -13. & 30 & 3.0 & -11.9  \\
\hline
\end{tabular}
\end{center}
\end{table*}

\subsection{FASTWIND coarse grid}

Before analyzing the impact of clumping by means of a comparison
between synthetic and actual spectra, we will test such impact for a
small set of template models. To this end, we construct a coarse grid
of models representing dwarfs, giants, and supergiants at different
temperatures (hot, mid, and cool), resulting in nine models covering
the O-star parameter range. In Fig.~\ref{fig:coarse_grid} we display 
these models in the $\log \Teff - \log g$ plane, to illustrate the
corresponding evolutionary stages. Table \ref{tbl:small_clf_grid}
lists these coarse grid models.
All models have the same (solar) helium abundance and $\beta$
velocity-field exponent. In the following sections, we will discuss
all coarse grid models and corresponding synthetic spectra resulting
from the application of the various clumping laws, and investigate and
compare their specific impact. 

%

\subsection{Clumping versus no clumping}
\label{clumpvsnoclump}

First, we explore some general clumping effects by means of our coarse
grid. Clumping modifies both radiative transfer and atomic occupation
numbers (because of the altered density and radiation field), thus
affecting the ionization equilibrium of all elements and consequently
the stellar/wind parameters derived from model fits. Even though in our
approach (at least) the subsonic stratification remains smooth, also
the photospheric lines might become affected by clumping, to a various
extent. This change is caused by the modified occupation numbers
resulting from a modified inward directed radiation field, and
particularly because of a modified filling of the absorption cores due
to a different wind structure.

\begin{figure*}
\begin{center}
\includegraphics[width=12cm,angle=90]{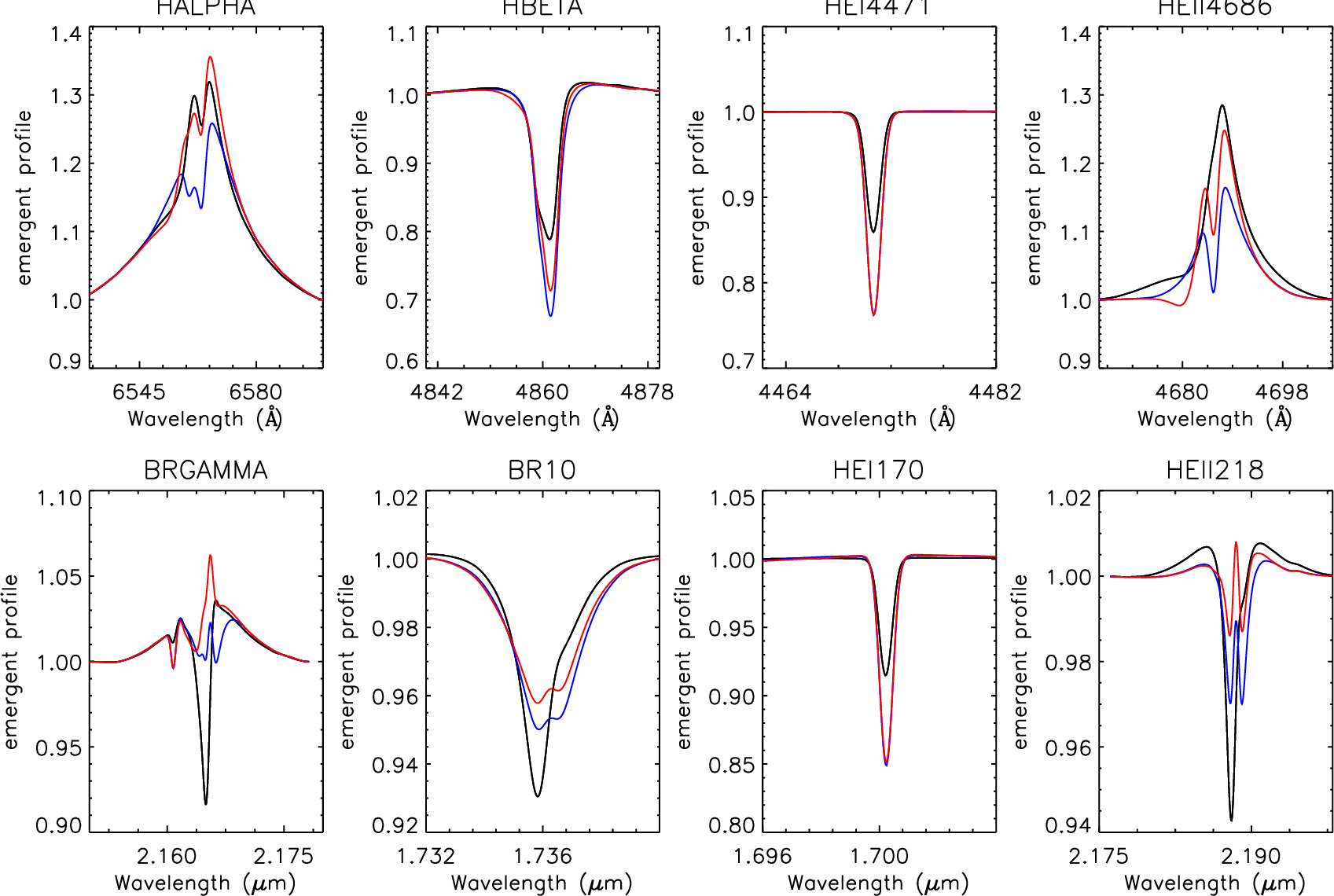}
\caption{Clumping effects for selected optical and NIR lines, for a
subset of the models from Fig.~\ref{fig:mcompensate}} (same broadening
parameters). Here, we compare the smooth model (in black) with the
clumped models with decreased (scaled) mass-loss rate, in red for the
Linear$_{10-025}$ law, and in blue for the Linear$_{10-050}$ one.
\label{fig:cleffects}
\end{center}
\end{figure*}

As already indicated, the most prominent effect is an increase of the
emission in lines such as H$_\alpha$. To obtain a similar emission in the
clumped and unclumped case (to provide us with a similar fit quality when performing a hypothetical fit), we need to divide the unclumped
mass-loss rate by $\sqrt{\fcl}$ (see Eq.~\ref{Mdotclvsuncl}); clearly, 
this is only an approximation, because of the radial
dependence of \fcl. This means that the wind strength parameter $Q$
for an unclumped wind will be (roughly) equivalent to a value
$Q'=Q/\sqrt{\fcl}$ for the clumped case, where in our approach we
approximate $\fcl$ by its maximum value, \fclmax. 

\begin{figure}
\begin{center}
\includegraphics[width=6.5cm,angle=90]{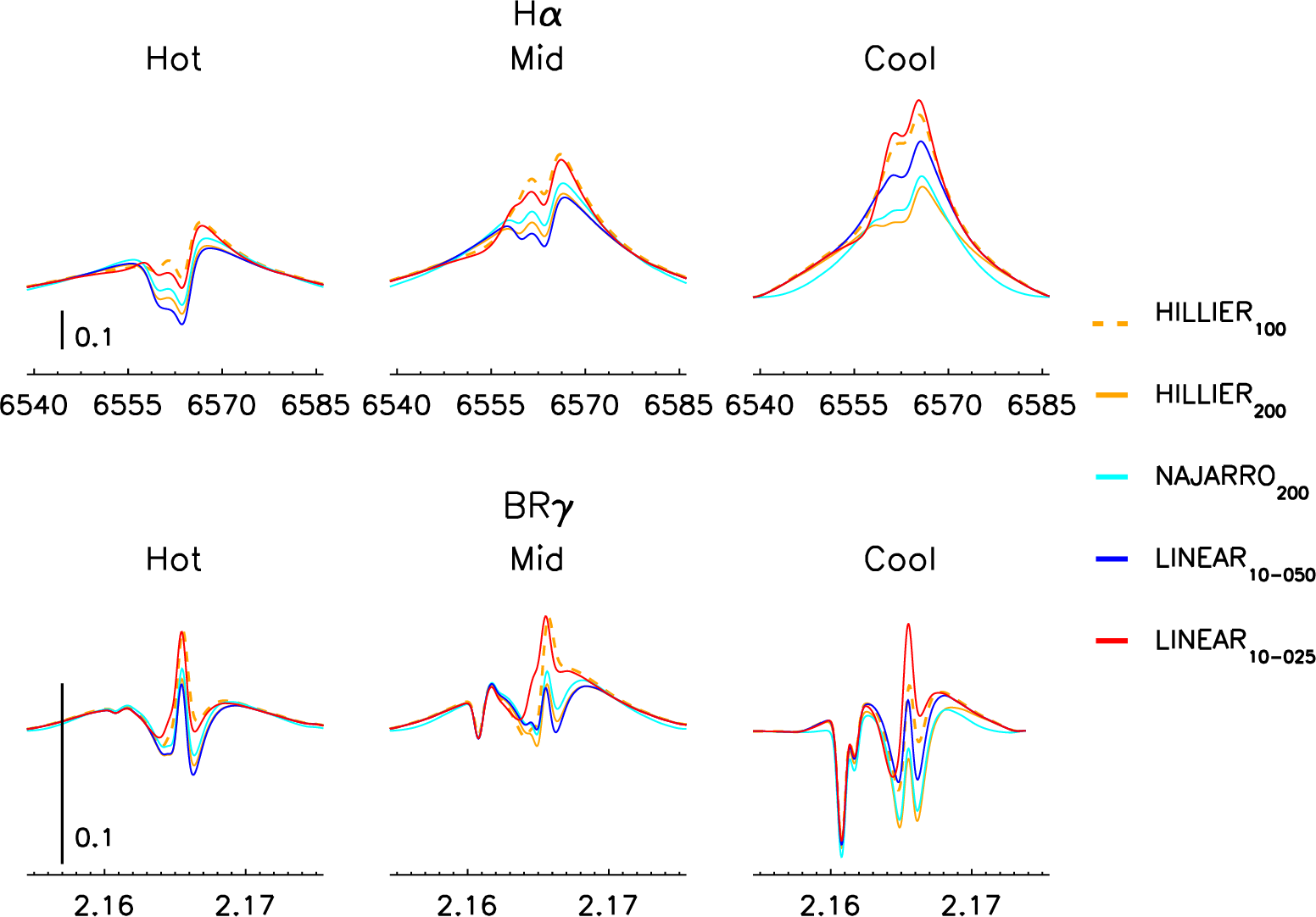}
\caption{H$_\alpha$ and Br$_\gamma$ profiles for clumped supergiant models 
 using different laws (see legend).
Mass-loss rates have been scaled according to \fclmax, and
profiles have been broadened as in Fig.~\ref{fig:mcompensate}. 
Wavelengths are given in \AA~ for H$_\alpha$ and in microns for Br$_\gamma$.
\label{fig:Hi_Pa_habg}}
\end{center}
\end{figure}

Figure~\ref{fig:mcompensate} illustrates the potentially strong impact
of clumping on the H$_\alpha$ line emission, by means of our grid
model with \Teff\ = 36\,000~K, $\log g = $3.40 and an (unclumped)
$\log Q = -11.90$ (corresponding to the "mid-temperature supergiant"
model). This unclumped model (profile in black) is compared with four
clumped ones. For two of those, we have used both the
Linear$_{10-025}$ (in red) and the Linear$_{10-050}$
law (in blue, for designations see
Tab.~\ref{tbl:clf_mod}) together with a reduced mass-loss rate, $\log
Q^\prime$ = -12.4 (because  of \fclmax\ = 10), to obtain a roughly equivalent
emission. As visible, all three H$_\alpha$ lines are fairly similar
indeed. The blue one (with $v_2 = 0.5 \vinf$)
displays a somewhat lower emission close to the core, because a large
part of the lower/intermediate wind has a lower {effective}
mass-loss rate than the model underlying the red
profile, where \fclmax\ is reached already at $0.25 \vinf$. The other
two profiles (in green and orange) have been calculated from clumped
models with identical clumping properties as above, but now with the
same mass-loss rate as in the unclumped case. The large difference 
is obvious, with an H$_\alpha$ emission
roughly corresponding to that of a smooth model with wind strength
parameter $\log Q + \sqrt{\fcl^{\rm max}} = -11.4$. Here, both
clumping laws deliver almost identical profiles, since due to the
larger densities the line formation zone shifts to the outer wind,
where both clumping laws are identical ($\fcl \approx \fclmax$).

Figure~\ref{fig:cleffects} shows, for the same mid-supergiant
parameters, the differences between the unclumped and clumped (scaled
\Mdot!) models, in selected optical and NIR spectral lines. Though, as
discussed above, the H$_\alpha$ emission remains almost identical,
other lines react differently. Br$_{10}$ (and also Br$_{11}$, not
shown) displays a weaker (and broader) absorption core for
the clumped models, and Br$_\gamma$ is affected even stronger: whereas
the unclumped model displays a slightly blue-shifted absorption, the
clumped ones show a narrow central emission (red profile), or only
weak absorption plus emission in the core region (blue profile).

Unlike these NIR H-lines, H$_\beta$ (clumped) presents {more}
absorption in the core, which is also true for the \ion{He}{i} lines
in both wavelength regimes. Since in the considered parameter range
the dominant helium ion is \ion{He}{iii} (for the main part of the
wind), \ion{He}{ii} lines behave similar to H lines when they are
dominated by recombination processes: in the NIR, \ion{He}{ii}
$\lambda$1.69 (not shown) and $\lambda2.18\mu$m show increased
emission in the core ( though on different scales), whilst \ion{He}{ii}
$\lambda$4686 remains mostly unchanged for the Linear$_{10-025}$ law,
in analogy to H$_\alpha$. For Linear$_{10-050}$ the emission is
clearly weaker, because of the lower effective mass-loss rate. 
In cooler winds, when \ion{He}{iii} is no longer dominant,
\ion{He}{ii} $\lambda$4686 will behave differently from H$_\alpha$
(see \citealt{Kudritzki2006}).

For most lines, the line formation regions will be altered as a
consequence of the different density structure in the clumped models.
In particular, the increased absorption of many lines can be explained
by their formation in the inner layers, before clumping plays a
decisive role. In those cases, the dominant effect will be the
decreased mass-loss rate in the clumped, \Mdot-scaled models, resulting
in a deeper absorption (less refilling than in the smooth models with
larger \Mdot). 

As well, the line emission at the cores of Br$_\gamma$ and \ion{He}{ii}
$\lambda$2.18$\mu$m is a (indirect) consequence of the modified
formation depth. Concentrating on Br$_\gamma$, we see at first that
the wind emission in the line wings is almost identical
for all three wind structures\footnote{except for
the \ion{He}{i} component blueward from line center, which is stronger in the
clumped, low-\Mdot\ model, see above.}, implying that such emission
forms in the intermediate/outer wind where our scaling via $\Mdot
\sqrt{\fcl^{\rm max}} = {\rm const}$ is applicable. 

The differences at line center, on the other hand, relate to different
NLTE conditions in the upper photosphere/lower wind. For the unclumped
model (with larger \Mdot), the apparent absorption results from a
comparatively low source function, when the lower level, $n=4$,
becomes overpopulated compared to the upper one, $n=7$. For the
clumped models, with lower \Mdot\ in the still smooth transonic
region, we find a similar effect as observed and modeled for
Br$_\alpha$ from weak-winded O-stars (\citealt{Najarro2011}). Also
here, the lower level becomes underpopulated compared to the upper one
in the transonic region, increasing the source function considerably,
and resulting in a narrow emission peak close to line center. 
From test calculations with an analogous unclumped model with identical,
low mass-loss rate as the clumped models analyzed here, we find a
similar emission peak (now inside a broad photospheric absorption --
no wind emission in this case). 
To summarize, the central emission observed in various lines is often not
directly related to clumping, but occurs from specific NLTE effects in
the upper photosphere when the line is formed in the transonic region,
where its strength is highly dependent on mass-loss rate.

Finally, we note that the models presented in Fig.~\ref{fig:mcompensate} and
Fig.~\ref{fig:cleffects} show the overall strongest effects within all
models of our coarse grid. In general, the supergiant models (hot,
mid, and cool) display the most pronounced effects, whilst for giants
we find smaller changes, becoming negligible for dwarf models.

\subsection{Which clumping law to use?}

The calculation of a full model grid is a numerically expensive task. Thus, before
analyzing the real spectra, we performed a series of tests using the
coarse grid to evaluate the differences among the clumping laws
provided in Table~\ref{tbl:clf_mod}. Our aim is to minimize the computational
effort when considering the full grid. Fig.~\ref{fig:Hi_Pa_habg} visualizes
the changes in the H$_\alpha$ and Br$_\gamma$ profiles from the most sensitive
supergiant models (see Tab.~\ref{tbl:small_clf_grid}) when applying the different clumping laws.

\subsubsection{Hillier vs. Najarro}
\label{HilliervsNajarro}

We first compare the Hillier$_{200}$ with the Najarro$_{200}$ clumping
laws (orange and cyan in Fig.~\ref{fig:Hi_Pa_habg}; 
see Table~\ref{tbl:clf_mod}, Fig.~\ref{fig:cl3law} and
Eqs.~\ref{hillierlaw}, \ref{najarrolaw}).  The main difference between
both laws concerns the outer wind layers, after the maximum clumping
factor in the Najarro$_{200}$ law has been reached. Thereafter, the
clumping factor decreases toward unity (no clumping) in case of
Najarro$_{200}$, whilst it continues to increase asymptotically for
Hillier's law, reaching its maximum at the outer wind boundary. 

In Fig.~\ref{fig:Hi_Pa_habg} we can see the impact of these two laws on
H$_\alpha$
and Br$_\gamma$.
Indeed, the line profiles are very similar, and
corresponding giant and dwarf models display even lower, almost
invisible differences. This is not only true for the above two lines,
but also for all H and He lines considered in the current study (not
shown here for brevity).
We conclude that there are no significant differences when using
either the Hillier or the Najarro law for the analysis of optical and
NIR H/He spectra of typical O-stars. 

The simple reason for these almost identical line shapes is that the
lines have already formed when both laws begin to deviate\footnote{The (small) differences between both laws in the inner wind
(Fig.~\ref{fig:cl3law}) do not play any role.}. Both NIR and optical
lines are formed below 2$R_{*}$, and the influence of the clumping law
beyond this point is irrelevant, contrasted to wavelengths in the UV,
far-IR or radio regimes where corresponding diagnostics might form at
much larger radii for sufficiently strong mass loss. For the purpose
of our present work, however, we can restrict ourselves to one of
these clumping laws, which, because of its higher simplicity, is the
one suggested by Hillier.
        

\subsubsection{Hillier vs. Linear}

We now compare the profiles obtained from the Hillier$_{200}$ and the
Linear$_{10-050}$ laws (orange vs. blue in Fig.~\ref{fig:Hi_Pa_habg}; 
see Table~\ref{tbl:clf_mod}). The differences
between both laws (see Fig.~\ref{fig:cl3law}) are larger than those
considered in the previous subsection, though in the inner layers, where
most of the optical and NIR lines are formed, they are quite similar.
It is thus not surprising that the largest differences between the
resulting line profiles, shown in Fig.~\ref{fig:Hi_Pa_habg}, are moderate.
Again, the largest differences are found for the supergiant models,
particularly at ``cool'' temperatures, whereas the giant and dwarf
models display no significant differences at all.

The already small discrepancies between the (supergiant) line profiles
might become even smaller when the clumping-law is altered. 
In the same figure, we also display the results for the 
Hillier$_{100}$ and Linear$_{10-025}$ laws
(dashed orange vs. red), i.e., when using
lower values for $v_2$ (in both cases, a factor of two lower than
before). Now, the profile differences have almost vanished.

Summarizing, the prime differences between clumped and unclumped
models mostly relate to the region of line formation and the overall
clumping distribution, though not on the details of the specific
clumping law (as long as there are enough parameters to describe the
essential behavior).

Consequently, we conclude that for a first study, it is sufficient to
consider only one family of clumping laws, and we decided to use the
simple linear one.



\begin{figure}
\begin{center}
\includegraphics[height=8cm,angle=90]{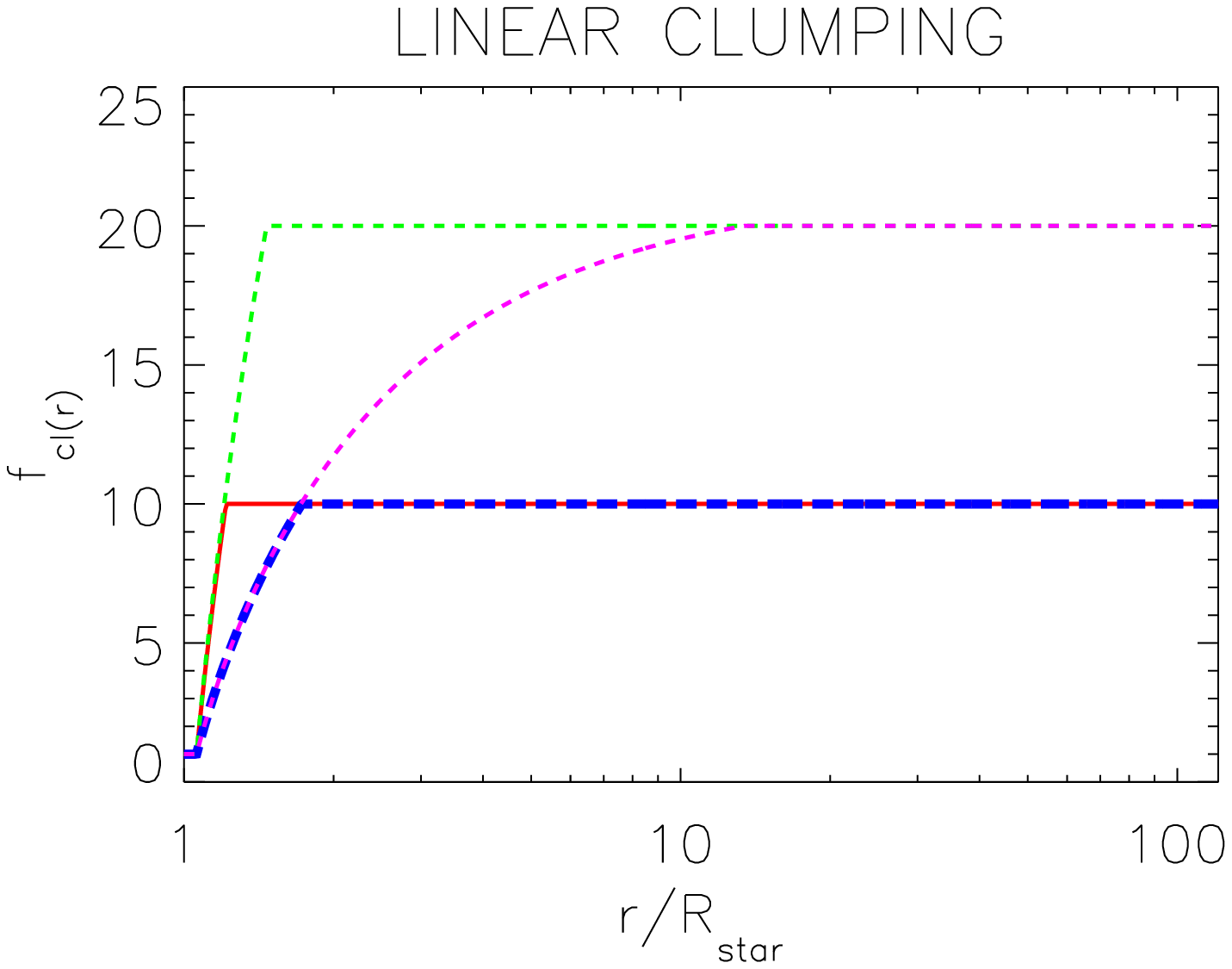}
\caption{Four different linear clumping laws considered in our study,
with the clumping factor as a function of stellar radius (in
units of the photospheric radius, $R_*$, with $\beta$= 0.8). Red solid line:
Linear$_{10-025}$; dashed blue line: Linear$_{10-050}$; dashed green line:
Linear$_{20-040}$; dashed magenta line: Linear$_{20-094}$. See text.}
\label{fig:clfactors}
\end{center}
\end{figure}

\begin{figure}
\begin{center}
\includegraphics[width=6.5cm,angle=90]{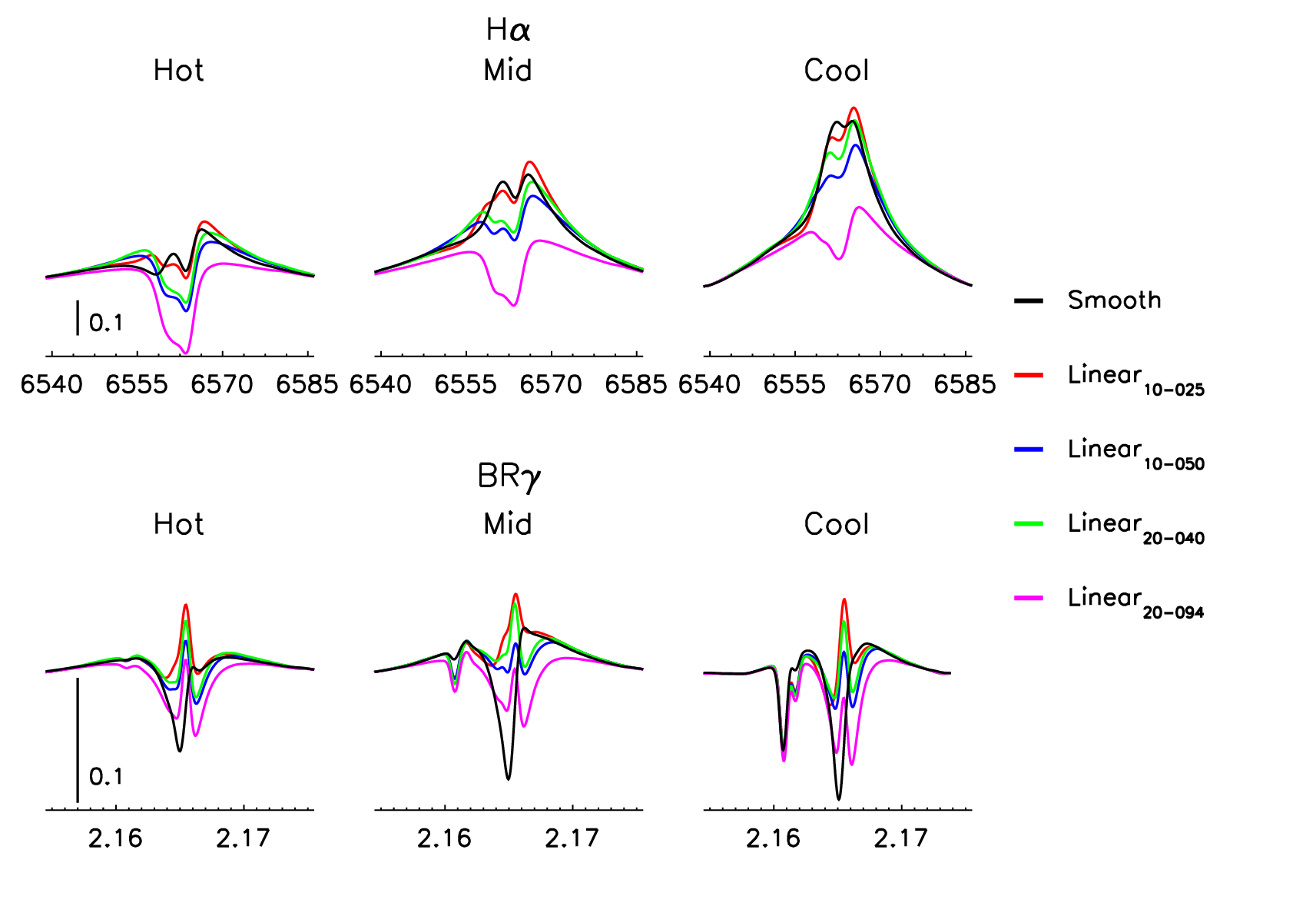}
\caption{H$_\alpha$ and Br$_\gamma$ profiles for the different clumping laws as
indicated, including smooth winds}. Mass-loss rates of clumped models have been scaled
according to \fclmax, and profiles have been broadened as in
Fig.~\ref{fig:mcompensate}.
\label{fig:profinner}
\end{center}
\end{figure}

\subsection{The Linear law: Varying the parameters}

In the following, we explore the changes introduced when modifying the
parameters of such linear laws.  We concentrate on the maximum
clumping factor, \fclmax, and the point where this maximum is reached,
$v_2$. We fix the point of clumping onset, $v_1/\vinf$ = 0.1, since
this value has only a weak impact on the results as long as it is
sufficiently small (0.1 {\ldots} 0.2), but larger than the speed of
sound to keep the photosphere unclumped. This latter condition might
need to be relaxed in forthcoming studies, given the possibility that
also the photosphere might be affected by inhomogeneities (e.g.,
\citealt{Puls2006, Cantiello2009}).

Figure~\ref{fig:clfactors} shows the four linear laws. 
For \fclmax, we consider two typical values, \fclmax = 10 and 20.
For \fclmax=10, we choose two values for the point where this maximum
is reached, namely $v_2/\vinf$ = 0.25 (Linear$_{10-025}$) and 0.5
(Linear$_{10-050}$), to simulate a rather steep and a moderate increase.
To investigate the impact of clumping also in the outer wind (in
addition to our considerations from Sect.~\ref{HilliervsNajarro}) and
in a systematic way, we proceed as follows. The $v_2$ values for 
the  \fclmax=20-laws (first two of the corresponding
entries in Table~\ref{tbl:clf_mod}) are chosen such that the
specific clumping factors are identical to their
\fclmax=10-counterparts in the inner wind, until \fcl= 10 is reached,
and then continue to increase until their maximum value, \fclmax=20.
This results in $v_2/\vinf$ = 0.40 (Linear$_{20-040}$) and 0.94
(Linear$_{20-094}$), respectively.


Again, in Fig.~\ref{fig:profinner} we only display the line profiles obtained for the supergiant
models  (with mass-loss rates scaled by the corresponding
factor, $(\fclmax)^{-1/2}$. At first we compare the H$_\alpha$ and Br$_\gamma$
profiles for the Linear$_{10-025}$ and Linear$_{20-040}$ laws
(red vs. green), i.e., when 
the maximum clumping factor is reached in the inner wind layers,
together with the profiles from the corresponding unclumped models
(in black).

We see that the H$_\alpha$ profiles are similar for the unclumped and
Linear$_{10-025}$ models, whereas the profiles for the Linear$_{20-040}$
law are somewhat different for the hot and mid-temperature
supergiants, with less emission at lower velocities in the latter
cases. This indicates that H$_\alpha$ is formed in a region where 
clumping fully compensates the lower mass-loss rate in
Linear$_{10-025}$ (i.e., beyond $v(r)/\vinf = 0.25$), but where
this is not yet the case for the Linear$_{20-040}$ law. We
conclude that the differences between the two clumped models are due to
the formation of H$_\alpha$ between $v(r)/\vinf = 0.25 {\ldots} 0.40$.

Contrasted to this behavior, the Br$_\gamma$ profiles of both clumped
models show a strong central emission, very similar to each other, and
differing from the (partly blue-shifted) absorption of
the unclumped wind. Again, however, all emission wings are identical. 
In agreement with our argumentation from Sect.~\ref{clumpvsnoclump},
we conclude that the wind emission in Br$_\gamma$ is mostly formed in
layers where \fcl\ has already reached its maximum value (i.e., beyond
$v(r)/\vinf= 0.4$). The more central absorption or emission is
controlled by the behavior of level $n=4$ vs. level $n=7$ in the
transonic regime, with absorption for larger and emission for lower
mass loss rates. Obviously, also the redward Stark-absorption wing  becomes
visible for the lowest mass-loss rate (Linear$_{20-040}$),

A second comparison refers to Linear$_{10-050}$ (blue) vs.
Linear$_{20-094}$ (magenta). Here the clumping degree increases more slowly with
radius than above, and H$_\alpha$ is majorly formed before the maximum
clumping factor is reached. As a consequence, the decrease in
mass-loss rate produces a lower wind emission in both clumped models. 
The effect is stronger for Linear$_{20-094}$, because of the larger
decrease in \Mdot. Now, also for Br$_\gamma$ the line wings deviate
from each other, with decreasing impact of wind emission, and
particularly Linear$_{20-094}$ displays a line profile dominated by
photospheric absorption. Consistent with our previous argumentation,
the extent of the central emission remains fairly unaffected by the
differences in clumping (though it depends on the actual mass-loss
rate).

\begin{table}
\begin{center}
\caption{Range of parameters used to produce the grids of synthetic
profiles for clumped winds with different clumping stratifications.
As for the smooth wind grid, the metallicity composition is solar.
$\log Q$ values refer to unclumped winds. Units as in
Tab.~\ref{tbl:range_grid}.}
\label{tbl:range_grid_clump}
\begin{tabular}{ll} \hline \hline
Parameter & Range of values \\ \hline
\Teff [K]  & [22000--55000]  (stepsize 1000 K) \\
\grav~ [$g$ in cgs] &[2.6--4.3] (stepsize 0.1 dex)\\
$\vmic$ [km s$^{-1}$] &5,10,15,20\\
\yhe  &  0.06, 0.10, 0.15, 0.20, 0.25, 0.30\\
\logq       & -15.0, -14.0, -13.5, -13.0, -12.7, -12.5, \\
                      & -12.3, -12.1, -11.9, -11.7\\
$\beta$       & 0.8, 1.0, 1.3\\ \hline
\end{tabular} 
\end{center}
\end{table}


Comparing now all five models in parallel, we conclude that
\begin{enumerate}
\item 
   the wind emission increases when the maximum clumping factor is
   reached in the inner wind layers. In such models, the lines are
   formed in regions when clumping already fully compensates the
   decrease in mass-loss rate.
\item the maximum value $\fcl^{\rm max}$ is of less relevance
   whenever the clumping factor increases over an extended region.
   What actually matters is the value of the clumping factor in the
   line-forming region, together with the global mass-loss rate.
\end{enumerate}

\begin{table*}
\begin{center}
\caption{Stellar parameters obtained from the optical analysis using
the Linear$_{10-025}$ clumping law. Upper and lower limits refer to the corresponding parameter ranges of our
model grids only (see Table~\ref{tbl:range_grid_clump}).}
\label{tbl:iacob_OP_cl1}
\begin{small}
\begin{tabular}{|c|c|c|c|c|c|c|} \hline
Star       &     \Teff\ (kK)   &     $\log g$ (dex) & $ \log Q $   & \yhe  & \vmic\ (\kms)    & $\beta$ \\ \hline  
   HD46223 & 43.4 $\pm$0.9  &3.83 $\pm$ 0.07 & -13.1 $\pm$ 0.1 & 0.10 $\pm$ 0.03  &10.2 $\pm$ 5.2  & $>$ 1.0  \\         
   HD15629 & 42.3 $\pm$1.8  &3.78 $\pm$ 0.10 & -13.1 $\pm$ 0.1 & 0.12 $\pm$ 0.03  &12.4 $\pm$ 7.4  & $>$1.0    \\         
   HD46150 & 40.0 $\pm$0.8  &3.80 $\pm$ 0.08 & -13.4 $\pm$ 0.2 & 0.10 $\pm$ 0.03  &$<$11.8         & $>$0.8    \\         
  HD217086 & 37.0 $\pm$1.0  &3.60 $\pm$ 0.10 & -13.5 $\pm$ 1.1 & 0.11 $\pm$ 0.03  &12.4 $\pm$ 7.4  & 1.0 $\pm$0.2  \\         
  HD149757 & 32.5 $\pm$0.9  &3.82 $\pm$ 0.17 & -14.0 $\pm$ 1.0 & 0.11 $\pm$ 0.03  &12.0 $\pm$ 7.0  & $>$0.8        \\         
  HD190864 & 37.2 $\pm$0.8  &3.60 $\pm$ 0.10 & -13.1 $\pm$ 0.1 & 0.12 $\pm$ 0.03  &10.4 $\pm$ 5.4  & $>$0.8        \\         
  HD203064 & 35.0 $\pm$0.5  &3.50 $\pm$ 0.06 & -13.1 $\pm$ 0.1 & 0.10 $\pm$ 0.03  &$>$13.7         & 1.0 $\pm$0.2  \\         
   HD15570 & 39.8 $\pm$0.6  &3.48 $\pm$ 0.07 & -12.4 $\pm$ 0.1 & 0.10 $\pm$ 0.03  &$<$19.9         & 1.1 $\pm$0.1  \\         
   HD14947 & 38.0 $\pm$0.2  &3.50 $\pm$ 0.03 & -12.5 $\pm$ 0.1 & 0.14 $\pm$ 0.03  &$<$11.3         & $>$1.2        \\         
   HD30614 & 29.1 $\pm$0.2  &$<$2.83         & -12.6 $\pm$ 0.1 & $>$0.20          &$>$18.4         & 1.1 $\pm$0.1  \\         
  HD210809 & 31.0 $\pm$0.8  &3.05 $\pm$ 0.12 & -12.7 $\pm$ 0.1 & $>$ 0.13         &$>$14.9         & 1.1 $\pm$0.2        \\         
  HD209975 & 31.5 $\pm$0.6  &3.26 $\pm$ 0.09 & -13.1 $\pm$ 0.2 & 0.10 $\pm$ 0.03  &$<$12.1         & 1.0 $\pm$0.2  \\ \hline  
\hline
\end{tabular} 
\end{small}
\end{center}
\end{table*}                                                                                                                 
\begin{table*}
\begin{center}
\caption{As Table~\ref{tbl:iacob_OP_cl1}, but for the NIR analysis
using the Linear$_{10-025}$ clumping law. For upper and lower limits see caption of Table~\ref{tbl:iacob_OP_cl1}.}
\label{tbl:iacob_IR_cl1}
\begin{small}
\begin{tabular}{|c|c|c|c|c|c|c|}\hline
Star       &     \Teff\ (kK)   &     $\log g$ (dex) & $  \log Q $   & \yhe  & \vmic\ (\kms)    & $\beta$ \\ \hline     
   HD46223 &42.7 $\pm$1.7 &3.83 $\pm$ 0.10 &-14.1 $\pm$ 1.4 &$<$ 0.10       & $>$5.0   &$<$ 1.3            \\         
   HD15629 &40.8 $\pm$1.2 &3.85 $\pm$ 0.10 &-13.0 $\pm$ 1.3 & 0.10 $\pm$ 0.03& $<$19.9  &$>$ 0.8            \\         
   HD46150 &39.5 $\pm$0.8 &3.85 $\pm$ 0.11 &-13.1 $\pm$ 0.2 &$<$0.08         & 12.1 $\pm$ 7.1& $>$0.9       \\         
  HD217086 &36.8 $\pm$1.1 &3.88 $\pm$ 0.11 &-14.2 $\pm$ 1.3 &0,13 $\pm$ 0.07 & $>$5.0   &$>$ 0.8            \\         
  HD149757 &32.5 $\pm$1.6 &3.52 $\pm$ 0.22 &$<$-13.3         &0.13 $\pm$ 0.07 & $>$9.4   &$<$ 1.3            \\         
  HD190864 &37.5 $\pm$1.0 &3.85 $\pm$ 0.10 &-12.9 $\pm$ 0.1 &$>$0.16        & $>$11.8  &$>$ 1.1            \\         
  HD203064 &35.6 $\pm$0.9 &3.87 $\pm$ 0.06 &-12.8 $\pm$ 0.1 &0.15 $\pm$ 0.05 & $>$8.4   &$>$ 1.2            \\         
   HD15570 &39.1 $\pm$0.3 &3.52 $\pm$ 0.03 &-12.4 $\pm$ 0.1 &$<$ 0.08        & $<$19.9  & 1.1 $\pm$0.1      \\         
   HD14947 &41.9 $\pm$1.0 &$>$3.94         &-12.6 $\pm$ 0.1 &$>$ 0.15        & $>$5.0   &$>$ 1.1            \\         
   HD30614 &28.5 $\pm$0.6 &$<$2.90         &-12.7 $\pm$ 0.1 &$<$ 0.09        & $>$10.1  & 1.0 $\pm$ 0.2     \\         
  HD210809 &34.6 $\pm$1.3 &$>$3.43         &-12.6 $\pm$ 0.1 &$>$ 0.08        & $>$5.0   &$>$ 0.8            \\         
  HD209975 &32.5 $\pm$1.0 &3.39 $\pm$ 0.13 & $<$-13.6       &0.15 $\pm$ 0.05 & $>$14.8  &$>$ 0.8            \\ \hline  
\hline
\end{tabular} 
\end{small}
\end{center}
\end{table*}
\begin{figure*}[!htb]
\begin{center}
\includegraphics[width=13cm,angle=90]{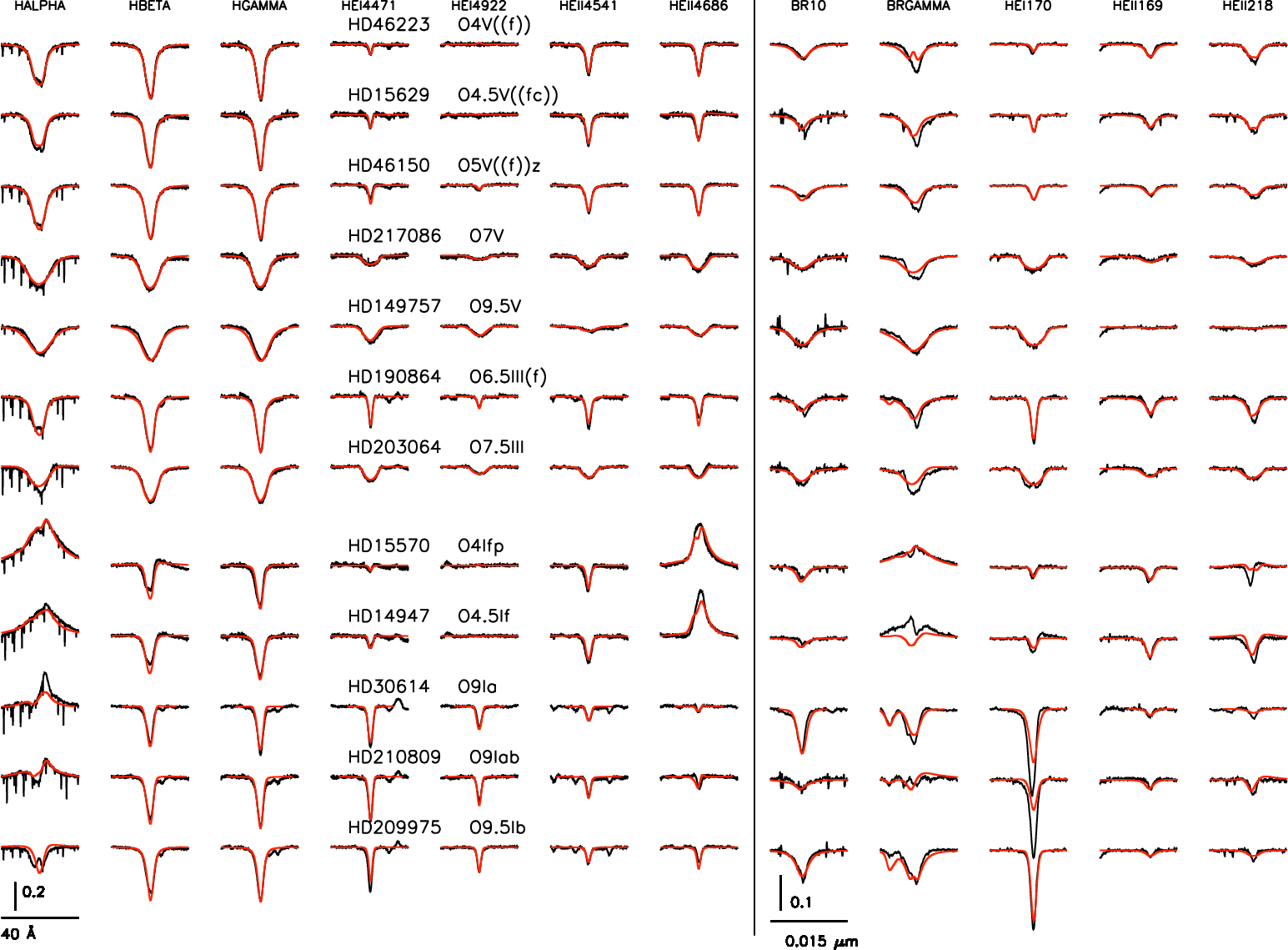}
\caption{As Fig.~\ref{fig:OP_profiles} using the clumping law Linear$_{10-025}$.}
\label{fig:OP_cl1_profiles}
\end{center}
\end{figure*}

For the rest of our current study, and given its exploratory
character, we will restrict our analysis to the linear
clumping description. On the one hand, we will use the same
Linear$_{10-025}$ and Linear$_{10-050}$ laws considered above. The two laws
with \fclmax = 20 as discussed in this section, however, are ``only''
linear extensions of these laws toward larger radii, studied to
investigate potential effects from a highly clumped outermost wind.
Since we argued that the decisive quantity is the value of the
clumping factor {in the line-forming region} (often dominated
by the lower and intermediate wind), in the next two sections we will
use two alternative \fclmax = 20-laws  (see below).
In this way, we are able to simulate a larger diversity of potential
line shapes and physical conditions, although this is still
a severe
simplification. For example, the most recent optical + UV analysis by
\cite{Hawcroft23} indicates a (maximum)
clumping factor that increases with \Teff, and in future work a more
extended parameter range (with respect to $\fcl^{\rm max}$ and $v_2$)
needs to be examined also in the NIR. 


\section{FASTWIND clumping grid}
\label{FWgrid_clump}
For a (re)analysis of our optical and NIR observations
using clumped models, we have calculated a full model grid
and restricted ourselves to four clumping laws in total: 
the Linear$_{025}$ laws, with [$v_1/\vinf, v_2/\vinf$] =
[0.1, 0.25], and
the Linear$_{050}$ laws with [$v_1/\vinf, v_2/\vinf$] = [0.1, 0.50] (see
Tab.~\ref{tbl:clf_mod}), applying \fclmax\ = 10 and \fclmax\ = 20 in both
cases.  Table \ref{tbl:range_grid_clump} shows the
parameter ranges for the grids.

\begin{figure*}
\begin{center}
\includegraphics[width=12cm,
angle=90]{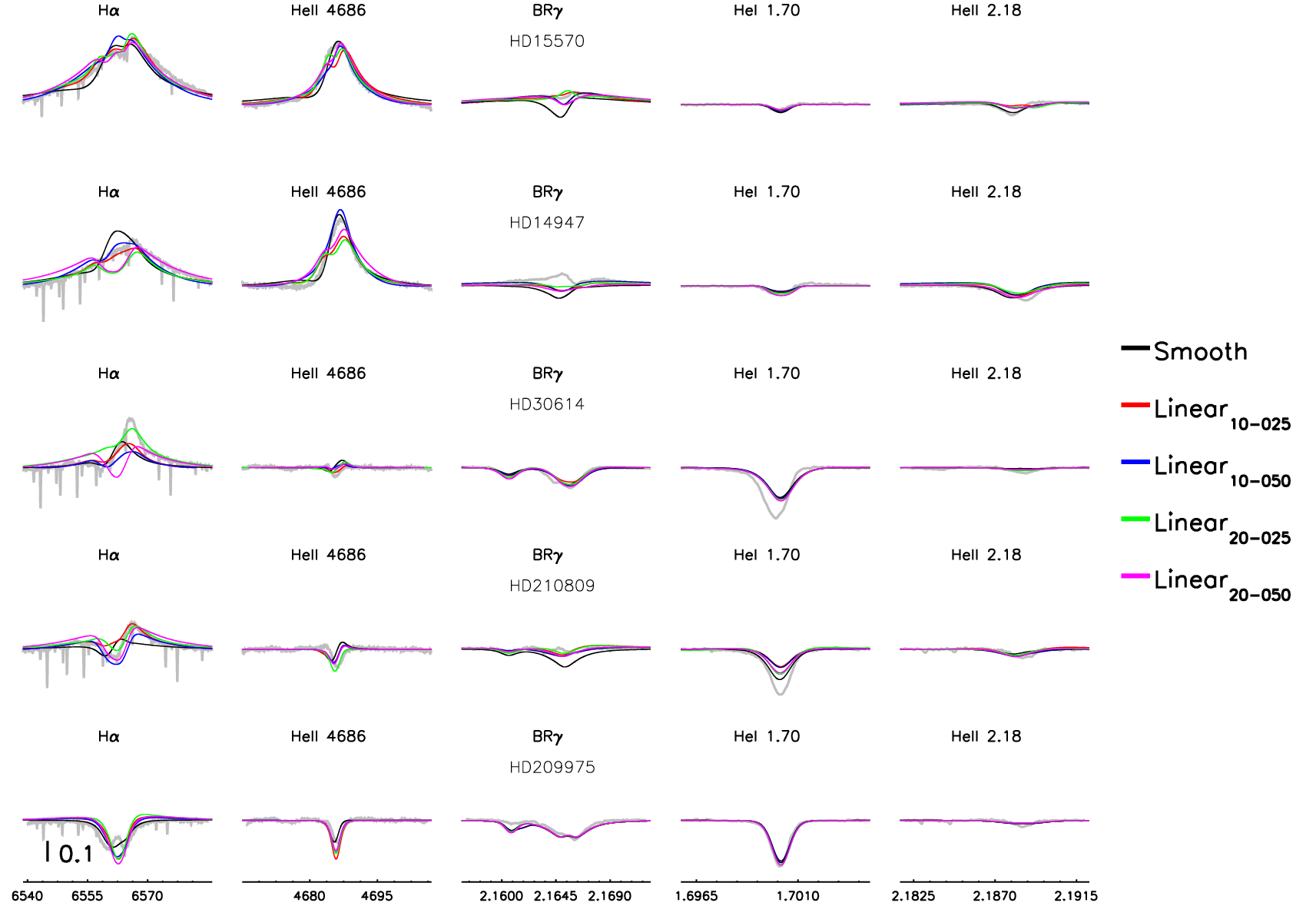} 
\caption{ Comparison of spectral fits to selected optical and NIR lines
from the supergiants of our sample. Observations: gray; synthetic
profiles from best-fitting models without (black) and with clumping
using various clumping laws (for color-coding, see legend). The
stellar and wind parameters of the individual best-fitting models are
provided in Tables~\ref{tbl:iacob_OP}, \ref{tbl:iacob_OP_cl1},
\ref{tbl:iacob_OP_cl2},\ref{tbl:iacob_OP_cl3}, \ref{tbl:iacob_OP_cl4}
for the optical lines, and in Tables~\ref{tbl:iacob_IR_bli},
\ref{tbl:iacob_IR_cl1}, \ref{tbl:iacob_IR_cl2},
\ref{tbl:iacob_IR_cl3}, \ref{tbl:iacob_IR_cl4} for the NIR lines. }
\label{fig:sg_profiles_cl}
\end{center}
\end{figure*}
\begin{figure*}[h!]
\begin{center}
\includegraphics[height=12cm,width=17cm]{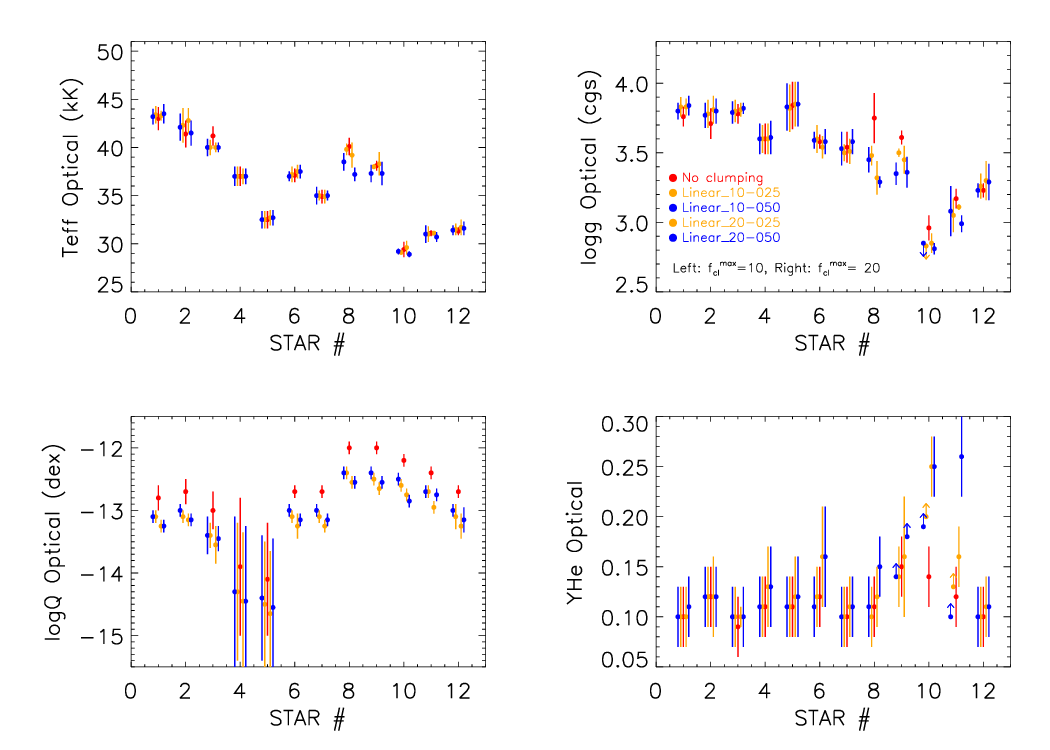}
\caption{Comparison of effective temperatures (upper left), gravity
(upper right), \logq\ (lower left) and \yhe\ (lower
right) obtained from the optical spectra with the different clumping
laws considered in this work. The abscissa gives the identification
number of the star. For each star, the results from the different clumping
laws (including the smooth wind model) are plotted (see legend).
Corresponding entries (except for the smooth wind results)
have been slightly displaced on the abscissa. Stars are ordered as in
Tab.~\ref{tbl:stars_sample}: $\#$1: HD\,46\,223; $\#$2: HD\,15\,629;
$\#$3: HD\,46\,150; $\#$4: HD\,217\,086; $\#$5: HD\,149\,757; $\#$6:
HD\,190\,864; $\#$7: HD\,203\,064; $\#$8: HD\,15\,570; $\#$9:
HD\,14\,947; $\#$10: HD\,30\,614; $\#$11: HD\,210\,809; $\#$12:
HD\,209\,975. Rapid rotators are stars \#4, \#5, and
\#7.
\label{fig:op_clumping}}
\end{center}
\end{figure*}

\subsection{Analysis with the Linear$_{10-025}$ clumping law}

The results of the analyses with the \texttt{iacob\_gbat} tool for the
Linear$_{10-025}$ law can be found in Tables \ref{tbl:iacob_OP_cl1}
(for the optical spectrum) and \ref{tbl:iacob_IR_cl1} (for the near
infrared). The corresponding fits are displayed in
Fig.~\ref{fig:OP_cl1_profiles}.Moreover, in
Fig.~\ref{fig:sg_profiles_cl} we compare, for all supergiants of our
sample, the spectral fits for selected optical and NIR lines 
for all clumping laws discussed in the following (including the homogeneous wind).

From both figures, we can see that the fits have
a similar global quality as those for the unclumped models,
but there are specific differences worth mentioning. We stress already
here that the parameters of the globally best-fitting clumped and
unclumped models are different; thus, the changes will not only be due
to clumping, but also due to the parameter changes produced by it.

For the hot supergiants we observe two major changes in the optical. 
The first one is a distinct improvement in the fit of H$_\alpha$
(see Fig. \ref{fig:sg_profiles_cl}, red vs. black profiles). 
A similar improvement is not seen for H$_\beta$ 
(see Figs.~\ref{fig:OP_profiles} and \ref{fig:OP_cl1_profiles}), that fits slightly
better in the red wing, but clearly worse in the core, due to less
core-filling in the inner layers\footnote{At
least in this specific case, this might suggest a lower value for
$v_1$ than adopted throughout this work.}. 
Upper Balmer lines remain unaffected. 

The second one is a deeper absorption in \ion{He}{ii} $\lambda$4541
that improves the fit. 
However, the good fit for \ion{He}{ii} $\lambda$4686 without clumping 
slightly deteriorates,
again because of less emission in the forming layers.
The cool supergiants do not present the same global improvement in
H$_\alpha$, but there is a partial improvement. Moreover, the He
lines, particularly \ion{He}{ii} $\lambda$4686, also improve slightly,
including a correction in the apparent shift in the line core between
the observations and the unclumped profile. This differential
behavior in \ion{He}{ii} $\lambda$4686 in (dense) hot and cool winds
is expected because of the change in the dominant ionization stage of
helium, as explained earlier,
and strengthens our warning about the use of a single
clumping law for all stars. We conclude that the Linear$_{10-025}$
improves H$_\alpha$ for the hot supergiants and improves the agreement
between H$_\alpha$ and \ion{He}{ii} $\lambda 4686$ for the cool
supergiants (but without reaching a good fit).

In the NIR, the fits to the spectra of the hot supergiant HD\,15\,570
and the cool one HD\,210\,809 improve considerably for Br$_\gamma$.
The rest of the line fits also improve slightly in these stars, except
for \ion{He}{ii} $\lambda 2.18\mu$m that deteriorates significantly
in HD\,15\,570. Unlike for the optical spectra, we now find
changes also in the line fits for giants and dwarfs.
Finally, there is a remarkably bad fit to the \ion{He}{i}
$\lambda1.70\mu$m line in the cool supergiants HD\,30\,614 and
HD\,210\,809, both in the models with and without clumping. Thus, in
the NIR the major improvement of using the Linear$_{10-025}$ law
regards the Br$_\gamma$ line of some supergiants.



\subsection{Analysis with the Linear$_{10-050}$ clumping law}

The results from the analysis of our stellar sample with the
Linear$_{10-050}$ law (that reaches the maximum clumping factor,
\fclmax, further out than Linear$_{10-025}$ from the previous section)
can be found in Tables~\ref{tbl:iacob_OP_cl2} (for the optical
spectrum) and \ref{tbl:iacob_IR_cl2} (for the infrared). The
corresponding best fits can be inspected in
Figure~\ref{fig:OP_cl2_profiles}. For a comparison of supergiant
fits, we refer again to Fig.~\ref{fig:sg_profiles_cl}.

As for the Linear$_{10-025}$ law, there is an improvement in the fit
of H$_\alpha$ from the hot supergiants, that in fact show an excellent
agreement in all optical lines. The cool supergiants HD\,209\,975
and HD\,30\,614
do not change significantly. But now we see a much better fit for
H$_\alpha$ 
for the cool supergiant
HD\,210\,809,
indicating that the clumping distribution is more extended in this
star than represented by Linear$_{10-025}$. For the rest of the sample
(giants and dwarfs), we obtain similarly good fits with the
Linear$_{10-050}$ law as with an homogeneous wind model.

In the NIR, the fits to Br$_\gamma$ of HD\,15\,570 and HD\,210\,809 
improve again significantly compared to the unclumped models.
For all stars, the fits to Br$_\gamma$ and other H and He NIR lines (except for \ion{He}{ii}
$\lambda2.18\mu$m) improve slightly. An exception is \ion{He}{i}
$\lambda1.70\mu$m in the hot dwarf HD\,46\,223, where
the insufficient quality is a consequence of the hotter temperature
resulting from the global fit parameters when using Linear$_{10-050}$. 
This line remains also very badly fitted in the cool supergiants
HD\,30\,614 and HD\,210\,809. Summarizing and overall, the global line
profile fits in the NIR 
do not seem to be strongly affected by the different clumping distributions when
using \fclmax = 10.

\subsection{Analysis with the \fclmax\ = 20 clumping laws}

In this subsection, we analyze whether a higher (maximum) clumping
factor enables an improvement in the fit quality for our sample. We
compare here the fits of Linear$_{20-025}$ and Linear$_{20-050}$ with
their corresponding counterparts, Linear$_{10-025}$ and
Linear$_{10-050}$ as described above. The fits with these laws can be
seen in Figs.~\ref{fig:OP_cl3_profiles} -- ~\ref{fig:OP_cl4_profiles} and again 
in Fig.~\ref{fig:sg_profiles_cl} for the supergiants,
and the derived parameters in Tables~\ref{tbl:iacob_OP_cl3} --
~\ref{tbl:iacob_IR_cl4} 

The changes when using the Linear$_{20-025}$ law as compared to the
Linear$_{10-025}$ are mostly minor. The most affected line is of
course H$_\alpha$,
with significant changes seen in the supergiants, reflecting their
higher sensitivity to clumping: the fit improves for HD\,210\,809 and
worsens for HD\,14\,947. HD\,30\,614 shows significant changes,
improving the fit in the emission peak and the red wing but worsening
it in the blue wing. HD\,15\,570 does not show changes in H$_\alpha$,
but the blue wings of the other Balmer lines are slightly less well
reproduced with the Linear$_{20-025}$ law. Only the supergiant
HD\,209\,975 remains almost unchanged.
The optical wind lines of HD\,14\,947, HD\,30\,614 and HD\,210\,809
are the most sensitive to changes in the clumping law.

In the NIR there are few changes. The Br$_\gamma$ line of dwarfs and
giants is marginally affected in many cases.
Other lines do not change,
except \ion{He}{i} 1.70$\mu$m in HD\,203\,064 and in
HD\,217\,086, with an improvement in both cases.

For the supergiants, the Br$_\gamma$ line of HD\,15\,570
deteriorates when using
Linear$_{20-025}$. 
For HD\,14\,947, the fit to Br$_{10}$
and \ion{He}{i} 1.70$\mu$m improves, but Br$_{11}$ (not displayed) and the
\ion{He}{ii} 2.18$\mu$m become worse. The cool
supergiants are not significantly affected.

When comparing the profile fits in the optical using the laws with
larger $v_2$, Linear$_{20-050}$ vs. Linear$_{10-050}$, we also find
only small changes. H$_\alpha$ is the most affected line, usually with
more core absorption (or less emission).
Concentrating on the supergiants,
there is a clearly worse fit
in HD\,14\,947 and HD\,210\,809 when using the
Linear$_{20-050}$ law. HD\,15\,570 and HD\,30\,614 show an
improvement in the red and a worsening in
the blue wing, with the former object showing also an improvement in
the emission peak.
Other lines are not significantly affected.

In the NIR, Br$_\gamma$ shows small changes for nearly all stars,
The weak changes for the supergiants HD\,15\,570 and HD\,14\,947 are now similar to those 
in the dwarfs HD\,46\,223 and HD\,46\,150.
The other Brackett lines display a mixed behavior,
as do the \ion{He}{} lines.
Particularly, \ion{He}{ii} 1.69$\mu$m (not shown), which is usually not affected, changes
in HD\,210\,809, where it shows a better fit with the
Linear$_{20-050}$ law. Overall, this time a larger number of stars
present changes, but these are small compared to  differences with
homogeneous-wind profiles.

Taken all our findings together, we conclude that, globally, clumping
has sometimes positive impact on the fits to H$_\alpha$, Br$_\gamma$
and \ion{He}{ii} $\lambda4686$ in supergiants. The impact may depend
on the particular clumping law chosen, although the differences
between the clumping laws explored are small (or even not present for
most lines), and they do not offer a clear indication of which one
better represents the distribution of inhomogeneities in the stellar
wind.  While for most cases the \fclmax\ = 10 linear laws shows a
better fit, we also find many counter-examples. This indicates that
more work is needed to determine the actual clumping distribution in
these stars.


\section{Discussion of results: Impact of clumping laws on optical and
NIR anaylses} 
\label{discussion}

We are now able to compare the derived parameters, to see whether the
introduction of the different clumping laws modifies our
determinations or improves the agreement between optical and infrared
stellar parameters. We will not consider microturbulence and $\beta$
exponent, as they remain basically unrestricted in our analyses.

We begin by comparing the results obtained from the optical analysis
for the effective temperature. Fig.~\ref{fig:op_clumping} (upper left)
shows the comparison for all five clumping laws (homogeneous wind and
four linear laws, as discussed above).
We see that the values for all stars are
fully consistent for almost all laws within the uncertainties. The only
exception is for the $v_2/\vinf$ = 0.5 laws in HD\,15\,570, that give a
slightly lower \Teff.
Thus, the temperature determination in the optical is not
significantly affected by the presence of clumping or by differences
in the clumping distribution (as far as it concerns the laws used in
this work).

The comparison of the gravities obtained from the optical analyses is
presented in Fig.~\ref{fig:op_clumping} (upper right). The difference
between dwarfs and giants on the one side, and supergiants on the
other, is obvious. For giants and dwarfs we obtain similar values of
\grav, independent of the clumping laws used (including the absence of
clumping). Also the uncertainties are similar ( but the
uncertainties are significantly larger for the fast rotators, as could
be expected). However, and except for HD\,209\,975, the situation is
different for the supergiants. Here, the unclumped values are always
larger than the clumped ones, and depend on the specific clumping law;
in the cases of HD\,15\,570 and HD\,14\,947, quite significantly. This
is a consequence of the lower mass-loss rate implied by clumping that
renders the red wings of the Balmer lines and the core of \ion{He}{i}
lines deeper. A lower gravity (sometimes accompanied by a lower \Teff)
compensates for this effect.

We compare the results for the wind parameter \logq\ in
Fig.~\ref{fig:op_clumping} (lower left). As we have used the same wind
terminal velocity and stellar radius, this quantity is equivalent to
the mass-loss rate. We see that the unclumped models give higher
mass-loss rates, and that the correction increases with the maximum
clumping factor, as expected. The mean differences in \logq\ are
somewhat below the ``nominal'' values of 0.5 (\fclmax\ = 10) and 0.65
dex (\fclmax\ = 20), with actual differences of 0.39$\pm$0.05 and
0.33$\pm$0.05 for the \fclmax\ = 10 laws, and 0.54$\pm$0.05 and
0.48$\pm$0.08 for the \fclmax\ = 20 laws, indicating that the diagnostic
lines form before the maximum clumping factor is reached.

The helium abundances are compared in Fig.~\ref{fig:op_clumping}
(lower right). All determinations (unclumped and clumped models)
result in equal values within typical uncertainties, again except for the
supergiants where the dispersion is larger and lower limits become
frequent. Although most values are still consistent within their
uncertainties, in two cases they are not. The largest discrepancy is
found for the cool and bright supergiant HD\,30\,614, which gives a
higher He abundance when clumping is included.
This higher He abundance is the result of a better fit to the He lines after the
changes produced in H$_\alpha$. The second strong discrepancy is seen
in HD\,210\,809, where the He abundance obtained with the
Linear$_{20-050}$ law is much larger than any other value

Regarding an optical diagnostics, we conclude that only specific
stellar parameters might be affected by different assumptions on the
clumping conditions: gravities determined for supergiants, helium
abundances in peculiar cases like HD\,30\,614, and wind strengths
(beyond their explicit dependence on \fclmax) for supergiants and
rapidly rotating dwarfs.

\begin{figure*}[h!]
\begin{center}
\includegraphics[height=12cm,width=17cm]{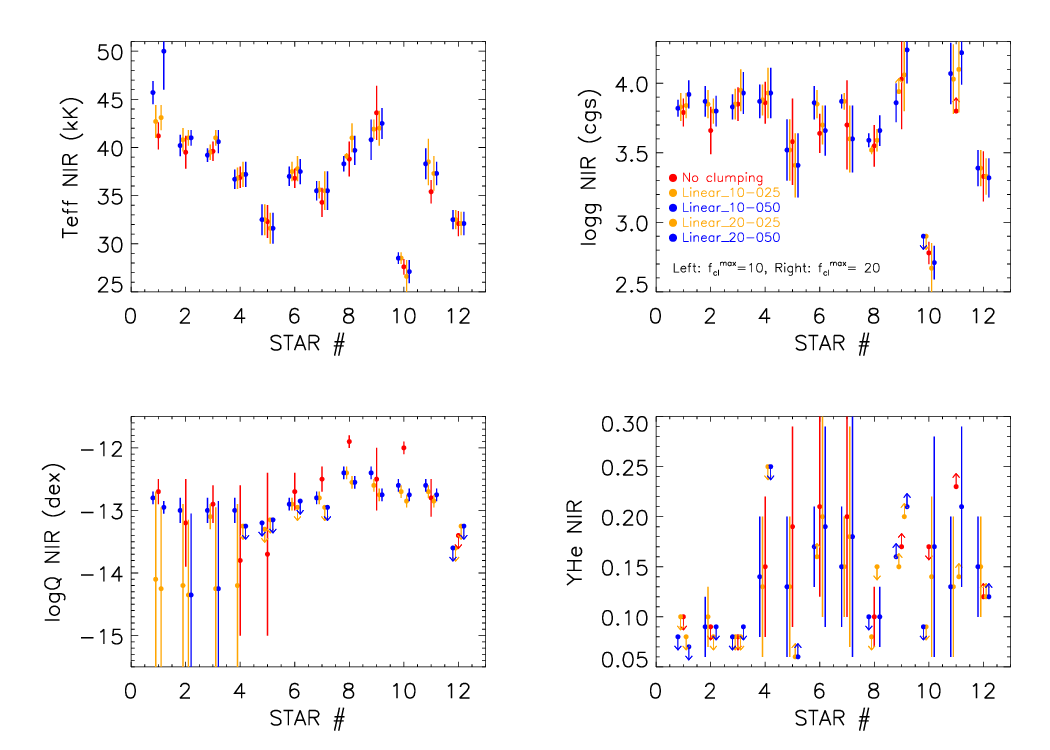}
\caption{Same as Fig.~\ref{fig:op_clumping}, but for the stellar parameters
derived from the NIR.}
\label{fig:ir_clumping}
\end{center}
\end{figure*}

The temperature values obtained from the analysis in the NIR are, in
general, again consistent when introducing clumping
(Fig.~\ref{fig:ir_clumping}, upper left). The exception is the
behavior of the hottest star, HD\,46\,223. Here, the models with $v_2
= 0.5\vinf$ give a temperature much higher than those with $v_2 =
0.25\vinf$ and the homogeneous wind models. The high temperature
produces a bad fit to the \ion{He}{i} lines compensated by a slightly
better fit to all other lines. This is then due to the lack of
sufficient constraints for such a case. Giving more weight to
 \ion{He}{i} 1.70$\mu$m (as the only diagnostics strongly constraining
the ionization balance) would reduce the discrepancy, as we have
convinced ourselves.

The gravity values show a large dispersion, even for dwarfs, though
they are all consistent within their 1-$\sigma$ uncertainties (which
are sometimes quite large, see Fig.~\ref{fig:ir_clumping}, upper
right).  This is a consequence of the poorer fits to the Brackett
lines (see Sect.~\ref{nir_extension}), together with a more limited
number of diagnostics.

This combination produces also a complicate behavior for the
\logq\ wind parameter. The general behavior is similar to the optical
case, namely the \logq\ values obtained with clumping are lower than
the ones derived from a homogeneous wind model, though with a larger
number of upper limits and larger error bars. This is particularly
obvious for the rapidly rotating dwarfs. In a few peculiar cases we
find clumped values that are larger than the unclumped ones. This
happens for the Linear$_{10-050}$ law in HD\,15\,629, HD\,14\,947 and
HD\,210\,809 (in the latter star, also for Linear$_{10-025}$ and
Linear$_{20-050}$). Though in all cases the values are consistent
within the uncertainties (and even consistent with the expected
behavior), the apparent problems result from the loss of information
produced by poorer fits.
In fact, only for supergiants, an actual determination of \logq\ for
all clumping laws is possible (except for HD\,209\,975 for which only
upper limits could be derived). With all these uncertainties, the
\ion{He}{} abundance remains nearly unconstrained.

The conclusion is that the impact of clumping on the derived
parameters is similar in the NIR and in the optical. The NIR shows a
larger scatter in the global trends and more upper/lower limits.
A second conclusion is that, compared to the optical, the H and K band
lines in the NIR do not offer us a clear advantage to characterize the
clumping. However, this conclusion depends (until further evidence) on
our assumptions about the shape of the clumping law and its
parameters; the impact on the different lines will depend on
the behavior of the clumping law in the line formation region.

\begin{figure*}
\begin{center}
\includegraphics[height=12cm,width=17cm]{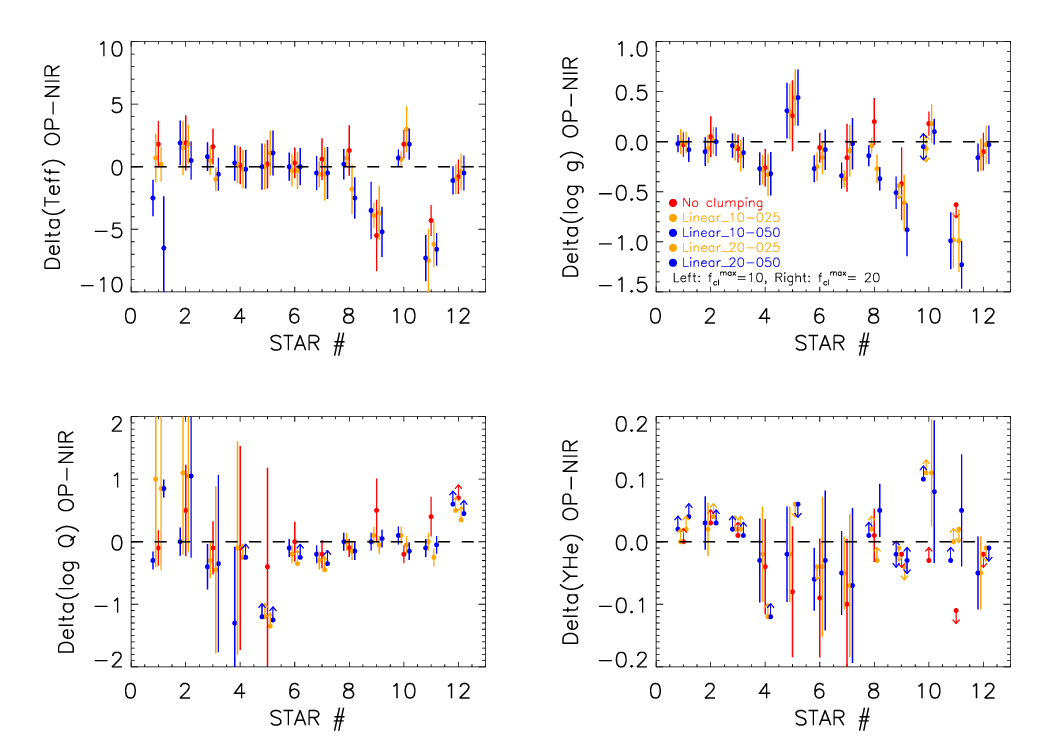}
\caption{Same as Fig.~\ref{fig:op_clumping}, but for the differences
between optical and NIR determinations. Temperature differences are
given in  kK, and differences in gravity and \logq\ in
dex.}
\label{fig:diff_clumping}
\end{center}
\end{figure*}

We are now in a position to finally address the question whether the
introduction of clumping improves the agreement between the optical
and the infrared parameter determinations, compared to the assumption
of a homogeneous wind. An improvement in this comparison would also
provide additional hints on the most appropriate clumping law to be
 adopted. 

\begin{figure}
\begin{center}
\includegraphics[height=8cm,width=7cm,angle=90]{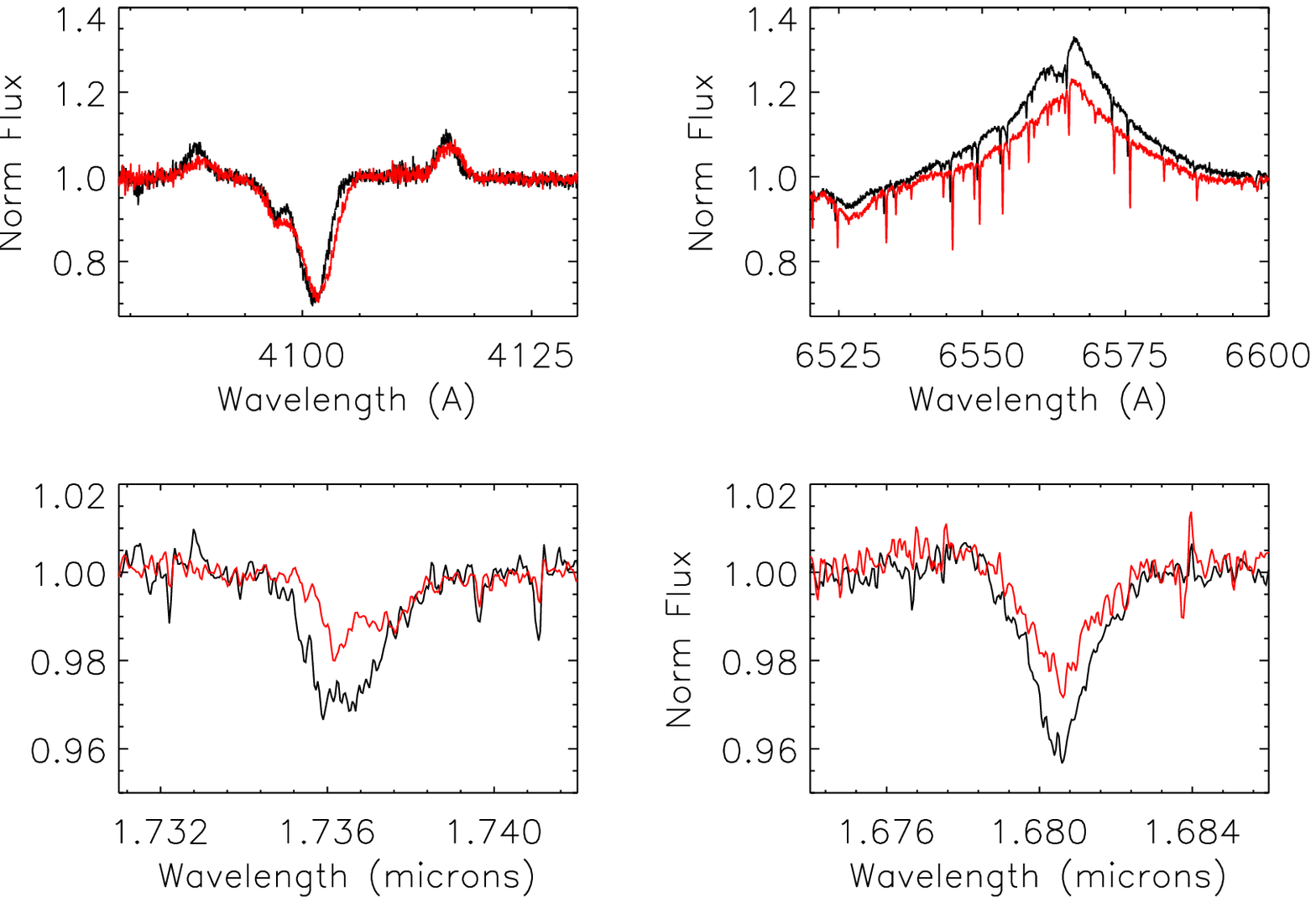}
\caption{Comparison of the observed Balmer and Brackett lines in HD\,15\,570
(black) and HD\,14\,947 (red). Whereas the Balmer lines are very
similar, the Brackett lines Br$_{10}$ and Br$_{11}$ are much shallower in
HD\,14\,947. See text.
}
\label{fig:br}
\end{center}
\end{figure}

Fig.~\ref{fig:diff_clumping}, upper left, shows the comparison of \Teff\
determinations in the optical and the infrared for the different
clumping conditions explored in our experiment. Most stars have
optical and NIR \Teff\ determinations consistent within the
uncertainties for all explored laws. However, there are also certain
outliers. The most important are the supergiants HD\,14\,947 and
HD\,210\,809, where the large discrepancies can be traced back to
their shallow Br$_{10}$ and Br$_{11}$ lines (shallow compared to those
from HD\,15\,570 and HD\,209\,975, which occupy similar parameter
ranges, respectively; see Fig.~\ref{fig:br}). As shown by \citet[their
Fig.~1]{Repolust2005} for models with low wind densities, the cores of
Br$_{10/11}$ strongly react to changes in gravity (an alteration of
gravity mostly affects the depth of the line cores, contrasted to the
behavior of the Balmer lines), where the depth decreases with
increasing \grav.  For our objects with substantial mass-loss rates,
shallow Br$_{10/11}$ lines can be only reproduced when in addition to
a high gravity also the effective temperatures and mass-loss rates lie
in a certain range. In particular, the mass-loss rates must not be too
large, since otherwise Br$_{10/11}$ would become severely asymmetric,
which is not observed. Taken all these constraints together, a fit of
the shallow Br$_{10/11}$ lines pushes the gravity and the temperature
toward values higher than derived from the optical, with the higher
temperature also required for compensating for the shift of the Helium
ionization equilibrium and the strong reaction of the \ion{He}{ii}
lines (see again Fig. 1 in \citealt{Repolust2005}). The somewhat lower
mass-loss rate (required to fit Br$_{10/11}$) also prevents
Br$_\gamma$ from entering into emission in the hot supergiant
HD\,14\,947, whereas the cooler supergiant HD\,210\,809 still partly
fills its Br$_\gamma$.  All these problems are, for example, not present in
HD\,15\,570, because here the Br$_{10/11}$ lines have a ``reasonable''
depth allowing for these lines to be fit at parameters that are mostly
compatible with the optical results (but see below). Whether these 
problems are related to the model calculation or to the reduction of NIR
spectra remains an open question.
Anyhow, and apart from a few individual cases that do not allow a
generalization, no clumping law (including the homogeneous wind) shows
a better agreement between the effective temperatures derived from the
optical and the NIR than the other laws.

The gravity differences reflect the larger scatter obtained for this
parameter in the NIR. Most stars are again close to the
zero-difference line, again with the exceptions of the supergiants
HD\,14\,947 and HD\,210\,809 already discussed above.

The difficulties to simultaneously fit the \ion{He}{i} and
\ion{He}{ii} lines in HD\,210\,809 could point to a
higher He abundance than considered in our grid (for this star and also for
HD\,14\,947, we obtain mostly lower limits for the He abundance;
however, this is not the case in the optical, cf.
Figs.~\ref{fig:op_clumping} and \ref{fig:ir_clumping}).

The optical/NIR differences in \logq\ show again a clear pattern: for
dwarfs, the large uncertainties dominate, whereas for supergiants,
results are consistent (although HD\,209\,975 is here an exception,
with a very low \logq\ value derived in the NIR). The \ion{He}{}
abundances also scatter around the zero-line, with large uncertainties
and numerous limits reflecting mainly the behavior in the NIR.
Nevertheless, there are no obvious outliers, except for the unclumped
values for HD\,14\,947 and HD\,210\,809, both
suffering from the mass-loss dependence of Br$_{10/11}$ (see above).
Interestingly, however, the \logq\ values from the optical and the NIR
agree when clumping is considered.

Finally, when combining the optical and near NIR lines in
a joint analysis, the results are dominated by the fits 
to the optical lines. This is a consequence of the larger number of
optical lines and the better fit quality in this wavelength range.

The most important conclusion from our comparison is
that whatever the differences between the optical and the NIR, the
inclusion of the different clumping laws as explored in our work, does
not contribute to a globally better agreement between the parameters
derived from either wavelength range. For example, the average value
of the differences for \Teff\ is $-0.1\pm2.4$~kK for the
unclumped values, and ranges from $-0.7\pm2.5$~kK to $+1.6\pm2.9$~kK for
the various clumping laws, with no star showing a clearly better
agreement when introducing clumping. Such a better agreement would
require either even higher than current quality NIR observation with more diagnostic spectral lines
or different types of clumping laws. 

Alternatively, it is also possible that clumping has a different
behavior in different stars, not only because of spectral type (cf.
Hawcroft et al. 2023, as already discussed in Sect.~\ref{sec_clump})
and luminosity class, but also because of additional differences like
pulsations and wind variability (not to mention ocassional mass
ejections or local magnetic fields). This is particularly relevant for
the two most extreme outliers, HD\,14\,974 and HD\,210\,809 (the
latter well known for its notorious wind variability, see
\citealt{Markova2005}), where the discrepancy of the optical and NIR
results is rooted in the weak Br$_{10}$ and Br$_{11}$ lines (see above).
But also for the other objects analyzed here we
are using single epoch observations, with the
optical and NIR spectra taken at different times, so that line
profile variability may play a role in the differences.

Besides the above possibility that the clumping
conditions in both stars deviate strongly from our current assumptions
(particularly regarding the lower wind), we cannot neglect the
possibility that the (complex) NIR data reduction (see appendix in
\citealt{Hanson1996} and \citealt{Hanson2005}) is free from any
problems, and that the actual line profiles might be stronger than
adopted here (see also \citealt{Repolust2005}). Another possibility
regards the question of (in)accurate hydrogen collision cross
sections. Using the most up-to-date, ab initio values from
\cite{PrzybillaButler2004} instead of the default values following 
\citet{Giovanardi1987} implemented into {\sc Fastwind} only
exacerbates the problem though, since the corresponding Brackett lines
become even stronger then (see Fig. 15 in \citealt{Repolust2005}).

\section{ Conclusions}
\label{conclusions}

We have carried out a determination of stellar parameters and a study
of the clumping effects in the optical and the NIR extending the
automatic methods developed in our group (see Sect.~\ref{sect_model_grid} 
and~\ref{iacob_gbat}). Our objectives were (a) to check
whether we can obtain stellar parameters from the infrared, with the
same or comparable accuracy to those in the optical; (b) to check whether
the parameters obtained were consistent; (c) to study the effects of
clumping in the determination of stellar parameters; and (d) to check
whether clumping improves the agreement between the infrared and the
optical parameters. 
To these ends, we have extended the automatic tools to include the NIR spectra. 


When analyzing the observed spectra in the optical and NIR
with unclumped models, we reached the following conclusions:

\begin{itemize}

\item 
In many cases, test calculations revealed a
problematic behavior of the Br lines. It was not possible to fit all
of the observed lines simultaneously. We decided not to use the highest
available line, Br$_{12}$, since this line deviates the most. However, Br$_{10/11}$ also frequently presented problems to achieve a
consistent fit. We conclude that the Br lines need to be studied in
more detail in the future. 

\item Globally, the quality of the fits to the optical spectrum is 
excellent. The only problems appear for supergiants, mostly related to
H$_\alpha$ and sometimes to \ion{He}{ii} 4686, with the fits
improving with decreasing luminosity class. In the infrared, again the
best fits are for dwarfs, and problems are concentrated in Br$_\gamma$ 
(and sometimes the other Br lines), which in some
supergiants appear in emission, while models still predict
absorption. \ion{Helium}{} lines in the NIR present a variety of fitting
problems, which might also be related to 
a lower number of available lines.

\item Both the optical and the NIR analyses without clumping show a good 
agreement with previous similar studies in the literature
\citep{Holgado2018, Repolust2005}.

\end{itemize}

When comparing the results in the optical and the NIR derived from unclumped
models, we find that:

\begin{itemize}
\item the rotational velocities derived from the NIR \ion{He}{i} $\lambda1.70\mu$m line agree in most cases well with those derived with a higher accuracy from the optical metal lines (with the known limitations due to the larger intrinsic (Stark-) broadening).

\item There is a good agreement between the parameters derived in the 
optical and the NIR, with some deviating individual cases
(particularly HD\,14\,947 and HD\,210\,809). The uncertainties in the
NIR are larger, mostly due to poorer fits
and, to a lesser extent, to the low number of diagnostic lines.
Helium abundances from the NIR frequently show 
upper and lower limits, indicating a lack of sensitivity to this
parameter. 

\item We could thus derive stellar parameters from the infrared 
with an almost similar accuracy to the optical. The
uncertainties are 
larger for the reasons given in the item above.

\end{itemize}

We then explored the effects of clumping using different clumping
laws. We considered a Najarro-type law, two Hillier-type laws, and
up to six different linear laws. We compared the behavior of the
different laws and their impact on the line profiles. The main
conclusions are:

\begin{itemize}

\item Using a coarse model grid, we show that clumping only had significant 
effects on the synthetic spectra of supergiants
once we accounted for the corresponding mass-loss scaling relations
as a function of the (maximum) clumping factor, \fclmax (or minimum volume filling
factor). For giant stars, effects are very modest and they are negligible for dwarfs.

\item We find only small differences in the synthetic wind lines 
based on the various clumping laws, which indicates
that these lines are formed in layers where the differences between
these laws are not critical. Differences can also be present in the
absorption cores of lines that are mainly formed in the photosphere,
because of a different refilling by wind emission when the mass-loss
rates have been appropriately scaled as a function of \fclmax.

\item Together with \fclmax, the second relevant 
parameter is the extent of the region where the clumping factor
increases until that maximum is reached. Both quantities define the
distribution of the clumping factor in the line formation region.

\item Primary differences between clumped and unclumped models are
related to the modified density structure in the line-forming region. The different laws explored in this work do not
trigger significant differences in the corresponding spectra (after
scaling the mass-loss rates), since they all share the same general behavior in that region: clumping is adopted to start close to (but
above) the photosphere, and increases more 
or less
rapidly to a maximum. 
The wind emission increases when \fclmax is reached in the
inner and intermediate wind layers, and the behavior of the clumping laws
in the outer wind does not affect the line formation region
relevant for this work.

\item Because of the rather weak differences raised by the three kinds of 
clumping laws investigated here, 
an analysis with the linear clumping law is sufficient for the exploratory
character of this work, because if its conceptual simplicity.

\item The central emission seen in some lines 
(either on top of an emission or absorption profile) is primarily
related to a NLTE effect in the transonic region, affecting the
occupation numbers of the upper and/or lower atomic levels. It is thus
(almost) independent of the specific clumping stratification, though it
depends on the actual mass-loss rate.

\end{itemize}

Subsequent to the above study of principal effects, we
compared the fits obtained from four model grids with different linear
clumping laws, discriminated by different combinations in $v_2$/\vinf (=
0.25, 0.50) and \fclmax (= 10, 20). We find that:

\begin{itemize}

\item Clumping usually has positive effects for the 
fit of H$_\alpha$, \ion{He}{ii} 4686, and Br$_\gamma$ in supergiants
(particularly in hot supergiants), sometimes improving the consistency
between the former two optical lines (for cool supergiants).
However, there is a trend to worse fits in \ion{He}{ii} 2.18 $\mu$m. 

\item The laws with $v_2$/\vinf= 0.50 imply a larger number of 
changes when comparing the fits for the two \fclmax values. This is a
consequence of a more pronounced variation of the clumping factor
along the line-forming regions.

\item The actual impact on the line profiles depends on the 
specific clumping law, although differences between the laws are
small in many cases. We note, however, that the best fit to
individual lines in a given star may be reached with different clumping laws,
pointing to a potentially more
complex distribution than the one considered here.

\end{itemize}

We finally compared the stellar parameters obtained with the different
clumping laws, to see whether the parameters change significantly and
whether the agreement between optical and NIR parameters is
better for a particular law. Our main conclusions are:

\begin{itemize}

\item In the optical, only \grav~ and \logq~ in supergiants are 
affected by the use of different clumping laws (except for particular
cases, such as \yhe~ in HD\,30\,614 or the wind strength in rapidly
rotating dwarfs).

\item In the NIR, the Br lines are often responsible for
problems in accurately determining \grav~ and consequently \logq.

\item As in the unclumped case, we obtain similar stellar parameters 
in the optical and the NIR, although with a larger scatter
and more upper and lower limits in the latter. HD\,14\,947 and
HD\,210\,809 (both supergiants) are outliers in this respect, mainly
due to problems with Br lines. 

\item In our analysis, the H and K bands did not offer a clear advantage 
over the optical wavelengths to characterize clumping.

\item Regarding the consistency between optical and NIR parameters,
none of the specific clumping laws displayed a better global agreement nor do
clumped models agree better than unclumped ones.
Results for \logq~ are mostly
consistent (the larger \fclmax, the lower the derived wind-strength),
particularly for strong winds (supergiants).

\end{itemize} 

Taking everything together, we reach the somewhat
disappointing conclusions that the inclusion of the NIR ( 
as done here) still does not allow actual mass-loss
rates to be derived. There is still the dichotomy between \Mdot\ and \fclmax, which
might be only broken by including lines that react in a different way
than typical recombination lines such as H$_\alpha$. However, 
including UV P~Cygni lines (when available) is difficult, because of
the impact of X-ray emission, optically thick clumping, and
saturation, though first analyses in such a respect have already been
undertaken (\citealt{Hawcroft21}, \citealt{Brands22}, \citealt{Hawcroft23}).
One might question whether an analysis of the predicted central
emission in, for example, Br$_\gamma$ might help, since this should depend on
the actual \Mdot\ alone, in the same spirit as Br$_\alpha$ for
weak-winded stars \citep{Najarro2011}. Unfortunately, the predicted
emission peak is quite narrow and small, much smaller than in the case
of Br$_\alpha$, and most likely not useful for \Mdot\ determinations.
Finally, at least for late-type O supergiants and early-type
B supergiants, constraints on the clumping properties and actual
mass-loss rates might be feasible, because of the different behavior
of H$_\alpha$ and \ion{He}{ii} 4686 \citep{Kudritzki2006, Holgado2018}.

Our study indicates that future work requires some improvement in the
treatment of the Br lines. We need to analyze a larger sample of stars
considering a wavelength range as large as possible
to find patterns among them that can be used to characterize the
clumping laws. The positive view is that our models give consistent results between
the optical and infrared wavelength regions, that the use of different
clumping laws does not result in significant differences in the derived
stellar parameters (although the use of a common clumping law may
introduce some extra uncertainty for individual cases), and that the
infrared contains enough information for a spectroscopic analysis
with an accuracy that is quite similar to the optical.   

\begin{acknowledgements}
This research has been supported by the Generalitat Valenciana under
grant PROMETEO/2019/041 and Spanish Ministerio de Ciencia e Innovación (MCIN) 
with funding from the European Union NextGenerationEU and Generalitat Valenciana 
in the call Programa de Planes Complementarios de I+D+i (PRTR 2022), 
(Project HIAMAS, reference ASFAE/2022/017) and also MCIN through the Spanish 
State Research Agency through grants PID2021-122397NB-C21/C22 and the Severo 
Ochoa Programe 2020-2023 (CEX2019-000920-S) (MICINN/AEI/FEDER, UE).
\end{acknowledgements}

\bibliographystyle{aa}
\bibliography{biblio}


\appendix
\section{Additional tables and figures}
\begin{table*}
\begin{center}
\caption{Stellar parameters obtained from the optical analysis using the Linear$_{10-050}$ clumping law. 
 For upper and lower limits see caption of Table~\ref{tbl:iacob_OP_cl1}.}
\label{tbl:iacob_OP_cl2} 
\begin{small}
\begin{tabular}{|c|c|c|c|c|c|c|} \hline
           Star               &       Teff(kK)  &        $\log g$ (cgs) &        $\log Q$ (cgs)     &         \yhe     &       micro (km/s) &         beta  \\ \hline  
   HD46223 &43.2$\pm$ 0.8  &3.80 $\pm$0.06  &-13.1 $\pm$ 0.1 &0.10$\pm$0.03  & $>$5.0          & $>$1.0                                     \\         
   HD15629 &42.1$\pm$ 1.4  &3.77 $\pm$0.09  &-13.0 $\pm$ 0.1 &0.12$\pm$ 0.03 & 12.3 $\pm$ 7.3  & $>$1.2                                     \\         
   HD46150 &40.0$\pm$ 0.9  &3.79 $\pm$0.08  &-13.4 $\pm$ 0.3 & 0.10$\pm$ 0.03 & $<$19.9         & 1.0 $\pm$ 0.2                                     \\         
  HD217086 &37.0$\pm$ 1.0  &3.60 $\pm$0.11  &-13.5 $\pm$ 1.2 &0.11$\pm$ 0.03 & 12.4 $\pm$ 7.4  & 1.0 $\pm$ 0.2                                     \\         
  HD149757 &32.5$\pm$ 0.9  &3.83 $\pm$0.17  &-13.5 $\pm$ 1.0 &0.11$\pm$ 0.03 & 12.1 $\pm$ 7.1  & $<$1.3                                       \\         
  HD190864 &37.0$\pm$ 0.5  &3.59 $\pm$0.06  &-13.0 $\pm$ 0.1 &0.11$\pm$ 0.03 & 14.7 $\pm$ 3.8  & $>$1.0                                       \\         
  HD203064 &35.0$\pm$ 0.9  &3.53 $\pm$0.12  &-13.0 $\pm$ 0.1 &0.10$\pm$ 0.03 & $>$ 13.9        & $>$0.8                                       \\         
   HD15570 &38.5$\pm$ 0.9  &3.45 $\pm$0.09  &-12.4 $\pm$ 0.1 &0.11$\pm$ 0.03 & 12.4 $\pm$ 7.4  & $>$1.3                                       \\         
   HD14947 &37.3$\pm$ 0.9  &3.35 $\pm$0.08  &-12.4 $\pm$ 0.1 &$>$0.14       & 12.4 $\pm$ 7.4  & $>$1.0                                       \\         
   HD30614 &29.2$\pm$ 0.3  &$<$2.85         &-12.5 $\pm$ 0.1 &$>$0.19       & $>$17.9         & $>$1.1                                       \\         
  HD210809 &31.0$\pm$ 0.9  &3.08 $\pm$0.18  &-12.7 $\pm$ 0.1 &$>$0.10        & 14.1 $\pm$ 5.8  & 1.1 $\pm$ 0.2                                     \\         
  HD209975 &31.4$\pm$ 0.5  &3.23 $\pm$0.05  &-13.0 $\pm$ 0.1 &0.10$\pm$ 0.03 & 8.5  $\pm$ 3.5  & $>$1.0                                     \\ \hline  
  \hline
\end{tabular} 
\end{small}
\end{center}
\end{table*}                                                                                                                 

\begin{table*}
\begin{center}
\caption{As Table~\ref{tbl:iacob_OP_cl2}, but for the NIR.  \label{tbl:iacob_IR_cl2}}
\begin{small}
\begin{tabular}{|c|c|c|c|c|c|c|}\hline
Star               &       Teff(kK)  &        $\log g$ (cgs)  &        $\log Q$ (cgs)     &         \yhe     &       micro (km/s) &         beta  \\ \hline                                                                                                        
   HD46223& 45.7 $\pm$ 1.2  & 3.82$\pm$ 0.06 &  -12.8$\pm$ 0.1  & $<$0.08         &   $<$19.9       & $>$1.1       \\         
   HD15629& 40.2 $\pm$ 1.1  & 3.87$\pm$ 0.11 &  -13.0$\pm$ 0.2  & 0.09 $\pm$ 0.03  &   $<$19.9       & $>$1.1       \\         
   HD46150& 39.2 $\pm$ 0.7  & 3.83$\pm$ 0.09 &  -13.0$\pm$ 0.2  & $<$0.08         &   9.9 $\pm$ 4.9 &$>$ 0.9       \\         
  HD217086& 36.7 $\pm$ 1.0  & 3.87$\pm$ 0.12 &  -13.0$\pm$ 0.2  & 0.14$\pm$ 0.06  &  $>$ 5.0        & $>$1.1       \\         
  HD149757& 32.5 $\pm$ 1.6  & 3.52$\pm$ 0.22 &  $<$-13.2        & 0.13$\pm$ 0.07  &  $>$ 9.4        & $<$1.3       \\         
  HD190864& 37.0 $\pm$ 1.0  & 3.86$\pm$ 0.12 &  -12.9$\pm$ 0.1  & 0.17$\pm$ 0.04  &  $>$ 9.4        & $>$1.2       \\         
  HD203064& 35.5 $\pm$ 1.0  & 3.87$\pm$ 0.05 &  -12.8$\pm$ 0.1  & 0.15$\pm$ 0.06  &   $>$5.0        & $>$1.2       \\         
   HD15570& 38.3 $\pm$ 0.8  & 3.59$\pm$ 0.05 &  -12.4$\pm$ 0.1  & $<$0.10         &  $<$19.9        & $>$1.2       \\         
   HD14947& 40.8 $\pm$ 2.1  & 3.86$\pm$ 0.14 &  -12.4$\pm$ 0.1  & $>$0.16        &  $>$ 5.0        & 1.0 $\pm$0.1     \\         
   HD30614& 28.5 $\pm$ 0.6  & $<$2.90        &  -12.6$\pm$ 0.1  & $<$0.09         &  $>$ 6.9        & $>$0.8       \\         
  HD210809& 34.4 $\pm$ 1.5  & $>$3.34        &  -12.6$\pm$ 0.2  &$>$ 0.08        &  $>$ 7.7        & $>$0.8       \\         
  HD209975& 32.5 $\pm$ 1.0  & 3.39$\pm$ 0.13 &  $<$-13.6        & 0.15$\pm$ 0.05 &  $>$14.8        & $>$0.8       \\ \hline  
\hline
\end{tabular} 
\end{small}
\end{center}
\end{table*}

  \begin{table*}
    \begin{center}
    \caption{As Table~\ref{tbl:iacob_OP_cl2}, but using the Linear$_{20-025}$ clumping law. \label{tbl:iacob_OP_cl3}}
    \begin{small}
    \begin{tabular}{|c|c|c|c|c|c|c|}\hline
    Star               &       Teff(kK)  &        $\log g$ (cgs)  &        $\log Q$ (cgs)     &         \yhe     &       micro (km/s) &         beta  \\ \hline                                                                                                        
       HD46223   & 43.3 $\pm$ 0.5  & 3.83 $\pm$ 0.06   & -13.3 +/- 0.1   & 0.10 $\pm$ 0.03   &       $<$ 12.8    &       $>$ 0.9 \\
       HD15629   & 42.8 $\pm$ 1.3  & 3.81 $\pm$ 0.10   & -13.2 +/- 0.1   & 0.12 $\pm$ 0.04   &       $<$ 14.9    &       $>$ 1.0 \\
       HD46150   & 40.0 $\pm$ 0.5  & 3.81 $\pm$ 0.05   & -13.6 +/- 0.3   &  0.10 $\pm$ 0.03   &       $<$ 16.1    & 1.0 $\pm$ 0.2 \\
      HD217086   & 37.0 $\pm$ 0.8  & 3.60 $\pm$ 0.11   & -14.5 +/- 1.1   & 0.13 $\pm$ 0.04   & 12.4 $\pm$ 7.4    &       $<$ 1.2 \\
      HD149757   & 32.7 $\pm$ 0.8  & 3.85 $\pm$ 0.16   & -14.7 +/- 1.0   & 0.12 $\pm$ 0.04   & 12.0 $\pm$ 7.0    &       $<$ 1.2 \\
      HD190864   & 37.5 $\pm$ 0.7  & 3.54 $\pm$ 0.08   & -13.3 +/- 0.2   & 0.16 $\pm$ 0.05   & 12.4 $\pm$ 7.4    & 1.0 $\pm$ 0.2 \\
      HD203064   & 34.9 $\pm$ 0.7  & 3.51 $\pm$ 0.09   & -13.3 +/- 0.1   & 0.11 $\pm$ 0.03   &       $>$ 13.7    &       $<$ 1.1 \\
       HD15570   & 39.2 $\pm$ 1.3  & 3.32 $\pm$ 0.12   & -12.6 +/- 0.1   & 0.12 $\pm$ 0.03   &       $>$ 5.0     &       $<$ 1.0 \\
       HD14947   & 38.3 $\pm$ 1.2  & 3.45 $\pm$ 0.10   & -12.7 +/- 0.1   & 0.16 $\pm$ 0.06   &       $<$ 17.3    &       $>$ 1.2 \\
       HD30614   & 29.6 $\pm$ 0.7  & 2.85 $\pm$ 0.07   & -12.8 +/- 0.1   & 0.25 $\pm$ 0.03   &       $>$ 14.9    & 1.0 $\pm$ 0.2 \\
      HD210809   & 31.1 $\pm$ 0.2  & 3.11 $\pm$ 0.02   & -13.0 +/- 0.1   & 0.16 $\pm$ 0.03   &       $>$ 15.4    &       $>$ 1.1 \\
      HD209975   & 31.7 $\pm$ 0.8  & 3.30 $\pm$ 0.14   & -13.3 +/- 0.2   & 0.11 $\pm$ 0.03   & 10.3 $\pm$ 5.3    & 1.0 $\pm$ 0.2 \\       \hline
    \end{tabular} 
    \end{small}
    \end{center}
    \end{table*}
    
    \begin{table*}
    \begin{center}
    \caption{As Table~\ref{tbl:iacob_OP_cl2}, but for the NIR using the Linear$_{20-025}$ clumping law.  \label{tbl:iacob_IR_cl3}}
    \begin{small}
    \begin{tabular}{|c|c|c|c|c|c|c|}\hline
    Star               &       Teff(kK)  &        $\log g$ (cgs)  &        $\log Q$ (cgs)     &         \yhe     &       micro (km/s) &         beta  \\ \hline                                                                                                        
       HD46223   & 43.1 $\pm$ 1.3  & 3.84 $\pm$ 0.09   & -14.3 $\pm$ 1.3  &        $<$ 0.08  &        $>$ 5.0    &       $<$ 1.2 \\
       HD15629   & 41.0 $\pm$ 0.8  & 3.80 $\pm$ 0.09   & -14.4 $\pm$ 1.2  &        $<$ 0.08  &        $<$ 19.9   &       $<$ 1.2 \\
       HD46150   & 41.0 $\pm$ 0.8  & 3.95 $\pm$ 0.15   & -14.3 $\pm$ 1.3  &        $<$ 0.08  & 12.4 $\pm$ 7.4    &       $<$ 1.2 \\
      HD217086   & 37.2 $\pm$ 1.3  & 3.93 $\pm$ 0.18   &       $<$ -13.3  &        $<$ 0.25 &        $>$ 5.0    & 1.0 $\pm$ 0.2 \\
      HD149757   & 31.6 $\pm$ 1.6  & 3.41 $\pm$ 0.23   &       $<$ -13.2  &        $>$ 0.06  & 12.5 $\pm$ 7.3    & 1.0 $\pm$ 0.2 \\
      HD190864   & 37.8 $\pm$ 1.3  & 3.70 $\pm$ 0.14   &       $<$ -13.0  & 0.20 $\pm$ 0.10 & 12.4 $\pm$ 7.4    &       $>$ 0.8 \\
      HD203064   & 35.5 $\pm$ 2.0  & 3.60 $\pm$ 0.24   &       $<$ -13.0  & 0.18 $\pm$ 0.11 &        $>$ 5.0    &       $>$ 0.8 \\
       HD15570   & 41.0 $\pm$ 1.5  & 3.59 $\pm$ 0.08   & -12.6 $\pm$ 0.1  &        $<$ 15.1 & 12.4 $\pm$ 7.4    &       $<$ 1.1 \\
       HD14947   & 42.0 $\pm$ 1.8  & 4.06 $\pm$ 0.26   & -12.8 $\pm$ 0.1  &        $>$ 20.2 & 12.4 $\pm$ 7.4    &       $>$ 1.1 \\
       HD30614   & 26.6 $\pm$ 1.7  & 2.67 $\pm$ 0.18   & -12.9 $\pm$ 0.1  &  0.14 $\pm$0.08 &        $>$ 5.0    &       $<$ 1.2 \\
      HD210809   & 37.3 $\pm$ 1.8  & 4.10 $\pm$ 0.31   & -12.9 $\pm$ 0.1  &        $>$ 14.1 &        $>$ 5.0    &       $>$ 1.1 \\
      HD209975   & 32.1 $\pm$ 1.2  & 3.33 $\pm$ 0.13   &       $<$ 13.3  &        $>$ 12.3 &        $>$ 13.6   &       $<$ 1.2 \\    \hline
    \end{tabular} 
    \end{small}
    \end{center}
    \end{table*}
    
    \begin{table*}
    \begin{center}
    \caption{As Table~\ref{tbl:iacob_OP_cl2}, but using the Linear$_{20-050}$ clumping law. \label{tbl:iacob_OP_cl4} }
    \begin{small}
    \begin{tabular}{|c|c|c|c|c|c|c|} \hline
               Star               &       Teff(kK)  &        $\log g$ (cgs) &        $ \log Q$ (cgs)     &         \yhe     &       micro (km/s) &         beta  \\ \hline  
       HD46223   & 43.5 $\pm$ 1.0  & 3.84 $\pm$ 0.07   & -13.3 $\pm$ 0.1   & 0.11 $\pm$ 0.03   & 12.4 $\pm$ 7.4  &       $>$ 1.0   \\
       HD15629   & 41.5 $\pm$ 1.3  & 3.80 $\pm$ 0.09   & -13.2 $\pm$ 0.1   & 0.12 $\pm$ 0.03   &       $<$ 11.4  &       $>$ 1.1   \\
       HD46150   & 40.0 $\pm$ 0.5  & 3.82 $\pm$ 0.04   & -13.5 $\pm$ 0.2   & 0.10 $\pm$ 0.03   & 9.9 $\pm$ 4.9   &       $>$ 0.8   \\
      HD217086   & 37.0 $\pm$ 0.8  & 3.61 $\pm$ 0.12   & -14.5 $\pm$ 1.2   & 0.13 $\pm$ 0.04   & 12.4 $\pm$ 7.4  &       $<$ 1.2   \\
      HD149757   & 32.7 $\pm$ 0.8  & 3.85 $\pm$ 0.16   & -14.6 $\pm$ 1.1   & 0.12 $\pm$ 0.04   & 12.3 $\pm$ 7.3  &       $<$ 1.2   \\
      HD190864   & 37.5 $\pm$ 0.7  & 3.58 $\pm$ 0.09   & -13.2 $\pm$ 0.1   & 0.16 $\pm$ 0.05   & 13.4 $\pm$ 6.4  &       $>$ 1.1   \\
      HD203064   & 35.0 $\pm$ 0.5  & 3.58 $\pm$ 0.09   & -13.2 $\pm$ 0.1   & 0.11 $\pm$ 0.03   &       $>$ 14.3  &       $>$ 0.9   \\
       HD15570   & 37.2 $\pm$ 0.7  & 3.29 $\pm$ 0.04   & -12.6 $\pm$ 0.1   & 0.15 $\pm$ 0.03   &       $<$ 19.9  &       $>$ 1.1   \\
       HD14947   & 37.3 $\pm$ 1.2  & 3.36 $\pm$ 0.11   & -12.6 $\pm$ 0.1   &     $>$ 0.18     &       $<$ 19.9  & 1.1 $\pm$ 0.1   \\
       HD30614   & 28.9 $\pm$ 0.3  & 2.81 $\pm$ 0.04   & -12.9 $\pm$ 0.1   & 0.25 $\pm$ 0.03   &       $>$ 17.6  &       $<$ 0.9   \\
      HD210809   & 30.7 $\pm$ 0.5  & 2.99 $\pm$ 0.06   & -12.8 $\pm$ 0.1   & 0.26 $\pm$ 0.04   &       $>$ 15.8  & 1.0 $\pm$ 0.2   \\
      HD209975   & 31.6 $\pm$ 0.7  & 3.29 $\pm$ 0.13   & -13.2 $\pm$ 0.2   & 0.11 $\pm$ 0.03   & 10.1 $\pm$ 5.1  &       $>$ 0.9   \\      
    \hline
    \end{tabular} 
    \end{small}
    \end{center}
    \end{table*}                                                                                                                 
    
    \begin{table*}
    \begin{center}
    \caption{As Table~\ref{tbl:iacob_OP_cl2}, but for the NIR using the Linear$_{20-050}$ clumping law.  \label{tbl:iacob_IR_cl4}}
    \begin{small}
    \begin{tabular}{|c|c|c|c|c|c|c|}\hline
    Star               &       Teff(kK)  &        $\log g$ (cgs)  &      $\log Q$ (cgs)     &         \yhe     &       micro (km/s) &         beta  \\ \hline                                                                                                        
       HD46223   & 50.0 $\pm$ 4.0  & 3.92 $\pm$ 0.10   & -13.0 $\pm$ 0.1  &          $<$ 0.07   &  12.4 $\pm$ 7.4    &       $>$ 0.8 \\
       HD15629   & 41.0 $\pm$ 0.8  & 3.80 $\pm$ 0.11   & -14.4 $\pm$ 1.3  &          $<$ 0.09   &         $<$ 19.9   &       $<$ 1.2 \\
       HD46150   & 40.6 $\pm$ 1.2  & 3.93 $\pm$ 0.15   & -14.3 $\pm$ 1.4  &          $<$ 0.09   &         $>$ 5.0    &       $<$ 1.2 \\
      HD217086   & 37.2 $\pm$ 1.3  & 3.93 $\pm$ 0.18   &       $<$ -13.3  &          $<$ 0.25  &         $>$ 5.0    & 1.0 $\pm$ 0.2 \\
      HD149757   & 31.6 $\pm$ 1.6  & 3.41 $\pm$ 0.23   &       $<$ -13.2  &          $>$ 0.06   &  12.5 $\pm$ 7.3    & 1.0 $\pm$ 0.2 \\
      HD190864   & 37.5 $\pm$ 1.3  & 3.66 $\pm$ 0.18   &       $<$ -12.9  &     0.19 $\pm$ 0.10 &  12.4 $\pm$ 7.4    &       $<$ 1.2 \\
      HD203064   & 35.5 $\pm$ 2.0  & 3.60 $\pm$ 0.24   &       $<$ -13.0  &   0.18 $\pm$ 0.12  &         $>$ 5.0    &       $>$ 0.8 \\
       HD15570   & 39.7 $\pm$ 1.5  & 3.66 $\pm$ 0.11   & -12.6 $\pm$ 0.1  &   0.10 $\pm$ 0.03   &         $>$ 5.0    &       $>$ 1.0 \\
       HD14947   & 42.5 $\pm$ 1.6  & 4.24 $\pm$ 0.24   & -12.8 $\pm$ 0.1  &         $>$ 0.21   &        $<$ 19.9    &       $>$ 1.1 \\
       HD30614   & 27.1 $\pm$ 1.2  & 2.71 $\pm$ 0.12   & -12.8 $\pm$ 0.1  &  0.17 $\pm$ 0.11   &         $>$ 5.0    &       $>$ 0.8 \\
      HD210809   & 37.3 $\pm$ 1.2  & 4.22 $\pm$ 0.23   & -12.8 $\pm$ 0.1  &   0.21 $\pm$ 0.08   &         $>$ 5.0    &       $>$ 0.9 \\
      HD209975   & 32.1 $\pm$ 1.2  & 3.32 $\pm$ 0.14   &       $<$ -13.3  &         $>$ 0.12   &        $>$ 13.6    &       $<$ 1.2 \\    \hline
    \end{tabular} 
    \end{small}
    \end{center}
    \end{table*}
    
\begin{figure*}
\begin{center}
\includegraphics[width=13cm,angle=90]{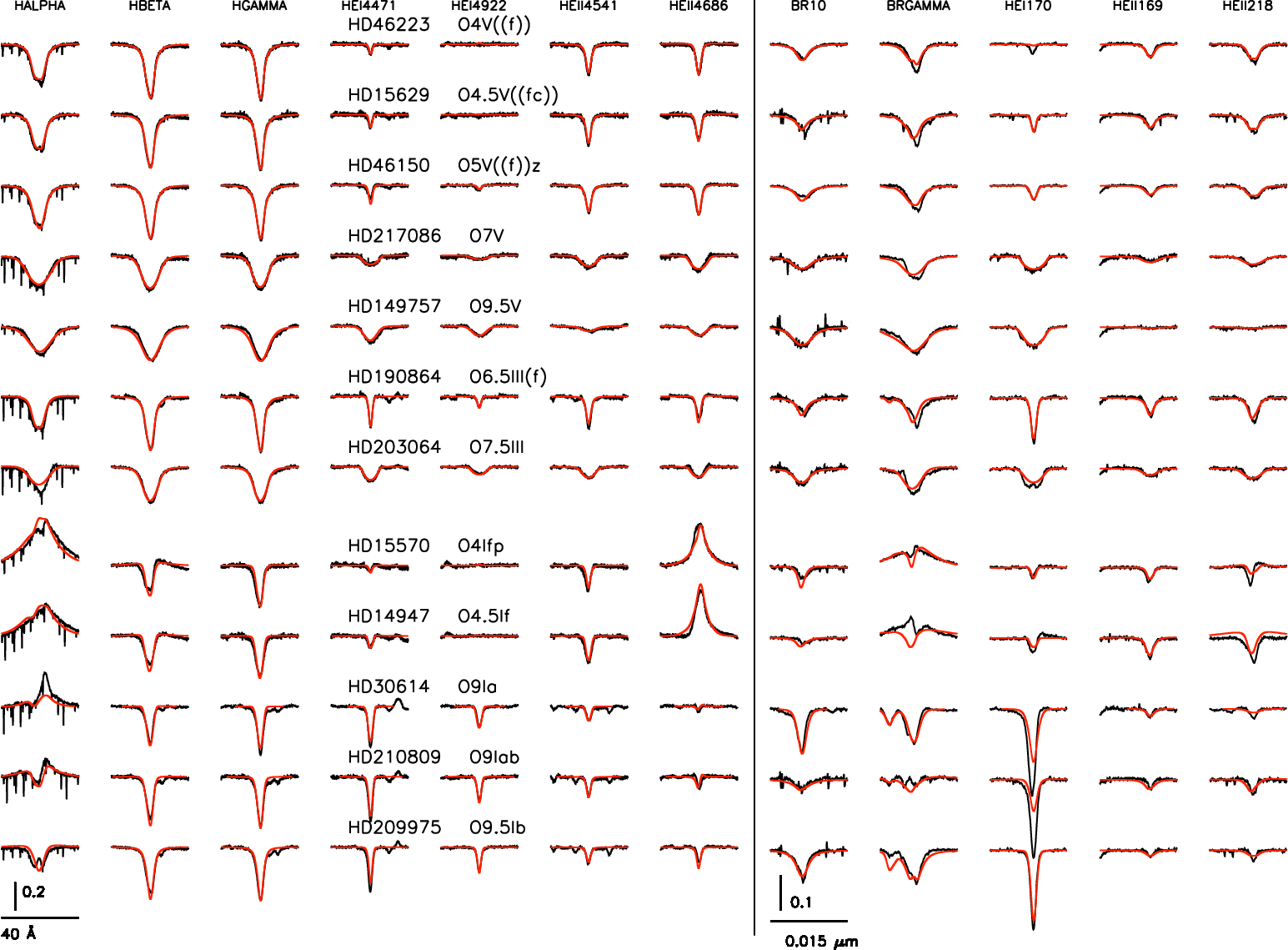}
\caption{Spectral fits for selected optical (left) and NIR (right) lines using the clumping law
Linear$_{10-050}$.
Observations are shown in black, and best fit model profiles in red.
We stress that the individual model parameters for the best fitting
optical and NIR profiles differ (to various extent) since the analyses have been
performed separately for both ranges (cf. Table~\ref{tbl:iacob_OP_cl2} vs.
Table~\ref{tbl:iacob_IR_cl2}). The horizontal bar gives the wavelength scale
for each range, and the scale of the ordinate axis is given by the
vertical bar (at the bottom of the H$_\alpha$
column for the optical range, and at the bottom of
the Br$_{10}$ column for the NIR.)}
\label{fig:OP_cl2_profiles}
\end{center}
\end{figure*}


\begin{figure*}[t!]
\begin{center}
\includegraphics[width=13cm,angle=90]{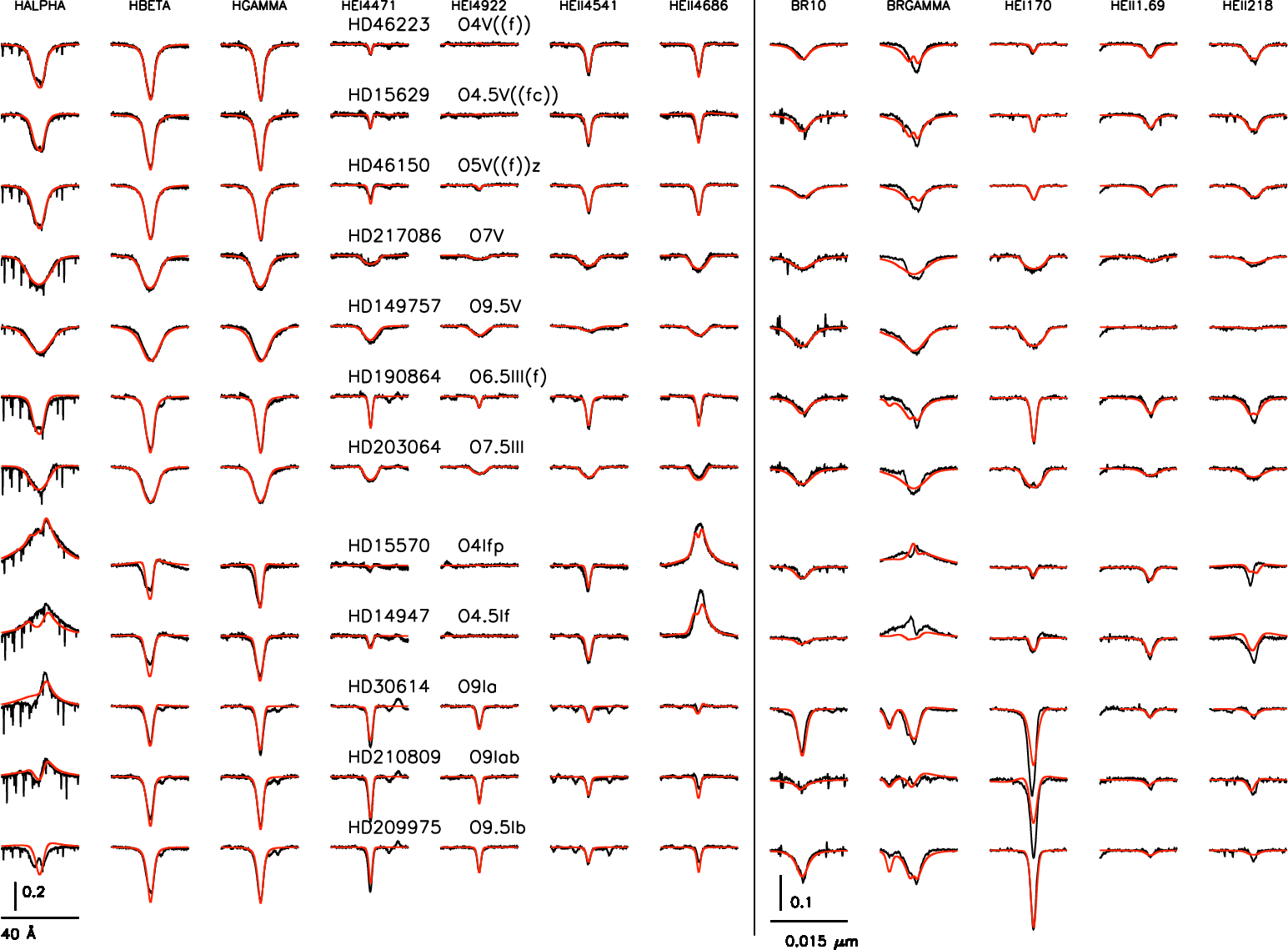}
\caption{As Fig.~\ref{fig:OP_cl2_profiles}, but using the clumping law
Linear$_{20-025}$.}
\label{fig:OP_cl3_profiles}
\end{center}
\end{figure*}


\begin{figure*}[t!]
\begin{center}
\includegraphics[width=13cm,angle=90]{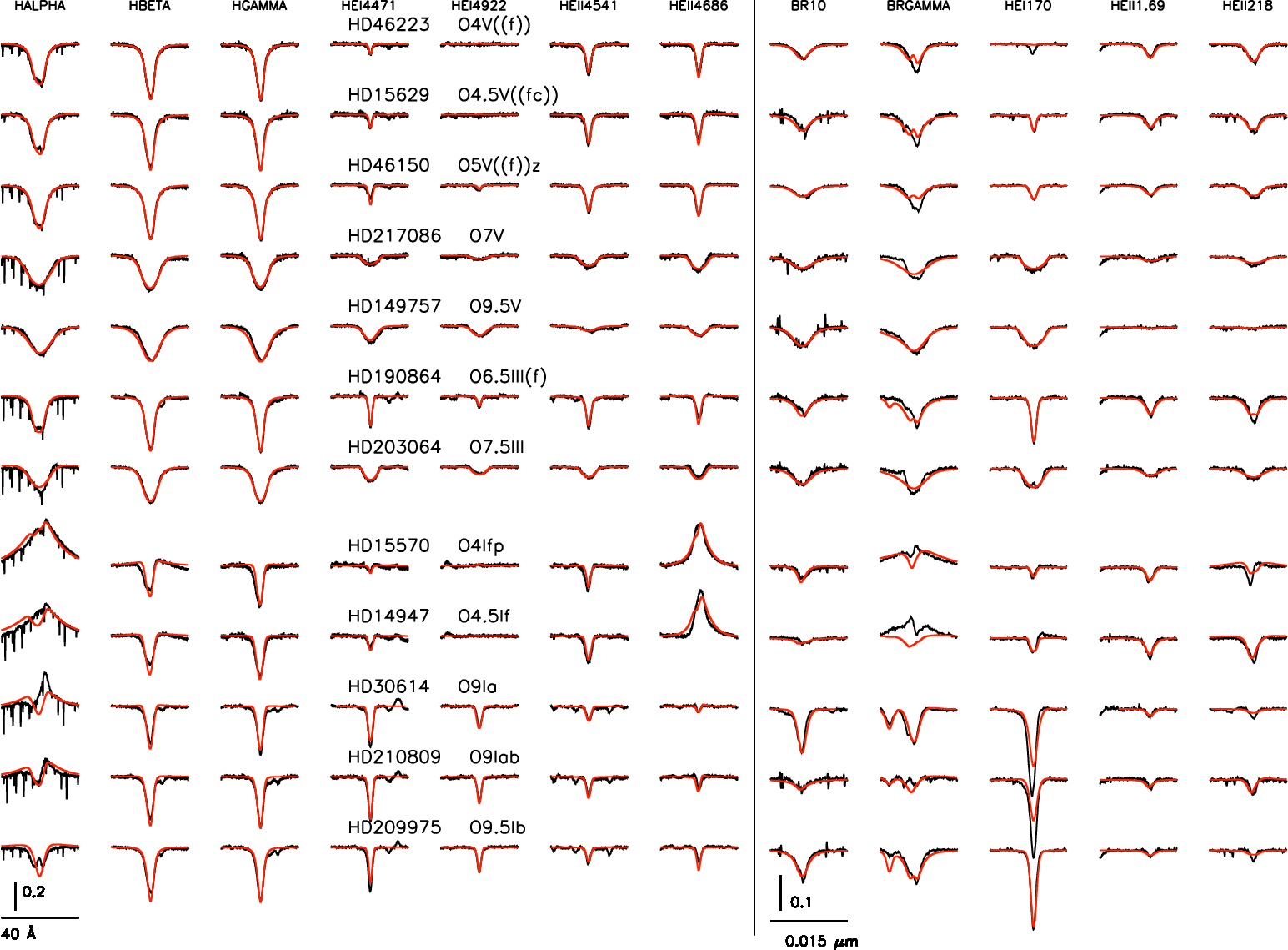}
\caption{As Fig.~\ref{fig:OP_cl2_profiles}, but using the clumping law
Linear$_{20-050}$.}
\label{fig:OP_cl4_profiles}
\end{center}
\end{figure*}


\end{document}